%% file: main.tex
\documentclass[a4paper,twocolumn,11pt,accepted=2025-11-30]{quantumarticle}
\pdfoutput=1
\usepackage{etoolbox}

\usepackage[utf8]{inputenc}
\usepackage[english]{babel}
\usepackage[T1]{fontenc}
\usepackage{amsmath}
\usepackage{hyperref}
\usepackage{braket}
\usepackage{dsfont}
\usepackage[numbers,sort&compress]{natbib}
\usepackage{etoolbox}
\usepackage{tikz}
\usetikzlibrary{quantikz2}
\usepackage{lipsum}

\definecolor{worange}{HTML}{FD8F29} 
\definecolor{wblue}{HTML}{3D8EDD}

\makeatletter
\@ifundefined{UseOneTimeHook}{%
  \newcommand\UseOneTimeHook[1]{}%
}{}%

\@ifundefined{@afterenddocumenthook}{%
  \let\@afterenddocumenthook\@empty
}{}%
\makeatother

\begin{document}
  \title{Ladder Operator Block-Encoding}

  \author{William A. Simon}
  \affiliation{Department of Physics and Astronomy, Tufts University, Medford, MA 02155, USA}
  \email{william.andrew.simon@gmail.com}
  \orcid{0000-0002-1832-4690}

  \author{Carter M. Gustin}
  \affiliation{Department of Physics and Astronomy, Tufts University, Medford, MA 02155, USA}
  \orcid{0009-0007-7853-057X}

  \author{Kamil Serafin}
  \affiliation{Department of Physics and Astronomy, Tufts University, Medford, MA 02155, USA}
  \orcid{0000-0002-9644-2588}

  \author{Alexis Ralli}
  \affiliation{Department of Physics and Astronomy, Tufts University, Medford, MA 02155, USA}
  \orcid{0000-0001-8953-1235}

  \author{Gary R. Goldstein}
  \affiliation{Department of Physics and Astronomy, Tufts University, Medford, MA 02155, USA}
  \orcid{0000-0002-0920-8540}

  \author{Peter J. Love}
  \affiliation{Department of Physics and Astronomy, Tufts University, Medford, MA 02155, USA}
  \orcid{0000-0002-8344-0532}

  \maketitle

  \input{text/abstract}
  \input{text/introduction}
  \input{text/theory}

  \input{text/block-encoding}
  \input{text/lobe/main}

  \input{text/results/main}

  \input{text/conclusions}
  \input{text/acknowledgements}

  \onecolumn\newpage
  \appendix

  \input{text/appendices/glossary}
  \input{text/appendices/usp}

  \input{text/appendices/compiling_toffolis}

  \input{text/appendices/addition}
  \input{text/appendices/pauli_transform}
  \input{text/appendices/uniformly_controlled_rotations}
  \input{text/appendices/grover_rudolph}
  \input{text/appendices/qft}

\end{document}

%% file: text/abstract.tex
\begin{abstract}
\label{abstract}
We describe and analyze LOBE (Ladder Operator Block-Encoding), a framework for block-encoding ladder operators that act upon fermionic and bosonic modes.
In this framework, we achieve efficient block-encodings by applying the desired action of the operator onto the quantum state and pushing any undesired effects outside of the encoded subspace.
This direct approach avoids any overhead caused by expanding the operators in another basis.
We numerically benchmark these constructions using models arising in quantum field theories including the quartic harmonic oscillator, and $\phi^4$ and Yukawa Hamiltonians on the light front.
These benchmarks show that LOBE often produces block-encodings with fewer non-Clifford operations, fewer block-encoding ancillae and overall number of qubits, and lower rescaling factors for various operators as compared to frameworks that expand the ladder operators in the Pauli basis.
LOBE constructions also demonstrate favorable scaling with respect to key parameters, including the maximum occupation of bosonic modes, the total number of fermionic and bosonic modes, and the locality of the operators.
LOBE is implemented as an open-source python package to enable further applications.
\end{abstract}

%% file: text/introduction.tex
\label{sec:intro}

The simulation of many-body quantum systems is a promising application for future quantum computers. 
Following the original suggestion of Feynman~\cite{feynman2018simulating}, many general early algorithms were proposed~\cite{lloyd1996universal,meyer1996quantum,boghosian1997quantum,abrams1999quantum}.
Subsequently, detailed proposals were developed for applications in specific domains, particularly quantum chemistry~\cite{lidar1999calculating,terhal2000problem,wu2002polynomial,aspuru2005simulated,kassal2008polynomial} and the simulation of quantum field theories~\cite{wiese2014towards,jordan2012quantum}. 
Quantum simulation is now a broad area, with well developed algorithms for many applications~\cite{cao2019quantum,mcardle2020quantum,bauer2023quantum}.

Quantum simulation algorithms are composed as a series of unitary operations.
Accessing the information of non-unitary operators - such as the Hamiltonian or other Hermitian observables - within a quantum algorithm is a necessary subroutine for conducting such simulations.
This encoding task has been pursued through various means, resulting in methods such as Trotterization \cite{suzuki1976generalized,hatano2005finding,lie1893theorie,trotter1959product,childs2021theory} and Block-Encoding \cite{lin2022lecture, poulin2018quantum, low2019hamiltonian}.

Block-Encoding describes a general strategy for encoding a non-unitary matrix within a chosen subspace (block) of a larger unitary operator.
Two frameworks have been proposed for explicitly constructing block-encodings of different operators: Sparse-Oracle block-encodings \cite{berry2009black, childs2009universal, lin2022lecture} and Linear Combinations of Unitaries (LCU) \cite{childs2012hamiltonian}.
These frameworks allow us to compile algorithms using block-encodings into discrete gate sets.

Compiling algorithms into discrete gates leads to numerical resource estimates.
These resource estimates establish how big of a quantum computer is needed to be to run the algorithm.
Generating and reducing numerical resource estimates for fault-tolerant quantum algorithms is particularly important given recent experimental demonstrations of quantum error correction  \cite{bluvstein2024logical, acharya2024quantum}.
Resource estimates for quantum simulation of purely fermionic systems have been heavily investigated, with a particular emphasis on the simulation of molecules in quantum chemistry \cite{aspuru2005simulated, peruzzo2014variational, babbush2014adiabatic, o2016scalable, babbush2018encoding, google2020hartree, lee2021even, kivlichan2020improved, campbell2021early}.

Simulating quantum field theories is another promising domain \cite{Peskin:1995ev, jordan2012quantum, girgus2024, bender1969}, with applications in areas such as high-energy physics \cite{bauer2023quantum}.
These systems often include interactions between fermions, antifermions, and bosons and several works have produced resource estimates for simulating such systems \cite{Kirby_2021, camps2024explicit, liu2024efficient, rhodes2024exponential, hariprakash2025strategies, du2024systematicinputschememanyboson, halimeh2024universal, Peng_2025}. 

In particular, Kirby et al. \cite{Kirby_2021} detail a compact-encoding scheme for ladder-operators that produces near-optimal scaling for the number of qubits used to encode the system.
Rhodes et al. \cite{rhodes2024exponential} construct near-optimal methods for simulating lattice gauge theories by block-encoding the Hamiltonian written in its second-quantized form, leveraging the inherent structure in the system.
Lastly, Peng et al. \cite{Peng_2025} propose block-encoding constructions for Hamiltonians that model interactions between bosonic modes, fermionic modes, and spin modes.

In this work, we detail a general framework for constructing block-encodings of ladder operators, which we refer to as Ladder Operator Block-Encoding (LOBE).
Relevant observables in many formulations of quantum mechanics are expressed in terms of ladder operators, making them a particularly useful operator basis.
In LOBE, states are directly updated based on the inferred action of the operator. 
Any undesired effects on the state are rotated outside of the encoded subspace using the block-encoding ancillae, drawing inspiration from the Sparse-Oracle framework.
This direct method avoids any potential overhead caused by transforming fermionic \cite{jordan-wigner, bravyi2002fermionic, seeley2012bravyi} and bosonic \cite{somma2005quantum,standard-binary} ladder operators in the Pauli basis, leading to efficient circuit constructions.

We provide numerical resource estimates for implementing block-encodings of several classes of operators and Hamiltonians that arise in quantum field theories.
This includes classes of operators that model nontrivial interactions between fermionic, antifermionic, and bosonic modes. 
We analyze space-time quantum resources for LOBE and compare these estimates to techniques that require transforming ladder operators into the Pauli basis.
Our results show that LOBE constructions often require fewer resources and have favorable scaling for several physically-motivated parameters for many of the systems we examine.

This work is organized as follows.
In section \ref{sec:theory}, we review the definition of fermionic and bosonic ladder operators.
In section \ref{sec:block-encoding}, we review block-encodings and discuss frameworks for constructing block-encodings of different operators. 
In section \ref{sec:ladder-op-oracles}, we describe the LOBE framework, show compiled block-encodings for several classes of operators, and give analytical space-time costs of the associated constructions. 
In section \ref{sec:results}, we provide numerical resource estimates for block-encodings of various classes of operators and Hamiltonians.
In section \ref{sec:conclusions}, we summarize the results presented in this work and discuss future directions. 

%% file: text/theory.tex
\section{Ladder Operators}
\label{sec:theory}

Quantum operators in many-body quantum mechanics and quantum field theory are often described in second quantization \cite{berezin1966method}.
In this form, multiparticle states are written in terms of fermionic, antifermionic, and bosonic modes which can be occupied by different numbers of particles.
The occupation number basis is defined by the product of occupation numbers of each of the respective modes: $\ket{n_i} = \ket{n_{I-1}, \dots, n_1, n_0}$ where $n_i$ is the number of particles in the $i^\text{th}$ mode.

The space spanned by the occupation basis is called the \textit{Fock space} ($\mathcal{F}$) and the state vectors are referred to as \textit{Fock states}.
The Fock space is a direct sum of $n$-particle sectors \cite{Schwartz_2013}.
In quantum field theories, field operators act on second-quantized states to create and annihilate particles in the field.
This action is described in terms of ladder operators.

Ladder operators are quantum operators that act on fermionic, antifermionic, and bosonic modes to either increase (create) or decrease (annihilate) the number of particles occupying a particular mode.
Many observables of interest for these systems, such as the Hamiltonian, can be efficiently expressed as sums or products of ladder operators.

\subsection{Bosonic Ladder Operators}
\label{subsec:bosonic-ladder}

There is no physical limitation on the number of particles in a bosonic mode.
This makes the Hilbert space of a single bosonic mode countably infinite dimensional. 
The commutation rules for the bosonic annihilation ($a_i$) and creation ($a_i^\dagger)$ operators are given by:
\begin{equation}
    \label{eq:bosonic-commutation}
    \begin{split}
        &[a_i, a_j^\dagger] = \delta_{ij}\\
        & [a_i, a_j] = [a_i^\dagger, a_j^\dagger] = 0 \\
    \end{split}
\end{equation}
Bosonic ladder operators also commute with fermionic (and antifermionic) ladder operators.

For discretized computations, a cutoff on the bosonic occupancy ($\Omega$) is chosen to make the dimension of the Hilbert space finite: $\omega_i \in [0, 1, 2, \dots, \Omega]$.
The cutoff on the bosonic occupancy can introduce error as some physically allowable states become computationally inaccessible.
Since there is no finite dimensional realization of the bosonic ladder operators,  simulations of bosonic systems often proceed by increasing the occupation cutoff ($\Omega$) until the error induced by this cutoff is either negligible or well understood.

With the imposed bosonic cutoff, the action of bosonic creation operator ($a_i^\dagger$) is defined by:
\begin{equation}
    \label{eq:bosonic-creation}
    a_i^\dagger \ket{\omega_i} = 
    \begin{cases} 
        \sqrt{\omega_i + 1} \ket{\omega_i + 1}  & {\rm when} \ket{\omega_i} \neq \ket{\Omega} \\
        0 & {\rm when} \ket{\omega_i} = \ket{\Omega}
    \end{cases}
\end{equation}
where $a_i^\dagger$ denotes a bosonic creation operator on the $i^{th}$ mode, $\ket{\omega_i}$ is the occupation of the $i^{th}$ bosonic mode, and $\Omega$ is the maximum number of bosons allowed in a single mode.

The action of the bosonic annihilation operator acting on the $i^\text{th}$ mode ($a_i$) is defined by:
\begin{equation}
    \label{eq:bosonic-annihilation}
    a_i \ket{\omega_i} = 
    \begin{cases} 
        \sqrt{\omega_i} \ket{\omega_i - 1}  & {\rm when} \ket{\omega_i} \neq \ket{0} \\
        0 & {\rm when} \ket{\omega_i} = \ket{0}
    \end{cases}
\end{equation}

\subsection{Fermionic Ladder Operators}
\label{subsec:fermionic-operators}

Fermions (and antifermions) obey the Pauli-exclusion principle \cite{pauli1925zusammenhang}.
Therefore, fermionic (or antifermionic) occupation numbers are either $0$ or $1$.
Fermionic ladder operators act on a single mode as:
\begin{equation}
    \label{eq:fermionic-creation}
    b_i^\dagger \ket{n_{i}} = 
    \begin{cases} 
        p(n)\ket{1}  & {\rm when} \ket{n_{i}} = \ket{0} \\
        0 & {\rm when} \ket{n_{i}} = \ket{1}
    \end{cases}
\end{equation}
where $b_i^\dagger$ denotes a fermionic creation operator on the $i^{th}$ mode, $\ket{n_{i}}$ is the occupation of the $i^{th}$ fermionic mode, and $p(n)$ denotes a potential sign flip described below.

The action of the fermionic annihilation operator is given by:
\begin{equation}
    \label{eq:fermionic-annihilation}
    b_i \ket{n_{i}} = 
    \begin{cases} 
        p(n) \ket{0}  & {\rm when} \ket{n_{i}} = \ket{1} \\
        0 & {\rm when} \ket{n_{i}} = \ket{0}
    \end{cases}
\end{equation}
where $b_i$ denotes a fermionic annihilation operator on the $i^{th}$ mode

The potential sign flip caused by the fermionic ladder operators, $p(n)$, is determined by the parity of the occupation of the preceeding modes: 
\begin{equation}
    \label{eq:parity}
    p(n) = (-1)^{\sum_{j < i} n_{j}}
\end{equation}
If the number of occupied fermionic modes with index $j < i$ is odd, then the sign of the output state is flipped.

The fermionic ladder operators can be reordered arbitrarily with the introduction of additional terms due to the following anticommutation rules:
\begin{equation}
    \label{eq:fermionic-commutation}
    \begin{split}
        &\{b_i, b_j^\dagger\} = \delta_{ij}\\
        & \{b_i, b_j\} = \{b_i^\dagger, b_j^\dagger\} = 0 \\
    \end{split}
\end{equation}
where $\{\}$ denotes the anticommutator.

In addition to fermions, antifermions are often included in quantum field theories.
The properties and commutation rules for antifermions are the same as for fermions and the sign flip for antifermions includes the parity of the occupation of the fermionic modes.
For simplicity, we treat antifermionic modes as fermionic modes that are indexed after all native fermionic modes \cite{kreshchuk2022quantum}.

\subsection{Observables}
\label{subsec:observables}

Ladder operators create a natural basis for expressing many relevant operators, such as Hamiltonians derived in second quantization in both quantum chemistry and quantum field theory.
Observables can be expressed as linear combinations of products of ladder operators acting on both fermionic and bosonic modes.

Normal ordering is a convention for expressing products of ladder operators wherein all creation operators appear to the left of all annihilation operators.
We say an operator ($O$) is \textit{mode ordered} when all ladder operators acting on a single mode are grouped next to one another and normal ordered:
\begin{equation}
    \label{eq:mode-ordered}
    O = \Big( \prod_i (b_i^\dagger)^{\delta_{b_i}^{\dagger}} (b_i)^{\delta_{b_i}} \Big) \Big( \prod_{i^\prime} (a_{i^\prime}^\dagger)^{R_{i^\prime}}(a_{i^\prime})^{S_{i^\prime}} \Big) 
\end{equation}
where $i$ and ${i^\prime}$ index the fermionic and bosonic modes respectively, $\delta$ takes the value $0$ or $1$ to denote if the individual ladder operator is included in the term, and $R_{i^\prime}$ ($S_{i^\prime}$) is an integer in the range $[0, \Omega]$ denoting the exponent of the bosonic creation (annihilation) operator acting on the $i^{\prime^\text{th}}$ bosonic mode.

Any term may be normal ordered or mode ordered using the commutation rules (Eq. \eqref{eq:bosonic-commutation} and Eq. \eqref{eq:fermionic-commutation}).
In general, this will introduce more terms when reordering, e.g., $a_ia_i^\dagger = a_i^\dagger a_i + 1$.
We will default to expressing operators in their \emph{mode ordered} form.

%% file: text/block-encoding.tex
\section{Block-Encoding}
\label{sec:block-encoding}

Quantum algorithms are composed as a sequence of unitary operations.
However, it is often necessary to access information about various non-unitary operators within a quantum algorithm.
``Block-Encoding" refers to an access model where the information regarding non-unitary operators is encoded in a labeled subspace (block) of a larger unitary operator \cite{childs2012hamiltonian, chakraborty2019power}.

If $A$ represents some $N_A \times N_A$ non-unitary matrix, then a block-encoding of $A$ in matrix form is given by:
\begin{equation}
    U_A = 
    \begin{pmatrix}
    \bar{A} & * \\
    * & * 
    \end{pmatrix}
\end{equation}
where $U_A$ is a unitary operator, $\bar{A}$ is a rescaled form of $A$ such that the eigenvalues of $U_A$ have magnitude $1$, and matrix entries $*$ denote elements that ensure $U_A$ is unitary.

The action of $U_A$ on an arbitrary quantum state ($\ket{\psi}$) with dimension $N_A$ tensored with an ancilla register beginning in the all-zero state can be written as:
\begin{equation}
    \label{eq:general-block-encoding}
    U_A \ket{\psi} \ket{0}_{\text{anc}} = \bar{A} \ket{\psi} \ket{0}_{\text{anc}} + \beta_\psi \ket{\perp}
\end{equation}
where $\ket{\phi}_{\text{anc}}$ is a register of block-encoding ancillae used to generate $U_A$, $\ket{\perp}$ is a normalized quantum state that is orthogonal to any state in the encoded subspace, and $\beta_\psi$ is a complex coefficient.

In the above equations and for all block-encodings in this work, the encoded subspace is chosen (without loss of generality) to be the subspace where all block-encoding ancillae are in the zero state ($\ket{0}$).

It is important to note that $\beta_\psi$ is dependent on the initial state of the system register ($\ket{\psi}$).
The dependence of $\beta$ on $\ket{\psi}$ can be seen in a simple case where $\ket{\psi}$ is an eigenstate of $\bar{A}$.
If $\ket{\gamma_k}$ is an eigenstate of $\bar{A}$ with eigenvalue $k$ (where $|k| < 1$ such that $U_A$ can be made unitary), then Eq. \eqref{eq:general-block-encoding} leads to:
\begin{equation}
    U_A \ket{\gamma_k} \ket{0}_{\text{anc}} = k \ket{\gamma_k} \ket{0}_{\text{anc}} + \sqrt{1 - |k|^2} \ket{\perp}
\end{equation}
where $\sqrt{1 - |k|^2}$ is clearly dependent upon the eigenvalue associated with the eigenstate.

One way to interpret the action of the block-encoding operator ($U_A$) of a matrix ($A$) is that it produces a probabilistic application of the rescaled matrix ($\bar{A}$).
After the block-encoding is applied, a measurement of the block-encoding ancillae resulting in all block-encoding ancillae being in the zero state implies $\bar{A}$ has been applied to the system.
The probability for this to occur is given by $||\bar{A} \ket{\psi}||^2$.

The rescaling of $A$ is given by:
\begin{equation}
    A = \lambda \bar{A}
\end{equation}
where $\lambda$ is referred to as either the subnormalization constant \cite{gilyen2019quantum} or the rescaling factor \cite{obrien2022}, and we will use the latter term in this work.

The rescaling factor has important implications for the cost of quantum algorithms. 
Smaller rescaling factors generally lead to more efficient algorithms, though the impact is algorithm-dependent.
For example, the success probability of applying $\bar{A}$ as described above decreases as $\lambda$ increases.
If the block-encoding is used to construct a quantum walk operator \cite{low2019hamiltonian, poulin2018quantum} and used in Quantum Phase Estimation \cite{poulin2018quantum, babbush2018encoding, lee2021even}, then the output eigenvalue estimates are rescaled by $\lambda$ which will affect both the error and the precision of the estimated value.

In the following subsections, we describe some commonly used frameworks for constructing block-encodings.
We discuss both the Sparse-Oracle and Linear Combination of Unitaries (LCU) frameworks.
Then we discuss methods to combine block-encodings to produce block-encodings for linear combinations of operators or products of operators.

\subsection{Sparse-Oracle Framework}
\label{subsec:sparse-be}

The Sparse-Oracle framework \cite{berry2009black, childs2009universal, berry2015hamiltonian,berry2015simulating, low2017optimal,childs2017quantum,gilyen2019quantum} uses oracles that provide both the location and the values of the nonzero matrix elements in $A$.
In this subsection, we primarily follow the description of this framework given by Lin \cite{lin2022lecture}.

When an operator is sparse within some known basis, the Sparse-Oracle framework can use this structure to reduce the cost of the block-encoding.
Despite the use of the word ``oracle", a method for generating quantum circuits that implement the required oracles for general matrices is given in \cite{camps2024explicit, camps2022fable}.
Several works have followed these constructions to explore explicitly compiled Sparse-Oracle block-encodings of particular systems \cite{camps2022fable, liu2024efficient, sanavio2024explicit}.

Let $A$ be an $s$-sparse matrix, such that there are a maximum of $s$ nonzero entries in a single row and each matrix element ($A_{ij}$) is restricted to have magnitude $\leq 1$.
A Sparse-Oracle block-encoding for the matrix $A$ is defined in terms of three oracles: $D_s$, $O_A$, and $O_c$.

The ``diffusion operator'' $(D_s)$ is defined by:
\begin{equation}
    \label{eq:diffusion}
    D_s \ket{0^{\log_2 \mathcal{S}}} = \frac{1}{\sqrt{\mathcal{S}}} \sum_{l=0}^{\mathcal{S}-1} \ket{l}
\end{equation}
where $\mathcal{S} = 2^{\lceil \log_2{s} \rceil}$ and $\ket{0^{\log_2 \mathcal{S}}}$ is a block-encoding ancilla register that the diffusion operator acts on.
This oracle can be implemented by a tensor product of Hadamards: $H^{\otimes \log_2{\mathcal{S}}}$.

The ``column oracle'' ($O_c$) is defined by: 
\begin{equation}
    \label{eq:column-oracle}
    O_c \ket{j} \ket{l} = \ket{j}\ket{c(j, l)}
\end{equation}
where $c(j, l)$ is a function that returns the row-index of the $l^\text{th}$ nonzero matrix element in the $j^\text{th}$ column.

Lastly, the ``value oracle'' ($O_A$) is defined by:
\begin{equation}
    \label{eq:value-oracle}
    O_A \ket{0} \ket{j}\ket{i} = \big( A_{ij} \ket{0}  + \beta_{ij} \ket{1} \big) \ket{j}\ket{i}
\end{equation}
where $\beta_{ij} \equiv \sqrt{1 - |A_{ij}|^2}$.
The value oracle rotates the block-encoding ancilla such that the amplitude of the state left within the encoded subspace matches the desired coefficient following Eq. \eqref{eq:general-block-encoding}.

With these three oracles, a block-encoding for $A$ is given by:
\begin{equation}
    \label{eq:so-be}
    U_A = D_s O_c O_A D_s
\end{equation}
Simple proofs that $U_A$ block-encodes $A$ are given in Camps et al. \cite{camps2024explicit} and Gilyén et al. \cite{gilyen2019quantum}.

The rescaling factor for this approach is:
\begin{equation}
    \label{eq:rescaling-so}
    \lambda_\text{SO} = 2^{\log_2 \mathcal{S}} \max_{ij} {|A_{ij}|} 
\end{equation}
where the factor of $2^{\log_2 \mathcal{S}}$ comes from the diffusion operator and is limited by the sparsity of $A$.
The factor of $\max_{ij} {|A_{ij}|}$ comes from the constraint that all matrix elements ($A_{ij}$) are restricted to have magnitude $\leq 1$.

If $s$ is not a power of two, then the rescaling factor can be reduced to $\lambda_\text{SO} = 2^{\log_2 s} \max_{ij} {|A_{ij}|}$ by replacing the diffusion operator ($D_s$) with the generalized \textit{Uniform State Preparation} (USP) protocol, which is discussed in Appendix \ref{sec:usp}.
Implementing USP typically requires more nontrivial operations than the diffusion operator, creating a trade-off between reducing the rescaling factor and reducing the number of nontrivial operations.

\subsection{Linear Combination of Unitaries}
\label{subsec:lcu}

An alternative framework proposed by Childs et al. \cite{childs2012hamiltonian} called Linear Combination of Unitaries (LCU) generates block-encodings of operators that can be written in the following form:
\begin{equation}
    \label{eq:lcu}
    A = \sum_{l=0}^{L-1} \alpha_l U_l
\end{equation}
where $\alpha_l$ are positive, real-valued coefficients and each $U_l$ is a unitary operator.
The phase of a coefficient can be absorbed into the operator such that the coefficient is positive and real: $-i \alpha_l U_l = \alpha_l (-i U_l)$.

To generate an LCU block-encoding, two oracles, \textit{Prepare} and \textit{Select}, are required.
LCU block-encodings also require a block-encoding ancilla register with $\lceil \log_2 L \rceil$ qubits.
This register is often referred to as the \textit{index register} since the computational basis states of this register $(\ket{l})$ index the $L$ terms in Eq. \eqref{eq:lcu}.

The \textit{Prepare} oracle maps the all-zero state of the index register to a particular superposition state that encodes the coefficients of the terms in the operator:
\begin{equation}
    \label{eq:prep-state}
    \textbf{Prepare}\text{:} \ket{0^{\otimes \lceil \log_2 L \rceil}} \rightarrow \sum_{l=0}^{L-1} \sqrt{| \alpha_l | / \lambda_\text{LCU}} \ket{l}
\end{equation}
where $\lambda_\text{LCU}$ is the rescaling factor and is given by:
\begin{equation}
    \label{eq:lambda-lcu}
    \lambda_\text{LCU} = \sum_{l=0}^{L-1} | \alpha_l |
\end{equation}

The \textit{Select} oracle applies the $l^\text{th}$ unitary $(U_l)$ onto the system register when the index register is in the computation basis state $\ket{l}$:
\begin{equation}
    \label{eq:select}
    \textbf{Select}\text{:} 
    \begin{cases} 
        \text{$\ket{\psi}\ket{l} \rightarrow U_l\ket{\psi}\ket{l}$} & \text{when $0 \leq l < L$} \\
        \text{Undefined} & \text{Otherwise} \\
    \end{cases}
\end{equation}
The \textit{Select} oracle can apply any unitary for computational basis states outside of the range $[0, L)$, assuming the \textit{Prepare} oracle satisfies Eq. \eqref{eq:prep-state}.

With these two oracles, a block-encoding for an operator $A$ specified by Eq. \eqref{eq:lcu} is given by:
\begin{equation}
    \label{eq:lcu-be}
    U_A = (\textbf{Prepare}^\dagger) (\textbf{Select}) (\textbf{Prepare})
\end{equation}
A proof that $U_A$ block-encodes $A$ is given in section 7.3 of Lin \cite{lin2022lecture}.

\subsubsection{Implementing \textbf{Prepare}}

The matrix representation of the \textit{Prepare} oracle in the computational basis is given by:
\begin{equation}
    \textbf{Prepare} = \begin{bmatrix}
        \sqrt{|\alpha_0| / \lambda_\text{LCU}} & * & ... & * \\
        \sqrt{|\alpha_1| / \lambda_\text{LCU}} & * & ... & * \\
        \vdots & \vdots & \ddots & \vdots \\
        \sqrt{|\alpha_{L-1} |/ \lambda_\text{LCU}} & * & ... & * \\
        0 & * & ... & * \\
        \vdots & * & ... & * \\
    \end{bmatrix}
\end{equation}
Where the first column is padded with zeroes if $L$ is not a power of $2$.
There are infinitely many unitaries that implement the $\textit{Prepare}$ oracle since only the first column is fixed.

The \textit{Prepare} oracle is a state preparation routine where the target state is a specific superposition state that encodes a probability distribution along the computational basis states.
Many state preparation algorithms exist, however, the Grover-Rudolph algorithm \cite{grover2002creating} is a state preparation algorithm designed for target states of this form (Eq. \eqref{eq:prep-state}).
The cost of implementing Grover-Rudolph scales exponentially with the number of qubits in the register it acts upon.
Since the size of the index register is logarithmic in the number of terms ($L$), the number of operations required to implement Grover-Rudolph is linear in $L$.
A more detailed discussion of the Grover-Rudolph state preparation algorithm is given in Appendix \ref{sec:grover-rudolph}.

For operators that have structure among the coefficients of the terms, this structure can be exploited to reduce the required number of operations.
In certain cases, this can drastically reduce the space-time cost of the block-encoding construction \cite{babbush2018encoding}.

\subsubsection{Implementing \textbf{Select}}

The \textit{Select} oracle can be implemented by the operator:
\begin{equation}
    \label{eq:select-naive}
    U_\text{Select} = \sum_{l=0}^{L-1} \ket{l}\bra{l} \otimes U_l + \sum_{l \geq L} \ket{l}\bra{l} \otimes \mathds{1}
\end{equation}

Here we take the action of $U_\text{Select}$ to be the identity operator for computational basis states of the index register outside the range $[0, L)$.
If this constraint is relaxed or if there is structure in the operator, then a more efficient implementation of $\textit{Select}$ can be constructed \cite{babbush2018encoding}.  

As an aside, if the implementation of the \textit{Select} oracle is self-inverse, then the LCU block-encoding is also self-inverse.
The construction for \textit{Select} in Eq. \eqref{eq:select-naive} is naturally self-inverse if the unitaries themselves are self-inverse, which is true when the operator is decomposed in the Pauli basis.
This structure makes LCU block-encodings particularly well-suited for being applied in algorithms based on Qubitization \cite{low2019hamiltonian}.

\subsection{Linear Combination of Operators}
\label{subsec:lco}

A natural question to ask is if the structure of an LCU block-encoding can be generalized to a linear combination of operators ($O_l)$ where each operator is not necessarily unitary:
\begin{equation}
    \label{eq:lco}
    A = \sum_{l=0}^{L-1} \alpha_l O_l
\end{equation}
and we again restrict the coefficients $\alpha_l$ to be positive and real-valued, without loss of generality.

Following Gilyén et al. \cite{gilyen2019quantum}, an operator of this form (Eq. \eqref{eq:lco}) can be block-encoded using an LCU block-encoding of a linear combination of unitary operators ($U_l$) that each block-encode the operators $O_l$.

Let the set of unitary operators $\{U_l\}$ represent block-encodings of the operators $O_l$:
\begin{equation}
    \label{eq:applying-operator}
    U_l \ket{\psi}\ket{0}_\text{anc} = \bar{O}_l \ket{\psi} \ket{0}_\text{anc} + \beta_{\psi, l} \ket{\perp}
\end{equation}
where each block-encoding uses the same encoded subspace and has a rescaling factor $\lambda_l$ such that $\lambda_l \bar{O}_l = O_l$.
If an operator is unitary, then it trivially block-encodes itself.
Additionally, let the operator $\Tilde{A}$ be defined by the linear combination of the block-encoding unitaries:
\begin{equation}
    \Tilde{A} = \sum_{l} \Tilde{\alpha_l} U_l \hspace{1em} \Tilde{\alpha_l} = \frac{\alpha_l \lambda_l}{\Lambda}
\end{equation} 
where $\Lambda \geq \max_l \lambda_l$.

Then, an LCU block-encoding of $\Tilde{A}$ ($U_{\Tilde{A}}$) will also be a block-encoding of $A$.
We will refer to a block-encoding of this form as a Linear Combination of Operators (LCO).

The fact that $U_{\Tilde{A}}$ block-encodes $A$ can be seen from the linearity of the operators themselves:
\begin{equation}
    \begin{split}
        U_{\Tilde{A}} &= 
        \begin{pmatrix}
        \bar{\Tilde{A}} & * \\
        * & * 
        \end{pmatrix} \propto
        \sum_{l}
        \begin{pmatrix}
        \Tilde{\alpha_l} U_l & * \\
        * & * 
        \end{pmatrix} \\
        & =\sum_{l}
        \begin{pmatrix}
        \begin{pmatrix}
            \alpha_l O_l & * \\
            * & * 
        \end{pmatrix} & * \\
        * & * 
        \end{pmatrix} 
    \end{split}
\end{equation}
It is clear from this form that the block-encodings $\{U_l\}$ must all act in the same subspace as they would otherwise not yield the desired linear combination in the encoded subspace of $U_{\Tilde{A}}$.
It is also worth noting that the block-encodings $\{U_l\}$ may all use the same block-encoding ancillae as long as the implementation of the \textit{Select} oracle ensures that the block-encoding ancillae begin in the all-zero state in the subspace in which each individual block-encoding unitary is applied.

The overall rescaling factor of an LCO block-encoding is given by:
\begin{equation}
    \lambda_\text{LCO} = \sum_{l=0}^{L-1} |\Tilde{\alpha_l}|
\end{equation}
As noted by Jennings et al. \cite{jennings2023efficient}, if $\Lambda = \max_l \lambda_l$, then the overall rescaling factor becomes:
\begin{equation}
    \lambda_\text{LCO} = \sum_{l=0}^{L-1} |\alpha_l \lambda_l|
\end{equation}

The space-time cost for implementing an LCO block-encoding is similar to that of an LCU block-encoding.
The \textit{Prepare} and \textit{Select} oracles can be implemented as described in subsection \ref{subsec:lcu}, however, one must account for the cost of implementing the block-encoding unitaries $\{U_l\}$ when implementing \textit{Select}.
Additionally, if $m$ is the maximum number of block-encoding ancillae used for the block-encoding unitaries $\{U_l\}$, then an LCO block-encoding requires at least $m + \lceil \log_2 L \rceil$ block-encoding ancillae.

As noted in prior works \cite{berry2015simulating, childs2017quantum, gilyen2019quantum, lin2022lecture, jennings2023efficient}, an LCO can be thought of as a generalization of LCU to produce a block-encoding for a linear combination of matrices.
Alternatively, it may be enlightening to consider an LCU block-encoding as a special case of an LCO block-encoding wherein the operators in the linear combination are all unitary and thus block-encode themselves.

\subsection{Products of Operators}
\label{subsec:be-products}

Block-encoding a product of operators can be easily achieved, given access to block-encodings for each operator.
As noted by Gilyén et al. \cite{gilyen2019quantum}, a block-encoding for a product of operators can be constructed by the product of the block-encodings of the individual operators.

Let $A$ represent a product of operators $B$ and $C$:
\begin{equation}
    \label{eq:product}
    A = BC
\end{equation}
with block-encodings $U_B$ and $U_C$.
The product of $U_B$ and $U_C$ results in a block-encoding of $A$:
\begin{equation}
    \begin{split}
        &U_B U_C \ket{\psi} \ket{0}_\text{anc,C}\ket{0}_\text{anc,B} = \\
        &U_B \big(\bar{C} \ket{\psi} \ket{0}_\text{anc,C} + \beta_{\psi, C} \ket{\perp^*}\big)\ket{0}_\text{anc,B} = \\
        &\bar{B} \bar{C} \ket{\psi} \ket{0}_\text{anc,C} \ket{0}_\text{anc,B} + \beta_{\psi, C, B} \ket{\perp}
    \end{split}
\end{equation}
with an overall rescaling factor of $\lambda = \lambda_B \lambda_C$.

It is important to note that, in general, the block-encodings $U_B$ and $U_C$ require distinct block-encoding ancillae.
This ensures that each block-encoding unitary acts on an ancilla register beginning in the all-zero state, following Eq. \eqref{eq:general-block-encoding}.

A block-encoding of this form can be generated for any operator written as a product of $L$ operators with individual block-encodings $\{U_l\}$.
The number of operations for a block-encoding of this form is simply the sum of the operations required for each block-encoding unitary.
The total number of block-encoding ancillae required is the sum of the number of block-encoding ancillae required for each block-encoding unitary.
The overall rescaling factor is the product of the rescaling factors of the block-encoding unitaries.

%% file: text/lobe/main.tex
\section{Ladder Operator Block-Encoding}
\label{sec:ladder-op-oracles}

In this section, we provide a framework for compiling block-encodings of various physically motivated products and linear combinations of fermionic and bosonic ladder operators.

In this framework, circuits update the quantum state directly following the definition of the operator in the occupation basis.
The action of several ladder operators can also be applied simultaneously to the state, reducing the cost of the block-encoding.
Any unwanted updates to the quantum state are pushed outside of the encoded subspace by rotating the block-encoding ancillae.
This rotation of states outside the encoded subspace is similar in effect to the value oracle in the Sparse-Oracle framework.
However, by focusing on the collective action of subsets of terms, we obtain block-encodings that require fewer computational resources.

These compilations are optimized primarily to reduce the number of non-Clifford operations and secondarily to reduce the number of block-encoding ancillae at the expense of ``clean ancillae".
Clean ancillae refer to qubits that are used temporarily, such that they begin and end in the zero-state and can be reused for subsequent operations.

Although controlled block-encodings are not always required, we explicitly include control qubits to demonstrate that coherent control does not significantly increase the space-time cost.
For applications that do not require controlled applications, these controls can be omitted, and the cost will be reduced accordingly.

\input{text/lobe/encoding}

\input{text/lobe/fermionic}
\input{text/lobe/fermionic_products}
\input{text/lobe/fermionic_linear_combinations}

\input{text/lobe/bosonic}
\input{text/lobe/bosonic_products}

\input{text/lobe/bosonic_linear_combinations}

\input{text/lobe/interactions}

%% file: text/lobe/encoding.tex
\subsection{Occupation Encoding}
\label{subsec:encoding}

The Jordan-Wigner transformation \cite{jordan-wigner} maps fermionic ladder operators to Pauli operators acting on qubits.
Under this transformation, the occupation states of fermionic modes are represented by qubit states using the following map:
\begin{equation}
    \ket{n_{{I_b - 1}} \dots n_{1} n_{0}} \rightarrow \ket{q_{{I_b - 1}} \dots q_{1} q_{0}}
\end{equation}
where $I_b$ is the number of fermionic modes and $n_{i} = q_{i} \in [0, 1]$ depending on if the $i^\text{th}$ fermionic mode is occupied ($\ket{1}$) or unoccupied ($\ket{0}$).
We follow this encoding to represent the occupation states of fermionic modes.
The number of qubits required for the fermionic system is equal to the number of fermionic modes.

The encoding scheme for bosons must be able to represent a number of bosons in the range $\omega_i \in [0, \Omega_i]$ for each bosonic mode.
We represent the occupancy of a bosonic mode in the binary encoding following \cite{rhodes2024exponential}.

We set the maximum bosonic occupancy for each bosonic mode to be identical ($\Omega_i = \Omega)$ for ease.
However, this encoding scheme and the block-encoding constructions presented below can be adapted to allow for bosonic modes with different bosonic cutoffs.
The number of qubits needed to represent each mode is $W = \lceil \log_2{(\Omega + 1)} \rceil$.

Fock basis states that include the occupation of multiple fermionic and bosonic modes can be expressed as tensor products of the occupation states of the individual modes:
\begin{equation}
    \begin{split}
    \ket{n} &\rightarrow \ket{n_{{I_b - 1}}} \otimes \dots \otimes \ket{n_{0}} \otimes \ket{\omega_{I_a - 1}} \otimes \hdots \otimes \ket{\omega_{0}} \\
    &\rightarrow \ket{q_{{I_b - 1}}} \hspace{0.15em}  \otimes \dots \otimes \ket{q_{0}} \otimes \ket{q_{I_a - 1}^{W-1} \hdots q_{I_a - 1}^{1} q_{I_a - 1}^0} \\
        & \hspace{4.575em} \otimes \hdots \otimes \ket{q_{0}^{W-1} \hdots q_{0}^{1} q_{0}^0} \\
    \end{split}
\end{equation}
where $I_a$ is the number of bosonic modes.

There are many alternative encoding schemes \cite{Kirby_2021}, including the Bravyi-Kitaev encoding for fermionic modes \cite{bravyi2002fermionic, seeley2012bravyi}.
Circuit constructions for these alternative encoding schemes could be considered, and we leave this for future work. 

%% file: text/lobe/fermionic.tex
\subsection{Fermionic Ladder Operators}
\label{subsec:fermionic-be}

Here, we construct block-encodings for the individual fermionic creation ($b_j^\dagger$) and annihilation ($b_j$) operators.
As these are block-encodings of non-unitary operators, we define the associated unitary operators following Eq. \eqref{eq:general-block-encoding}:
\begin{equation}
    U_{b^\dagger_j} \ket{n_j} \ket{0}_\text{anc} = b^\dagger_j \ket{n_j}\ket{0}_\text{anc} + \beta \ket{\perp}
\end{equation}
The L2 norm of the fermionic ladder operators in the Fock basis is $1$, meaning these operators do not need to be rescaled.

The definition of the fermionic creation operator (Eq. \eqref{eq:fermionic-creation}) results in two implications.
If the mode is \textit{unoccupied}, then the mode should become occupied and the amplitude should be updated with a potential sign flip ($p(n)$, Eq. \eqref{eq:parity}).
If the mode is occupied, then the output state should have no amplitude in the encoded subspace. 
The non-trivial action is dependent on the occupation of the associated fermionic mode:
\begin{equation}
    U_{b^\dagger_j} \ket{n_j} \ket{0}_\text{anc} =
    \begin{cases} 
        p(n) \ket{1} \ket{0}_\text{anc} & \text{when $\ket{n_j}$ is $\ket{0}$} \\
        \ket{\perp} & \text{when $\ket{n_j}$ is $\ket{1}$} \\
    \end{cases}
\end{equation}

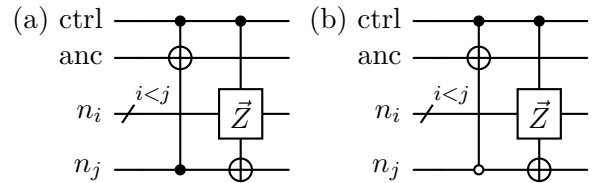
\begin{figure}[h]
    \parbox[t]{8cm}{
        \begin{quantikz}[column sep=0.35cm, row sep=0.25cm]
            \lstick{(a) ctrl} &                 & \ctrl{1} & \ctrl{3} &\\
            \lstick{anc}      &                 & \targ{} &&\\
            \lstick{$n_i$}    & \qwbundle{i<j}  & & \gate[1]{\vec{Z}}&\\
            \lstick{$n_j$}    &                 & \ctrl{-2} & \targ{} &\\
        \end{quantikz}
        \begin{quantikz}[column sep=0.35cm, row sep=0.25cm]
            \lstick{(b) ctrl} &                 & \ctrl{1} & \ctrl{3} &\\
            \lstick{anc}      &                 & \targ{} &&\\
            \lstick{$n_i$}    & \qwbundle{i<j}  & & \gate[1]{\vec{Z}}&\\
            \lstick{$n_j$}    &                 & \ctrl[open]{-2} & \targ{} &\\
        \end{quantikz}
    }
    \caption{
        \textbf{Fermionic Ladder Operator Block-Encodings}
        In (a), a block-encoding for the fermionic creation operator acting on the $j^\text{th}$ mode is given.
        In (b), a block-encoding for the fermionic annihilation operator acting on the $j^\text{th}$ mode is given.
        For a creation (annihilation) operator, the branch of the wavefunction will be flipped outside of the encoded subspace if the mode is occupied (unoccupied).
        The state is updated by applying Pauli $Z$ gates to the preceeding fermionic modes which result in the output state having the appropriate sign based on $p(n)$.
        Then a Pauli $X$ gate is applied to flip the occupation of the $j^\text{th}$ mode.
    }
    \label{fig:fermionic-be}
\end{figure}

An implementation for $U_{b^\dagger_j}$ is given in subfigure \ref{fig:fermionic-be}a.
The Toffoli gate flips the ancilla qubit to push the state entirely outside of the encoded subspace when the control qubit is on ($\ket{1}$) and the fermionic mode is occupied ($\ket{1}$).
The update of the amplitude corresponding to $p(n)$ can be applied using a series of controlled Pauli $Z$ operators applied to each of the fermionic modes with index $i < j$: $\vec{Z}_i$.
The occupation of the fermionic mode on which the ladder operator acts is updated using a controlled Pauli $X$ operator: $X_j$.

A block-encoding for the fermionic annihilation operator ($U_{b_j}$) can be constructed similarly and is shown in subfigure \ref{fig:fermionic-be}b.
The fermionic creation operator ($b^\dagger$) only acts nontrivially when the mode it acts upon is \textit{unoccupied}.
Inversely, the fermionic annihilation operator ($b$) will only act nontrivially if the mode is \textit{occupied}.
Therefore, the block-encoding ancilla is flipped outside of the encoded subspace if the control qubit is on ($\ket{1}$) and the fermionic mode is unoccupied ($\ket{0}$).

Each Toffoli gate can be implemented with four $T$ gates using one clean ancilla \cite{selinger2013quantum,jones2013low}.
Likewise, an $N$-controlled Toffoli gate acting on a clean ancilla can be decomposed into a series of Toffoli gates acting on several clean ancillae \cite{barenco1995elementary}.
The space-time costs for these decompositions are discussed in more detail in Appendix \ref{sec:elbows}, and we will assume these strategies are used when computing resource estimates.
Recent strategies for decomposing multi-controlled Toffoli gates have been proposed \cite{gosset2025}, which require exponentially fewer $T$ gates as a function of the number of controls.
We do not assume this decomposition strategy, though these techniques will apply to many of the circuits presented in this work.

In total, these block-encoding circuits have a rescaling factor of $\lambda = 1$, require one block-encoding ancilla and one clean ancilla, and use four $T$ gates.

%% file: text/lobe/fermionic_products.tex
\subsection{Products of Fermionic Ladder Operators}

In this subsection, we construct block-encodings for a product of fermionic ladder operators.
These constructions use fewer resources than would be required by combining the block-encodings of the individual operators using the strategy outlined in subsection \ref{subsec:be-products}.

We define the number of ``active modes" as the number of unique modes upon which a ladder operator is directly applied.
For example, the operator $b_i b_j^\dagger b_k b_l^\dagger b_l$ has $4$ active modes: $i, j, k, l$.
We will use $B$ to represent the number of active modes in a given operator.

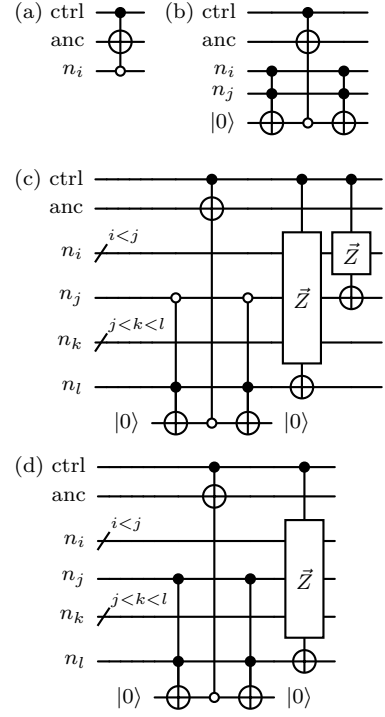
\begin{figure}[t]
    \begin{quantikz}[column sep=0.15cm, row sep=0.15cm, baseline=(current bounding box.north), font=\scriptsize]
        \lstick{(a) ctrl} & \ctrl{1} &\\
        \lstick{anc} & \targ{} &\\
        \lstick{$n_i$} & \ctrl[open]{-1} &\\
    \end{quantikz}
    \begin{quantikz}[column sep=0.15cm, row sep=0.15cm, baseline=(current bounding box.north), font=\scriptsize]
        \lstick{(b) ctrl} & & \ctrl{1} & &\\
        \lstick{anc} &  & \targ{} & &\\
        \lstick{$n_i$} & \ctrl{2} && \ctrl{2} &\\
        \lstick{$n_j$} & \ctrl{1} && \ctrl{1} &\\
        \lstick{$\ket{0}$} & \targ{} & \ctrl[open]{-3} & \targ{} &\\
    \end{quantikz}

    \begin{quantikz}[column sep=0.15cm, row sep=0.15cm, font=\scriptsize]
        \lstick{(c) ctrl} &                 &&&&&   & \ctrl{1} &&& \ctrl{5} & \ctrl{3} &\\
        \lstick{anc}      &                 &&&&&   & \targ{} &&&&&\\
        \lstick{$n_i$}    & \qwbundle{i<j}  &&&&&   & &&& \gate[3]{\vec{Z}} & \gate[1]{\vec{Z}} &\\
        \lstick{$n_j$}    &                 &&&&& \ctrl[open]{3} &   & \ctrl[open]{3} &&& \targ{} &\\
        \lstick{$n_k$}    & \qwbundle{j<k<l}&&&&&   &  &   &&&&\\
        \lstick{$n_l$}    &                 &&&&& \ctrl{1} &  &\ctrl{1}&& \targ{} &&\\
        &\wireoverride{}&\wireoverride{}&\wireoverride{}&\wireoverride{}&\wireoverride{}\lstick{$\ket{0}$}& \targ{}  & \ctrl[open]{-5} & \targ{}&\rstick{$\ket{0}$}&\wireoverride{}&\wireoverride{}&\wireoverride{}
    \end{quantikz}
    
    \begin{quantikz}[column sep=0.15cm, row sep=0.15cm, font=\scriptsize]
        \lstick{(d) ctrl} &                 &&&&&   & \ctrl{1} &&& \ctrl{5} &\\
        \lstick{anc}      &                 &&&&&   & \targ{}&&&&\\
        \lstick{$n_i$}    & \qwbundle{i<j}  &&&&&   & &&& \gate[3]{\vec{Z}} &\\
        \lstick{$n_j$}    &                 &&&&& \ctrl{3} &   & \ctrl{3} &&&\\
        \lstick{$n_k$}    & \qwbundle{j<k<l}&&&&&   &  &   &&&\\
        \lstick{$n_l$}    &                 &&&&& \ctrl{1} &  &\ctrl{1}&& \targ{} &\\
        &\wireoverride{}&\wireoverride{}&\wireoverride{}&\wireoverride{}&\wireoverride{}\lstick{$\ket{0}$}& \targ{}  & \ctrl[open]{-5} & \targ{}&\rstick{$\ket{0}$}
    \end{quantikz}
    
    \caption{
        \textbf{Block-Encoding Products of Fermionic Ladder Operators}
        In (a), a block-encoding for the fermionic number operator acting on the $i^\text{th}$ mode ($b_i^\dagger b_i$) is given.
        In (b), a block-encoding for the product of two fermionic number operators acting on the $i^\text{th}$ and $j^\text{th}$ modes ($b_i^\dagger b_i b_j^\dagger b_j$) is given.
        In (c), a block-encoding for the operator $b_j^\dagger b_l$ with $j \neq l$ is given.
        In (d), a block-encoding for the operator $b_j^\dagger b_j b_l$ with $j \neq l$ is given.
    }
    \label{fig:fermionic-products-be}
\end{figure}

Consider the action of the fermionic number operator ($b_i^\dagger b_i$):
\begin{equation}
    \begin{split}
        b_i^\dagger b_i \ket{n_i} = \begin{cases} 
            \ket{1} & \text{when $\ket{n_i}$ is $\ket{1}$} \\
            0 & \text{when $\ket{n_i}$ is $\ket{0}$} \\
                                        \end{cases}
    \end{split}
\end{equation}
If the $i^\text{th}$ mode is occupied, the state is left unchanged.
If the $i^\text{th}$ mode is unoccupied, the state should have zero amplitude in the encoded subspace.

This action can be applied using the circuit shown in subfigure \ref{fig:fermionic-products-be}a.
The Toffoli gate flips the block-encoding ancilla outside of the encoded subspace if the control is on and the $i^\text{th}$ mode is unoccupied.
Otherwise, the system and the block-encoding ancilla are left unchanged.

Block-encoding this operator as the product of the block-encodings for $b_i$ and $b_i^\dagger$ would require two block-encoding ancillae and use eight $T$ gates.
This construction has a rescaling factor of $\lambda = 1$, requires one block-encoding ancilla and one clean ancilla, and uses four $T$ gates.

When two number operators are applied to distinct modes ($b_i^\dagger b_i b_j^\dagger b_j$), their actions have shared structure.
The state is pushed outside the encoded subspace \textit{unless} both modes are occupied.
This implies that their action can be applied together for a more efficient block-encoding construction.
A block-encoding circuit for this operator is shown in subfigure \ref{fig:fermionic-products-be}b.
This circuit has a rescaling factor of $\lambda = 1$, requires one block-encoding ancilla and two clean ancillae, and uses eight $T$ gates.
This uses one fewer block-encoding ancilla than would be required by taking the product of the block-encodings for the two number operators individually.

A block-encoding circuit for the operator $b_j^\dagger b_l$ is given in subfigure \ref{fig:fermionic-products-be}c.
For this operator, the amplitude of the state in the encoded subspace should be zero \textit{unless} both the $j^\text{th}$ mode is unoccupied and the $l^\text{th}$ mode is occupied.
If the control qubit is on and these two conditions are not both true, then the block-encoding ancilla is flipped to push the state outside of the encoded subspace.
The system is updated based on the order in which the two operators would act on the quantum state: $\vec{Z}X_l$ ($b_l$) then $\vec{Z}X_j$ ($b_j^\dagger$).
For clarity, these system updates are shown individually for all circuit diagrams in this work.
In practice, the action of these Pauli operators should be merged to reduce the total number of Clifford operations.
This block-encoding circuit has a rescaling factor of $\lambda = 1$, requires one block-encoding ancilla and two clean ancillae, and uses eight $T$ gates.

A block-encoding circuit that combines the action of a number operator on the $j^\text{th}$ mode and an annihilation operator on the $l^{th}$ mode ($b_j^\dagger b_j b_l$) is shown in subfigure \ref{fig:fermionic-products-be}d.
Despite including an additional ladder operator compared to the previous operator ($b_j^\dagger b_l$), the resources required by this circuit are the same.
This is because the number of control conditions that dictate the action of the operator are determined by the number of active modes, not the number of ladder operators.

This construction can be generalized to an arbitrary product of ladder operators.
Each active mode will contribute a new control condition based on the occupation of the corresponding mode.
A block-encoding circuit of this form with $B$ active modes will have a rescaling factor of $\lambda = 1$, require one block-encoding ancilla and $B$ clean ancillae, and use $4B$ $T$ gates.

%% file: text/lobe/fermionic_linear_combinations.tex
\subsection{Linear Combinations of Fermionic Ladder Operators}
\label{sec:LCFLO}

In this subsection, we construct block-encodings of linear combinations of products of fermionic ladder operators.
These circuits use fewer resources than are required by the LCO construction described in subsection \ref{subsec:lco}.
We generalize this construction for a product of fermionic ladder operators plus its Hermitian conjugate, however, we note that the strategies we present here are not restricted to Hermitian conjugates.

\begin{figure}
    \begin{quantikz}[column sep=0.15cm, row sep=0.15cm, font=\scriptsize]
        \lstick{(a) ctrl} &                 &&&& \ctrl{2} &\\
        \lstick{$n_i$}    & \qwbundle{i<j}  &&&& \gate[1]{\vec{Z}}&\\
        \lstick{$n_j$}    &                 &&&& \targ{} &
    \end{quantikz}

    \begin{quantikz}[column sep=0.15cm, row sep=0.15cm, font=\scriptsize]
        \lstick{(b) ctrl} &                 &&&&&   & \ctrl{1} &&\ctrl{3}& \ctrl{5} & \ctrl{3} &\\
        \lstick{anc}      &                 &&&&&   & \targ{} &&&&&\\
        \lstick{$n_i$}    & \qwbundle{i<j}  &&&&&   & &&& \gate[3]{\vec{Z}} & \gate[1]{\vec{Z}} &\\
        \lstick{$n_j$}    &                 &&&&& \ctrl{2} &   & \ctrl{2} &\ctrl[open]{-2}&& \targ{} &\\
        \lstick{$n_k$}    & \qwbundle{j<k<l}&&&&&   &  &   &&&&\\
        \lstick{$n_l$}    &                 &&&&& \targ{} & \ctrl{-4}  &\targ{}&& \targ{} &&
    \end{quantikz}

    \begin{quantikz}[column sep=0.15cm, row sep=0.15cm, font=\scriptsize]
        \lstick{(c) ctrl} &                 &&&   & \ctrl{1} &&\ctrl{3}& \ctrl{5} & \ctrl{3} &\\
        \lstick{anc}      &                 &&&   & \targ{} &&&&&\\
        \lstick{$n_i$}    & \qwbundle{i<j}  &&&   & &&& \gate[3]{\vec{Z}} & \gate[1]{\vec{Z}} &\\
        \lstick{$n_j$}    &                 &&& \ctrl{2} &   & \ctrl{2} &\ctrl[open]{-2}&& \targ{} &\\
        \lstick{$n_k$}    & \qwbundle{j<k<l}&&&   &  &   &&&&\\
        \lstick{$n_l$}    &                 &&\targ{}& \targ{} & \ctrl{-4}  &\targ{}&\targ{}& \targ{} &&
    \end{quantikz}

    \begin{quantikz}[column sep=0.15cm, row sep=0.15cm, baseline=(current bounding box.north), font=\scriptsize]
        \lstick{(d) ctrl} &                 &&&   &  && \ctrl{1}&&& \ctrl{3} & \ctrl{5} & \ctrl{3} &\\
        \lstick{anc}      &                 &&&   &  && \targ{}&&&&&&\\
        \lstick{$n_i$}    & \qwbundle{i<j}  &&&   & &&&&&& \gate[3]{\vec{Z}} & \gate[1]{\vec{Z}}&\\
        \lstick{$n_j$}    &                 &&&  & \ctrl{2}  &&&& \ctrl{2} & \ctrl[open]{-1} &&\targ{}&\\
        \lstick{$n_k$}    & \qwbundle{j<k<l}&&&   &  &   &&&&&&&\\
        \lstick{$n_l$}    &                 &&& \targ{} & \targ{} & \ctrl[open]{2}&& \ctrl[open]{2} & \targ{} & \targ{} & \targ{} &&\\
        \lstick{$n_m$}    &                 &&& &   & \ctrl{1} && \ctrl{1} &&&&&\\
        &\wireoverride{}&\wireoverride{}&\wireoverride{}&\wireoverride{}&\wireoverride{}\lstick{$\ket{0}$}& \targ{} & \ctrl[open]{-7} & \targ{}&\rstick{$\ket{0}$}&\wireoverride{}&\wireoverride{}&\wireoverride{}&\wireoverride{}
    \end{quantikz}
    
    \caption{
        \textbf{Block-Encoding Product of Fermionic Operators Plus Hermitian Conjugate}
        In (a), a block-encoding for the operator $b_j + b_j^\dagger$ is given.
        In (b), a block-encoding for the operator $b_j b_l + b_l^\dagger b_j^\dagger$ is given.
        In (c), a block-encoding for the operator $b_j b_l^\dagger + b_l b_j^\dagger$ is given.
        In (d), a block-encoding for the operator $b_j b_l^\dagger b_m^\dagger b_m + b_m^\dagger b_m b_l b_j^\dagger$ is given.
    }
    \label{fig:fermionic-be-lc-small}
\end{figure}
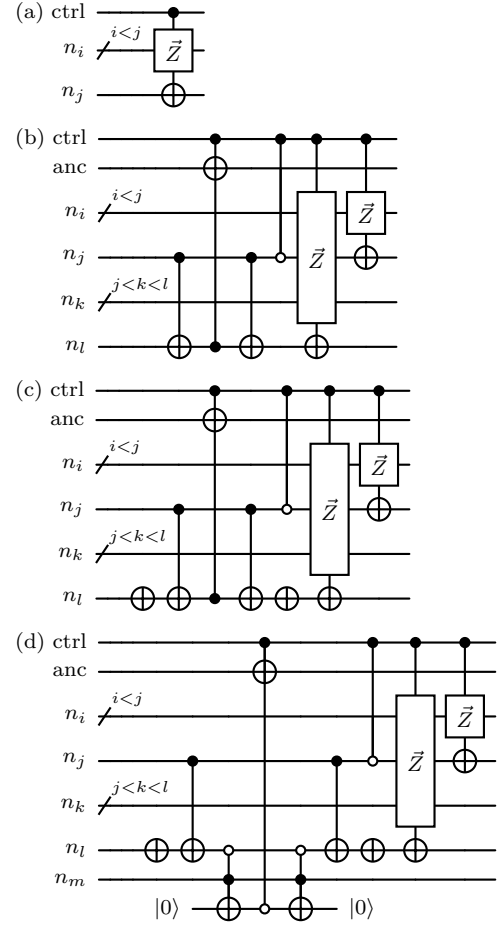

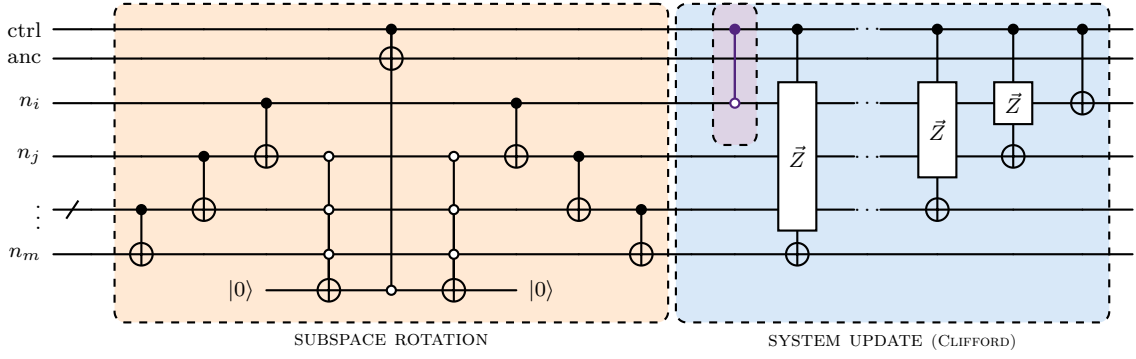
\begin{figure*}[t]
    \begin{center}
    \begin{quantikz}[column sep=0.5cm, row sep=0.15cm, baseline=(current bounding box.north), font=\scriptsize]
        \lstick{ctrl}     &            &\gategroup[7,steps=9,style={dashed,rounded corners,fill=worange!20, inner xsep=2pt},background,label style={label position=below,anchor=north,yshift=-0.2cm}]{{\sc subspace rotation }}&&&&\ctrl{1}&&&&&\gategroup[7,steps=7,style={dashed,rounded corners,fill=wblue!20, inner xsep=2pt},background,label style={label position=below,anchor=north,yshift=-0.2cm}]{{\sc system update {\tiny (Clifford)}}}&\ctrl[style={quantumviolet}]{2} \gategroup[3,steps=1,style={dashed,rounded corners,fill=quantumviolet!20, inner xsep=0.5pt},background]{} &\ctrl{5}&\ldots&\ctrl{4}&\ctrl{3}&\ctrl{2}&\\
        \lstick{anc}      &            &&&&&\targ{}   &  &&&&&&&&&&&\\
        \lstick{$n_i$}    &            &&&\ctrl{1}&&&   & \ctrl{1}&&&&\ctrl[open, style={quantumviolet}]{-2}&\gate[3]{\vec{Z}}&\ldots&\gate[2]{\vec{Z}}&\gate[1]{\vec{Z}}&\targ{}&\\
        \lstick{$n_j$}    &            &&\ctrl{1}&\targ{}& \ctrl[open]{3}  &&\ctrl[open]{3}&\targ{}&\ctrl{1}&&&&&\ldots&&\targ{}&&\\
        \lstick{$\vdots$} & \qwbundle{}&\ctrl{1}&\targ{}&&\ctrl[open]{2}&&\ctrl[open]{2} &  &\targ{}&\ctrl{1}&&&&\ldots&\targ{}&&&\\
        \lstick{$n_m$}    &            &\targ{}&&&\ctrl[open]{1}&&\ctrl[open]{1} &&&\targ&&&&\targ{}&&&&&\\
        &     \wireoverride{}       &\wireoverride{}&\wireoverride{}&\wireoverride{}\lstick{$\ket{0}$}&\targ{}& \ctrl[open]{-6}&\targ{} &\rstick{$\ket{0}$}&\wireoverride{}&\wireoverride{}&\wireoverride{}&\wireoverride{}&\wireoverride{}&\wireoverride{}&\wireoverride{}&\wireoverride{}&\wireoverride{}&\wireoverride{}
    \end{quantikz}
    \end{center}
    \caption{
        \textbf{Generalized Block-Encoding for Product of Fermionic Operators Plus Hermitian Conjugate}
        A block-encoding for the operator $b_i b_j ... b_m + b_m^\dagger ... b_j^\dagger b_i^\dagger$ is given.
        The $CZ$ gate highlighted in purple is present only if the swapping on the order of the operators in the second term induces a negative sign on the second term (an odd number of swaps are required to reorder the ladder operators).
        Block-encodings for similar operators such as those that include number operators or different arrangements of the creation and annihilation operators can be accounted for using the modifications shown in subfigures \ref{fig:fermionic-be-lc-small}c and \ref{fig:fermionic-be-lc-small}d.    }
    \label{fig:fermionic-be-lc}
\end{figure*}

Consider a linear combination of an individual fermionic ladder operator with its Hermitian conjugate: $b_j^\dagger + b_j$.
An LCO construction using the block-encodings for these two ladder operators presented in the previous section would have a rescaling factor of $\lambda = 2$, require two block-encoding ancillae and two clean ancillae, and use $12$ $T$ gates.

By considering the action of this operator on the active mode, a more efficient block-encoding can be constructed:
\begin{equation}
    \label{eq:action-of-fermionic-op-plus-hc}
    \begin{split}
        (b_j^\dagger + b_j) \ket{n_j} = \begin{cases} 
            p(n) \ket{1} & \text{when $\ket{n_j}$ is $\ket{0}$} \\
            p(n) \ket{0} & \text{when $\ket{n_j}$ is $\ket{1}$} \\
                                        \end{cases}
    \end{split}
\end{equation}

Following this definition, a Pauli $X$ gate can be applied to the $j^\text{th}$ mode to flip the occupation and a string of Pauli $Z$ gates can be applied on each fermionic mode with index $i < j$ to apply the potential sign flip induced by $p(n)$: ($\vec{Z}X_j$).
A circuit diagram for this block-encoding is shown in subfigure \ref{fig:fermionic-be-lc-small}a.
This block-encoding circuit has a rescaling factor of $\lambda = 1$, requires zero block-encoding ancillae and zero clean ancillae, and uses zero non-Clifford operations.
We note that this results in the same circuit one would arrive at following an LCU approach after applying the Jordan-Wigner transformation.

For a two-body fermionic operator and its Hermitian conjugate ($b_j b_l + b_l^\dagger b_j^\dagger$) with $j \neq l$ we arrive at a construction that differs from the Jordan-Wigner transformation.
Expanding this operator in the Pauli basis would result in a linear combination of two Pauli operators, which can be block-encoded using the LCU framework.

In LOBE, we focus on describing the action of the joined operator on the state.
We begin by reordering the ladder operators such that the active modes appear in the same order in each term: $b_j b_l - b_j^\dagger b_l^\dagger$.
This reordering ensures that the potential sign flips caused by the respective ladder operators in each term are consistent, since the order in which they are applied to the system in the circuit is fixed.

The action of this operator on the system can be determined by the parity of the occupation of the two active fermionic modes.
This can be seen by describing the action of the joined operator on the possible occupation states of the $j^\text{th}$ and $l^\text{th}$ fermionic modes:
\begin{equation}
    \begin{split}
        (b_j b_l - b_j^\dagger b_l^\dagger) \ket{n_l, n_j} = \\
        \begin{cases} 
            - p(n) \ket{11} & \text{when $\ket{n_l, n_j}$ is $\ket{00}$} \\
            \hspace{0.75em} p(n) \ket{00} & \text{when $\ket{n_l, n_j}$ is $\ket{11}$} \\
            \hspace{2em} 0 & \text{when $\ket{n_l, n_j}$ is $\ket{01}$} \\
            \hspace{2em} 0 & \text{when $\ket{n_l, n_j}$ is $\ket{10}$} \\
        \end{cases}
    \end{split}
\end{equation}

In subfigure \ref{fig:fermionic-be-lc-small}b, we give a circuit diagram for block-encoding this two-body operator.
The parity of the occupation of the active modes can be computed using a CNOT gate controlled on the $j^\text{th}$ mode, targeting the $l^\text{th}$ mode.
If the control qubit is on and $\ket{n_l}$ is in the $\ket{1}$ state (odd parity), then the block-encoding ancilla is flipped to push that branch of the wavefunction outside of the encoded subspace.
The parity can then be uncomputed, returning the qubit storing the occupation of the $l^\text{th}$ mode to its original state.
Next, a $CZ$ gate that is $0$-controlled on the $j^\text{th}$ fermionic mode and $1$-controlled on the control qubit applies the desired sign flip corresponding to the term $- b_j^\dagger b_l^\dagger$.
Lastly, a series of $\vec{Z}X$ operators is applied to each active mode in the order in which the operators would be applied onto the system (right to left).

This construction differs significantly from what one would arrive at by expanding in the Pauli basis and then using the LCU framework.
In LOBE, the circuit begins by directing each computational basis state to either remain in the encoded subspace or to be pushed outside.
The system is then updated appropriately for the states that remain in the encoded subspace.
States that are pushed outside may be updated arbitrarily since the desired action of the block-encoding is only defined within the encoded subspace.

This framework can also be applied to operators that are not simply a linear combination of a product of annihilation operators with its Hermitian conjugate.
For an operator of the form $b_j b_l^\dagger + b_l b_j^\dagger$, we can construct a similar block-encoding with a slight modification.
In this case, we simply flip the occupation of the $l^\text{th}$ mode using a Pauli $X$ gate before we compute the parity of the modes.
The circuit diagram for block-encoding this operator is given in subfigure \ref{fig:fermionic-be-lc-small}c.

Likewise, if a number operator is present, this can be accounted for by excluding that mode from the parity computations and including a control condition on that mode when determining if the state should remain in the encoded subspace.
A circuit diagram for an operator of this form is given in subfigure \ref{fig:fermionic-be-lc-small}d.

A generalized circuit diagram for an operator that is a product of fermionic ladder operators plus its Hermitian conjugate is shown in Figure \ref{fig:fermionic-be-lc}.
These block-encoding circuits will all have rescaling factors of $\lambda=1$, require one block-encoding ancilla and $B - 1$ clean ancillae, and use $4(B-1)$ $T$ gates.

The circuit begins by selecting which states should remain inside the encoded subspace.
For the case where the terms are either a product of annihilation operators or creation operators, this is determined by whether the active modes are either all occupied or unoccupied and can be computed by the parities of neighboring modes.
For terms that are not strictly of this form, the modifications shown in subfigures \ref{fig:fermionic-be-lc-small}c and \ref{fig:fermionic-be-lc-small}d can be applied.
After the states are placed in the appropriate subspace and the system is returned to the Fock state, the state of the system is updated appropriately.
The $CZ$ gate shown in purple is present only if $(C \text{ choose } 2)$ is odd, where $C$ is the number of active modes excluding modes where a number operator is present.
This accounts for cases where the sign of the Hermitian conjugate term becomes negative after reordering the ladder operators.

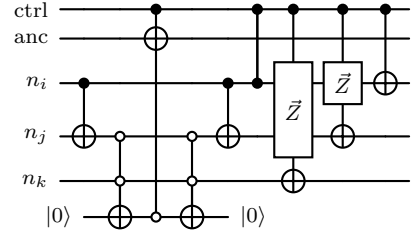
\begin{figure}
    \begin{quantikz}[column sep=0.15cm, row sep=0.15cm, font=\scriptsize]
        \lstick{ctrl} &          &&\ctrl{1} &&&\ctrl{2}& \ctrl{4} & \ctrl{3} &\ctrl{2} &\\
        \lstick{anc}  &          &&\targ{} &&&&&&&\\
        \lstick{$n_i$}& \ctrl{1} &&&&\ctrl{1}& \ctrl{-2}& \gate[2]{\vec{Z}} & \gate[1]{\vec{Z}} & \targ{}&\\
        \lstick{$n_j$}& \targ{}  &\ctrl[open]{2}&& \ctrl[open]{2} & \targ{}&  & &\targ{}& &\\
        \lstick{$n_k$}&          &\ctrl[open]{1}&& \ctrl[open]{1}&  &   &\targ{}&&&\\
        &\wireoverride{}\lstick{$\ket{0}$}&\targ{}&\ctrl[open]{-4}&\targ{} &\rstick{$\ket{0}$}&\wireoverride{}&\wireoverride{}&\wireoverride{}&\wireoverride{}&\wireoverride{}
    \end{quantikz}
    \caption{
        \textbf{Block-Encoding for Linear Combination of Non-Conjugate Fermionic Operators}
        A block-encoding for the operator $b_i b_j b_k^\dagger + b_j^\dagger b_i^\dagger b_k^\dagger$ is given.
    }
    \label{fig:fermionic-be-lc-not-conjugate}
\end{figure}

This framework can also account for the action of multiple terms that are not Hermitian conjugates.
Consider the operator $b_i b_j b_k^\dagger + b_j^\dagger b_i^\dagger b_k^\dagger$.
The desired action is to push the state outside of the encoded subspace \textit{unless} both $\ket{n_i \oplus n_j}$ and $\ket{n_k}$ are $\ket{0}$.
The block-encoding circuit for this operator is given in Figure \ref{fig:fermionic-be-lc-not-conjugate} and has a rescaling factor of $\lambda = 1$, requires one block-encoding ancilla and two clean ancillae, and uses $8$ $T$ gates.

%% file: text/lobe/bosonic.tex
\subsection{Bosonic Ladder Operators}
\label{subsec:bosonic-lobe}

In this section, we define unitaries that block-encode the bosonic creation ($a_i^\dagger$) and annihilation ($a_i$) operators.
Based on the definition of the bosonic creation operator (Eq. \eqref{eq:bosonic-creation}), a block-encoding unitary following the form of Eq. \eqref{eq:general-block-encoding} can be defined as follows: 
\begin{equation}
    \label{eq:be-bos-creation}
    \begin{split}
        &U_{a^\dagger_i} \ket{\omega_i} \ket{0}_\text{anc} = \\
        &\begin{cases}
            \sqrt{\frac{\omega_i + 1}{\Omega}} \ket{\omega_i + 1} \ket{0}_\text{anc} + \beta \ket{\perp} & \text{when } \omega_i < \Omega \\
            \ket{\perp} & \text{when } \omega_i \geq \Omega \\
        \end{cases}
    \end{split}
\end{equation}
where $\omega_i$ is the occupation of the $i^\text{th}$ bosonic mode and the operator is rescaled by a factor of $\sqrt{\Omega}$.

\begin{figure}
    \begin{quantikz}[column sep=0.15cm, row sep=0.15cm, font=\scriptsize]
        \lstick{(a) ctrl}  &             &&&\ctrl{2}&\ctrl{2}&\\
        \lstick{anc}       &             &&&&\gate{R_y(\theta_\omega)}&\\
        \lstick{$\omega_i$}& \qwbundle{W}&&& \gate{+1} &\gate{In_{\omega}}&\\
    \end{quantikz}
    \begin{quantikz}[column sep=0.15cm, row sep=0.15cm, font=\scriptsize]
        \lstick{(b) ctrl}  &             &\ctrl{2}&\ctrl{2}&\\
        \lstick{anc}       &             &\gate{R_y(\theta_\omega)}&&\\
        \lstick{$\omega_i$}& \qwbundle{W}& \gate{In_{\omega}} &\gate{-1}&\\
    \end{quantikz}

    \caption{
        \textbf{Bosonic Ladder Operator Block-Encoding}
        In (a), a block-encoding for the bosonic creation operator ($a_i^\dagger$) is given.
        In (b), a block-encoding for the bosonic annihilation operator ($a_i$) is given.
        The ``$+ 1$'' operation depicts an incrementer.
        The ``$- 1$'' operation depicts a decrementer.
        The ``$In_\omega - R_y(\theta_\omega)$'' operation depicts a series of uniformly controlled $R_y$ rotations.
    }
    \label{fig:bosonic-ladder-op-be}
\end{figure}
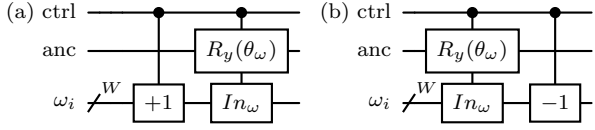

For the bosonic creation operator, the desired update to the system can be achieved by applying an incrementer circuit to the corresponding bosonic mode.
Then, the block-encoding ancilla can be rotated using a series of uniformly controlled $R_y$ gates that are controlled on the corresponding bosonic occupation to achieve the desired amplitude in the encoded subspace.
Efficient implementations for incrementing the occupation are given in Appendix \ref{sec:addition} and different implementations for a \textit{controlled} set of uniformly controlled rotations are given in Appendix \ref{sec:multiplexed-rotations}.

A circuit diagram for block-encoding the bosonic creation operator is given in subfigure \ref{fig:bosonic-ladder-op-be}a.
A block-encoding for the bosonic annihilation operator can be constructed similarly, but with the block-encoding ancilla rotated prior to the occupation of the mode being decremented (subfigure \ref{fig:bosonic-ladder-op-be}b).

The rotation angles can be determined classically:
\begin{equation}
    \label{eq:single-op-angles}
    \theta(\omega_i) = 
    \begin{cases} 
        2 \cos^{-1}\Big(\sqrt{\omega_i/\Omega}\Big) & \text{when } \omega_i < \Omega\\
        \pi & \text{when } \omega_i \geq \Omega
    \end{cases}
\end{equation}
where $\omega_i$ here in comparison with $\omega_i + 1$ in Eq. \eqref{eq:be-bos-creation} is due to the occupation of the state being updated prior to the controlled rotations.
The value of $\pi$ sets the state of the block-encoding ancilla to be $\ket{1}$, indicating the state is rotated entirely outside of the encoded subspace.

A controlled incrementer requires $W - 1$ clean ancillae and $3W$ $T$ gates \cite{Gidney_2015}. 
The controlled set of uniformly controlled rotations can be implemented using (at most) $\Omega + 3$ uncontrolled rotations and $4 \lceil{\log_2{\Omega}}\rceil$ $T$ gates following the protocol outlined in Appendix \ref{sec:multiplexed-rotations}.
In total, this block-encoding circuit has a rescaling factor of $\lambda = \sqrt{\Omega}$, requires one block-encoding ancilla and $\lceil{\log_2{\Omega}}\rceil$ clean ancillae, and uses $7 W$ $T$ gates and (at most) $\Omega + 3$ arbitrary rotations.

%% file: text/lobe/bosonic_products.tex
\subsection{Products of Bosonic Ladder Operators}

In this subsection, we discuss block-encodings for a product of bosonic ladder operators acting on the same mode.
Products of bosonic ladder operators acting on different modes can be block-encoded using the methods described in subsection \ref{subsec:be-products}.

Consider the desired action of a block-encoding of a product of $R$ bosonic creation operators acting on the $i^\text{th}$ bosonic mode:
\begin{equation}
    \label{eq:be-S-creation-ops}
    \begin{split}
        &U_{(a^\dagger_i)^R} \ket{\omega_i} \ket{0}_\text{anc} = \\
        &\begin{cases}
            f(\omega_i) \ket{\omega_i + R} \ket{0}_\text{anc} + \beta \ket{\perp} & \text{when } \omega_i \leq \Omega - R \\
            \ket{\perp} & \text{when } \omega_i > \Omega - R \\
        \end{cases}
    \end{split}
\end{equation}
where
\begin{equation}
    f(\omega_i) = \prod_{r=1}^{R} \sqrt{(\omega_i + r) / \Omega}
\end{equation}

\begin{figure}
    \begin{quantikz}[column sep=0.15cm, row sep=0.15cm, font=\scriptsize]
        \lstick{(a) ctrl}  &             &&&\ctrl{2}&\ctrl{2}&\\
        \lstick{anc}       &             &&&&\gate{R_y(\theta_\omega, R)}&\\
        \lstick{$\omega_i$}& \qwbundle{W}&&& \gate{+R} &\gate{In_{\omega}}&\\
    \end{quantikz}

    \begin{quantikz}[column sep=0.15cm, row sep=0.15cm, font=\scriptsize]
        \lstick{(b) ctrl}  &             &&&\ctrl{2}&\ctrl{2}&\\
        \lstick{anc}       &             &&&&\gate{R_y(\theta_\omega, S)}&\\
        \lstick{$\omega_i$}& \qwbundle{W}&&& \gate{-S} &\gate{In_{\omega}}&\\
    \end{quantikz}

    \begin{quantikz}[column sep=0.15cm, row sep=0.15cm, font=\scriptsize]
        \lstick{(c) ctrl}  &             &&&\ctrl{2}&\ctrl{2}&\\
        \lstick{anc}       &             &&&&\gate{R_y(\theta_\omega, R, S)}&\\
        \lstick{$\omega_i$}& \qwbundle{W}&&& \gate{+R-S} &\gate{In_{\omega}}&\\
    \end{quantikz}
    \caption{
        \textbf{Block-Encoding Product of Bosonic Ladder Operators}
        In (a), a block-encoding for the operator $(a_i^\dagger)^R$ is given.
        In (b), a block-encoding for the operator $(a_i)^S$ is given.
        In (c), a block-encoding for the operator $(a_i^\dagger)^R (a_i)^S$ is given.
        The ``$+ R$'' operation depicts an addition by the classical integer $R$.
        The ``$- S$'' operation depicts a subtraction by the classical integer $S$.
    }
    \label{fig:products-bosonic-operators}
\end{figure}
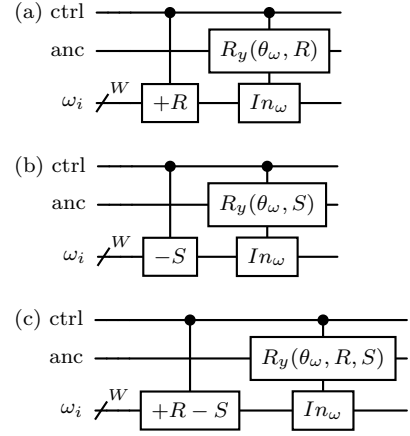

\begin{figure*}[t]
    \begin{tikzpicture} \node[scale=0.77]{
    \begin{quantikz}[column sep=0.15cm, row sep=0.15cm, baseline=(current bounding box.north), font=\scriptsize]
        \lstick{(a) ctrl}  &             &&\ctrl{3}&\ctrl{3}&\ctrl{3}&&\\
        \lstick{index}     &             & \gate{H} & \ctrl[open]{2}&&\ctrl{2}& \gate{H} &\\
        \lstick{anc}       &             &&&\gate{R_y(\theta_\omega, R,S)}&&&\\
        \lstick{$\omega_i$}& \qwbundle{W}&& \gate{+R-S} &\gate{In_{\omega}}&\gate{-R+S}&&
    \end{quantikz}
    \begin{quantikz}[column sep=0.05cm, row sep=0.15cm, baseline=(current bounding box.north), font=\scriptsize]
        \lstick{(b) ctrl}  &&&             &&\ctrl{7}&\ctrl{5}&\ctrl{6}&\ctrl{7}&\ctrl{7}&&&&\\
        \lstick{index}     &&&             & \gate{H} & \ctrl[open]{6}&&&&\ctrl{2}& \gate{H} &&&\\
        \lstick{$\text{anc}_0$}       &&&             &&&\gate{R_y(\theta_\omega, R_i,S_i)}&&&&&&&&\\
        \lstick{\vdots}       &&&             &&&&\gate{R_y(\theta_\omega, R_{\hdots},S_{\hdots})}&&&&&&&\\
        \lstick{$\text{anc}_\text{B-1}$}       &&&             &&&&&\gate{R_y(\theta_\omega, R_m,S_m)}&&&&&&\\
        \lstick{$\omega_i$}&&& \qwbundle{W}&& \gate{+R_i-S_i} &\gate{In_{\omega}}&&&\gate{-R_i+S_i} &&&&&\\
        \lstick{\vdots}&&& \qwbundle{W}&& \gate{+R_{\hdots} - S_{\hdots}} &&\gate{In_{\omega}}&&\gate{-R_{\hdots} + S_{\hdots}}&&&&&\\
        \lstick{$\omega_m$}&&& \qwbundle{W}&& \gate{+R_m-S_m} &&&\gate{In_{\omega}}&\gate{-R_m + S_m}&&&&&
    \end{quantikz}
    };\end{tikzpicture}
    \caption{
        \textbf{Block-Encoding Product of Bosonic Ladder Operators Plus Hermitian Conjugate}
        In (a), a block-encoding for the operator $\big((a_i^\dagger)^R (a_i)^S + (a_i^\dagger)^S (a_i)^R\big)$ is given.
        In (b), a block-encoding for the operator $\big((a_i^\dagger)^{R_i} (a_i)^{S_i}...(a_m^\dagger)^{R_m} (a_m)^{S_m} + h.c.\big)$ is given.
    }
    \label{fig:lc-bosonic}
\end{figure*}
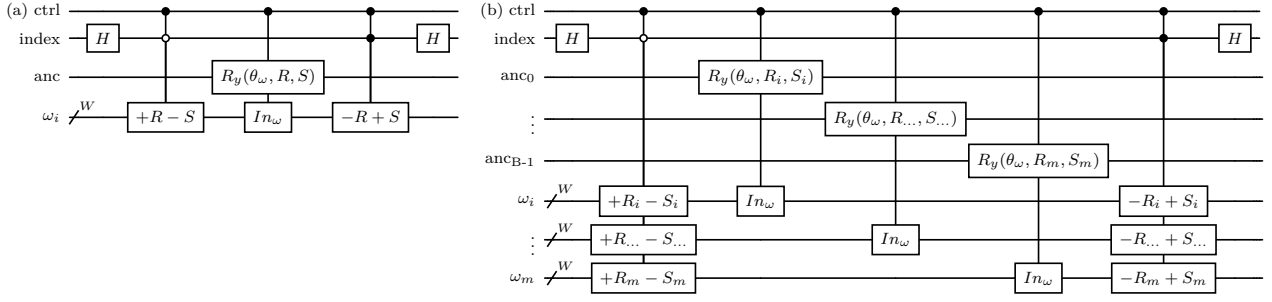

A block-encoding for this operator can be achieved by updating the occupation of the bosonic mode by $+R$ and then performing a single series of uniformly controlled rotations (subfigure \ref{fig:products-bosonic-operators}a).
The rotation angles are given by:
\begin{equation}
    \begin{split}
        &\theta(\omega_i, R) = \\
        &\begin{cases} 
            2\cos^{-1}\Big(\prod_{r=0}^{R-1}\sqrt{\frac{\omega_i - r}{\Omega}}\Big) & \text{when } \omega_i \leq \Omega - R \\
            \pi & \text{when } \omega_i > \Omega - R
        \end{cases}
    \end{split}
\end{equation}

A block-encoding for a bosonic annihilation operator being applied $S$ times can be achieved using a similar construction shown in subfigue \ref{fig:products-bosonic-operators}b.
The occupation of the mode is first decreased by $S$ and then the uniformly controlled rotations are applied to pick up the corresponding coefficient on the block-encoding ancilla.
The rotation angles are given by:
\begin{equation}
    \begin{split}
        &\theta(\omega_i, S) = \\
        &\begin{cases} 
            2\cos^{-1}\Big(\prod_{s=1}^{S}\sqrt{\frac{\omega_i + s}{\Omega}}\Big) & \text{when } \omega_i \geq S \\
            \pi & \text{when } \omega_i < S
        \end{cases}
        \\
        \\
    \end{split}
\end{equation}

A block-encoding for an operator of the form $(a_i^\dagger)^R (a_i)^S$ can be constructed following subfigue \ref{fig:products-bosonic-operators}c.
The occupation of the mode is updated by a value of $+ R - S$ and then the uniformly controlled rotations are applied to properly rotate the state outside the encoded subspace.
The rotation angles are given by:
\begin{widetext}
\begin{equation}
    \theta(\omega_i, R, S) = 
    \begin{cases} 
        2\cos^{-1}\Big( \big(\prod_{r=0}^{R-1}\sqrt{\frac{\omega_i - r}{\Omega}} \big) \big( \prod_{s=1}^{S}\sqrt{\frac{\omega_i - R + s}{\Omega}}\big)\Big) & \text{when } S \leq \omega_i \leq \Omega - R \\
        \pi & \text{Otherwise} 
    \end{cases}
\end{equation}
\end{widetext}

We discuss different implementations to increment a register by a classical value in Appendix \ref{sec:addition}.
We assume the compilation scheme that requires one uncontrolled addition circuit, using (at most) $3W$ $T$ gates \cite{gidney2018halving}.
In total, the block-encoding circuits shown in Figure \ref{fig:products-bosonic-operators} have a rescaling factor of $\lambda = \Omega^{(R+S)/2}$, require one block-encoding ancilla and $\lceil{\log_2{\Omega}}\rceil$ clean ancillae, and use (at most) $7W$ $T$ gates and $\Omega + 3$ arbitrary rotations.

%% file: text/lobe/bosonic_linear_combinations.tex
\subsection{Linear Combinations of Bosonic Ladder Operators}
\label{subsec:bosonic-lc}

In this subsection, we construct block-encodings for the sum of a product of bosonic ladder operators with its Hermitian conjugate.

An efficient block-encoding of the operator $a_i^\dagger + a_i$ can be constructed by noting the symmetry between the block-encodings for each ladder operator (Figure \ref{fig:bosonic-ladder-op-be}). 
The desired rotation angles are independent of which operator is being applied (Eq. \eqref{eq:single-op-angles}), so these rotations can be applied only once for both operators.
One additional block-encoding ancilla is required to index between the two operators.
This can be generalized to operators of the form: $(a_i^\dagger)^R (a_i)^S + (a_i^\dagger)^S (a_i)^R$.
A circuit diagram for this block-encoding is given in subfigure \ref{fig:lc-bosonic}a.

\begin{figure*}[t]
    \begin{quantikz}[column sep=0.15cm, row sep=0.1cm, baseline=(current bounding box.north), font=\scriptsize]
        \lstick{(a) ctrl}   &             &&&\ctrl{3}&\ctrl{3}&\ctrl{3}&\ctrl{2}&\\
        \lstick{anc$_a$}&             &&&& \gate{R_y(\theta_\omega)}&&&\\
        \lstick{$n_i$}      &             &&&\ctrl[open]{1}&&\ctrl{1}& \gate{\vec{Z}X_i}&\\
        \lstick{$\omega_j$} & \qwbundle{W}&&&\gate{+1} &\gate{In_{\omega}}&\gate{-1}&&
    \end{quantikz}
    \begin{quantikz}[column sep=0.15cm, row sep=0.1cm, baseline=(current bounding box.north), font=\scriptsize]
        \lstick{(b) ctrl}   &             &&&\ctrl{3}&\ctrl{5}&\ctrl{3}&&\ctrl{1}&&\ctrl{3}&\ctrl{4}&\ctrl{3}&\\
        \lstick{anc$_{b}$}&             &&&& &&&\targ{}&&&&&\\
        \lstick{anc$_{a_0}$}&             &&&& \gate{R_y(\theta_\omega)}&&&&&&&&\\
        \lstick{$n_i$}      &             &&&\ctrl{2}&&\ctrl[open]{2}& \ctrl{1} &&\ctrl{1}&\ctrl[open]{-3}& \gate[2]{\vec{Z}X_j}& \gate{\vec{Z}X_i}&\\
        \lstick{$n_j$}      &             &&&&&&\targ{}& \ctrl[open]{-3} & \targ{} & &&&\\
        \lstick{$\omega_k$} & \qwbundle{W}&&&\gate{+1} &\gate{In_{\omega}}&\gate{-1}&&&&&&&
    \end{quantikz}

    \begin{quantikz}[column sep=0.15cm, row sep=0.1cm, baseline=(current bounding box.north), font=\scriptsize]
        \lstick{(c) ctrl}   &             &&&\ctrl{7}&\ctrl{6}&\ctrl{7}&\ctrl{7}&&\ctrl{1}&&\ctrl{3}&\ctrl{4}&\ctrl{4}&\\
        \lstick{anc$_{b}$}&             &&&& &&&&\targ{}&&&&&\\
        \lstick{anc$_{a_0}$}&             &&&& \gate{R_y(\theta_\omega)}&&&&&&&&&\\
        \lstick{anc$_{a_1}$}&             &&&& &\gate{R_y(\theta_\omega)}&&&&&&&&\\
        \lstick{$n_i$}      &             &&&\ctrl{2}&&&\ctrl[open]{2}& \ctrl{1} &&\ctrl{1}&\ctrl[open]{-3}& \gate[2]{\vec{Z}X_j}& \gate{\vec{Z}X_i}&\\
        \lstick{$n_j$}      &             &&&&&&&\targ{}& \ctrl[open]{-4} & \targ{} & &&&\\
        \lstick{$\omega_k$} & \qwbundle{W}&&&\gate{+1} &\gate{In_{\omega}}&&\gate{-1}&&&&&&&\\
        \lstick{$\omega_l$} & \qwbundle{W}&&&\gate{+1} &&\gate{In_{\omega}}&\gate{-1}&&&&&&&
    \end{quantikz}
    \caption{
        \textbf{Block-Encoding Terms}
        In (a), a block-encoding for the operator $b_i^\dagger a_j^\dagger + a_j b_i$ is given.
        In (b), a block-encoding for the operator $ b_i^\dagger b_j^\dagger a_k + a_k^\dagger b_j b_i$ is given.
        In (c), a block-encoding for the operator $b_i^\dagger b_j^\dagger a_k a_l + a_l^\dagger a_k^\dagger b_j b_i$ is given.
    }
    \label{fig:be-term-example}
\end{figure*}
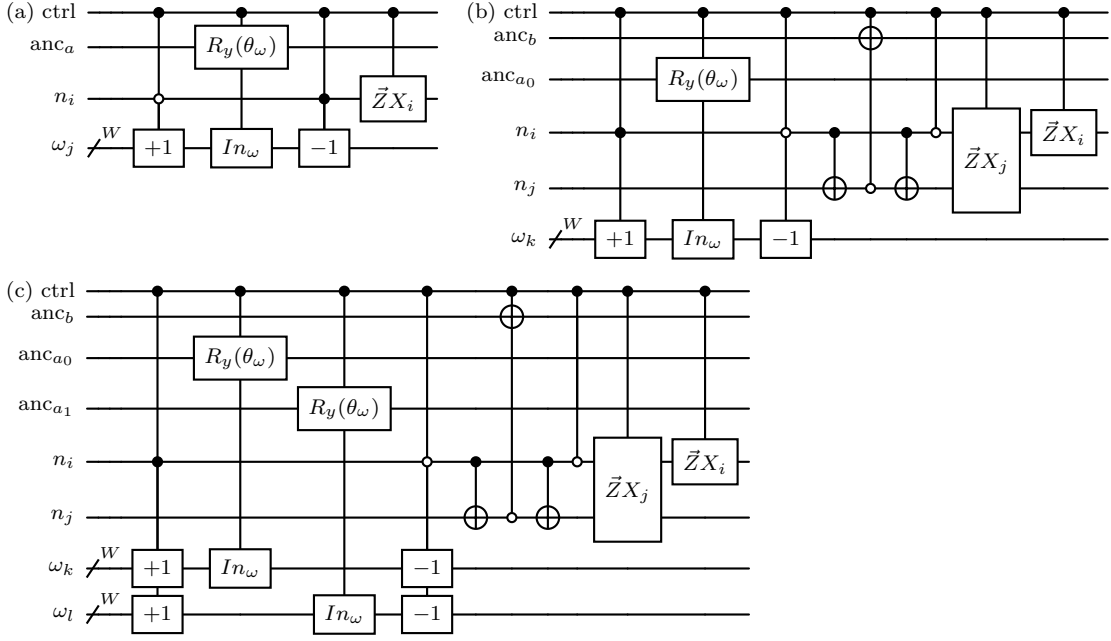

This construction can be generalized to a linear combination of a product of bosonic operators acting on \textit{different} modes plus its Hermitian conjugate.
A circuit diagram is shown in subfigure \ref{fig:lc-bosonic}b.
For an operator acting on $B$ bosonic modes, these block-encodings have a rescaling factor of $\lambda = 2 \Omega^{P/2}$ where $P$ is the sum of the exponents of the operators in a term: $P = \sum_{b=0}^{B-1}(R_b+S_b)$.
Additionally, they require $B+1$ block-encoding ancillae and $\lceil{\log_2{\Omega}}\rceil + 1$ clean ancillae, and use (at most) $12BW - 8B + 4$ $T$ gates and $B(\Omega + 3)$ arbitrary rotations.

%% file: text/lobe/interactions.tex
\subsection{Interactions Between Fermionic and Bosonic Modes}

In this subsection, we discuss block-encodings for operators that model interactions between both fermionic and bosonic modes.

Consider the desired action of the operator $b_i a_j + a_j^\dagger b_i^\dagger$.
When the fermionic mode is occupied (unoccupied), only $b_i a_j$ ($a_j^\dagger b_i^\dagger$) will act nontrivially.
The occupation of the fermionic mode can be used to dictate which bosonic operator should be applied to the system.
After the appropriate bosonic operator is applied, the fermionic system is updated.
A circuit diagram is given in subfigue \ref{fig:be-term-example}a.
This block-encoding circuit will have a rescaling factor of $\lambda = \sqrt{\Omega}$, requires one block-encoding ancillae and $W + 1$ clean ancillae, and uses $12W - 4$ $T$ gates and (at most) $\Omega + 3$ arbitrary rotations.

This strategy can be employed for other combinations of bosonic and fermionic ladder operators.
In Yukawa theory on the lightfront, the operator $b_i^\dagger b_j^\dagger a_k + a_k^\dagger b_j b_i$ is used to model the process of a boson being annihilated to form a fermion-antifermion pair and a fermion-antifermion pair being annihilated to form a boson \cite{kreshchuk2022quantum}.
A block-encoding for this operator is shown in subfigue \ref{fig:be-term-example}b.
This block-encoding circuit will have a rescaling factor of $\lambda = \sqrt{\Omega}$, requires two block-encoding ancillae and $W + 1$ clean ancillae, and uses $12W$ $T$ gates and (at most) $\Omega + 3$ arbitrary rotations.

A similar block-encoding circuit (subfigue \ref{fig:be-term-example}c) can be constructed for the operator $b_i^\dagger b_j^\dagger a_k a_l + a_l^\dagger a_k^\dagger b_j b_i$, which also appears in Yukawa theory on the lightfront \cite{kreshchuk2022quantum}.
This block-encoding circuit will have a rescaling factor of $\lambda = \Omega$, requires three block-encoding ancillae and $\lceil \log_2\Omega \rceil + 1$ clean ancillae, and uses $24W - 8$ $T$ gates and (at most) $2\Omega + 6$ arbitrary rotations.

%% file: text/results/main.tex
\section{Results}
\label{sec:results}

We numerically evaluate the quantum resources required to block-encode operators using LOBE.
The space-time costs that we analyze include: the number of $T$ gates, the number of non-Clifford single qubit rotations, the number of block-encoding ancillae, the maximum number of qubits required, and the rescaling factor imposed on the resulting block-encoding.

Non-Clifford rotations can be further decomposed into Clifford gates and $T$ gates using several methods \cite{Kitaev_1997,dawson_2006,gidney2018halving,campbell2021early}, which have different tradeoffs.
The number of $T$ gates for each non-Clifford rotation depends on the degree of accuracy of the rotation angle that is desired.
This degree of accuracy is highly dependent on the algorithm in which the block-encoding is being using and the degree of accuracy required for the overall computation. 
Additionally, the optimal method for decomposing rotations may be dependent on the algorithm since some methods are cheaper when the same rotations are applied multiple times within the algorithm.
For these reasons, we count non-Clifford rotations as a separate resource, allowing future works to determine the optimal compilation strategy.

We first compute the resources required to block-encode several classes of operators that appear in physically-motivated models.
These models include Hamiltonians derived from two non-relativistic models - the quartic harmonic oscillator \cite{bender1969, girgus2024, wojcik2012} and the static massive Yukawa model \cite{glazek2021} - and two fully relativistic models - $\phi^4$ theory \cite{vary2022} and the massive Yukawa model \cite{pauli_and_brodsky}.
Then, we compute the cost of block-encoding these Hamiltonians when using the LCO framework after generating block-encodings of the individual terms using LOBE.

The results are generated using several open-source software libraries.
LOBE \cite{lobe} provides methods to construct and validate block-encoding circuits, including those presented in this work. 
Circuits are implemented and simulated using Cirq \cite{cirq} and are numerically verified for block-encodings of up to 18 qubits.
OpenParticle \cite{openparticle} is used to construct and manipulate the operators in terms of fermionic, antifermionic, and bosonic ladder operators.
The Symmer software library \cite{Weaving_2025, PhysRevResearch.5.013095, doi:10.1021/acs.jctc.2c00910, Kirby2021contextualsubspace} is used for various subroutines, including expanding the ladder operators in the Pauli basis.
A separate software library \cite{grover-rudolph-github} is used to determine the rotation angles required for arbitrary state preparation using the Grover-Rudolph Algorithm \cite{grover2002creating}.

\input{text/results/pauli}
\input{text/results/components}

\input{text/results/qosc}
\input{text/results/static_yukawa}
\input{text/results/phi4}

\input{text/results/full_yukawa}

%% file: text/results/pauli.tex
\begin{figure*}[t]
    \centering
    \includegraphics[width=16cm]{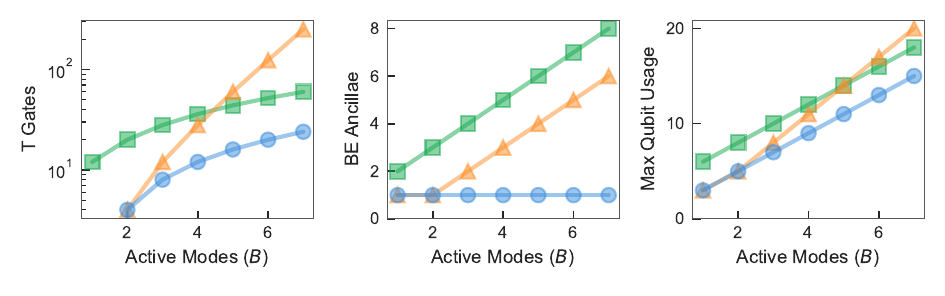}
    \caption{
        \textbf{Space-time Cost to Block-Encode $O = b_0 b_1 \hdots b_{B-1} + h.c.$}
        The number of $T$ gates (left), block-encoding ancillae (middle), and maximum number of qubits used (right) are shown as a function of the number of active modes ($B$).
        Results for Pauli Expansion are shown as the orange triangles, results for Piecewise Pauli are shown as the green squares, and results for LOBE are shown as the blue circles.
        All block-encodings use zero non-Clifford rotations.
        When $B = 1$ both Pauli Expansion and LOBE require zero $T$ gates.
        Pauli Expansion and LOBE have rescaling factors of $\lambda = 1$, while Piecewise Pauli has a rescaling factor of $\lambda = 2$.
    }
    \label{fig:fermionic-hc-comparison}
\end{figure*}

\subsection{Pauli Frameworks}

We benchmark the space-time cost of LOBE against LCU block-encodings, which expand the ladder operators in the Pauli basis.
Below, we outline two methods for constructing these LCU block-encodings that have different space-time costs. 

The Jordan-Wigner transformation \cite{jordan-wigner} is used to expand fermionic ladder operators in the basis of Pauli operators.
Under this transformation, a single fermionic ladder operator is mapped to a linear combination of two Pauli operators.

The Standard Binary encoding \cite{standard-binary} is used to expand bosonic ladder operators in the Pauli basis.
Under this transformation, a single bosonic ladder operator is mapped to a linear combination of $(\Omega+1) \lceil\log{\Omega}\rceil$ Pauli operators.

A straightforward way to apply the LCU framework to is to expand the entire operator being block-encoded in the Pauli basis.
This can be achieved by first transforming each ladder operator into a sum of Pauli operators and then expanding any products. 
This results in a single linear combination of Pauli operators.
We refer to this method as ``Pauli Expansion''.

In the worst-case, the number of Pauli operators scales exponentially with the number of active modes.
Expanding a product of $B$ fermionic ladder operators results in a total of $2^B$ Pauli operators.
Likewise, an expansion of bosonic ladder operators acting on $B$ active modes will result in $((\Omega+1) \lceil\log{\Omega}\rceil)^B$ Pauli operators.
The number of $T$ gates and non-Clifford rotations for LCU scales linearly with the number of unitaries, meaning these metrics will scale exponentially with the number of active modes.

When an operator is written as a sum of ladder operators acting the same modes, the ``Pauli Expansion'' method can result in cancellations of Pauli operators.
This is seen in the simple case of $b_i^\dagger + b_i$, which results in only a single Pauli operator.
This can significantly reduce the total number of Pauli operators, making this method favorable when there are a large number of terms, but only a small number of modes.

The second Pauli-based method we compare against avoids any exponential scaling with the number of active modes.
However, it does not allow for cancellations when an operator is written as a sum over multiple terms.
We refer to this method as ``Piecewise Pauli''.

In this method, each ladder operator is block-encoded individually by expanding it in the Pauli basis then using LCU.
Operators that are written as a product of ladder operators are then block-encoded by a product of their block-encodings following subsection \ref{subsec:be-products}.
Likewise, an operator written as a sum of a product of ladder operators can be block-encoded using the techniques discussed in subsection \ref{subsec:lco}.

%% file: text/results/components.tex
\subsection{Components}
\label{sec:components}

The first class of operators we examine are described by a product of fermionic annihilation operators acting on different modes plus its Hermitian conjugate: $(b_0 b_1 \hdots b_{B-1} + h.c.)$.
The LOBE constructions are described in subsection \ref{sec:LCFLO}.
The space-time costs to block-encode these operators are shown as a function of the number of active modes ($B$) in Figure \ref{fig:fermionic-hc-comparison}.

The space-time costs for LOBE and Pauli Expansion are identical for $B = 1$ and $B = 2$.
The number of required $T$ gates scales exponentially with $B$ for Pauli Expansion, yet scales linearly for both Piecewise Pauli and LOBE.
Additionally, the LOBE constructions only require a single block-encoding ancilla, while both Pauli methods require a number of block-encoding ancillae that scales linearly with $B$.

\begin{figure*}
    \centering
    \includegraphics[width=16cm]{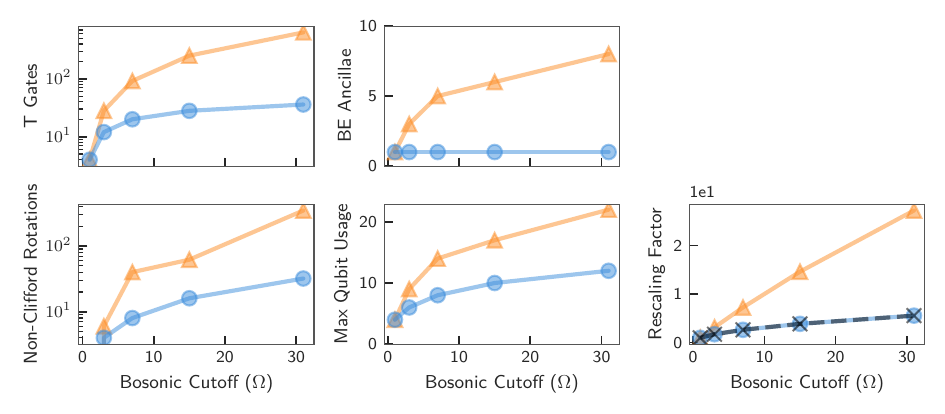}
    \caption{
        \textbf{Space-time Cost to Block-Encode Bosonic Annihilation Operator}
        The number of $T$ gates (upper-left), number of non-Clifford rotations (lower-left), block-encoding ancillae (upper-middle), maximum number of qubits used (lower-middle), and rescaling factor (lower-right) are shown as a function of the bosonic occupation cutoff ($\Omega$).
        Results for the Pauli method are shown as the orange triangles and results for LOBE are shown as the blue circles.
        The L2 norm of the matrix representing the operator is shown as the dashed black crosses. 
    }
    \label{fig:bosonic-comparison}
\end{figure*}

The second class of operators we consider is a single bosonic annihilation operator ($a$) with an increasing bosonic occupation cutoff ($\Omega$).
The LOBE construction is described in subsection \ref{subsec:bosonic-lobe}.
Since there is only a single operator, both Pauli methods result in the same construction, so we refer to them as ``Pauli''.
The space-time costs are shown in Figure \ref{fig:bosonic-comparison}.

LOBE results in block-encodings with fewer resources for all metrics we consider.
Notably, the number of required $T$ gates scales logarithmically with $\Omega$ for LOBE.
The number of $T$ gates required for the Pauli method scales roughly linearly, as it is dominated by the scaling of the number of Pauli operators in the expansion: $(\Omega+1) \lceil\log{\Omega}\rceil$.
The number of block-encoding ancillae for LOBE is constant ($1$), yet scales logarithmically for the Pauli method.
Lastly, LOBE results in a rescaling factor that matches the L2 norm, while the Pauli method results in a larger rescaling factor.

\begin{figure*}
    \centering
    \includegraphics[width=16cm]{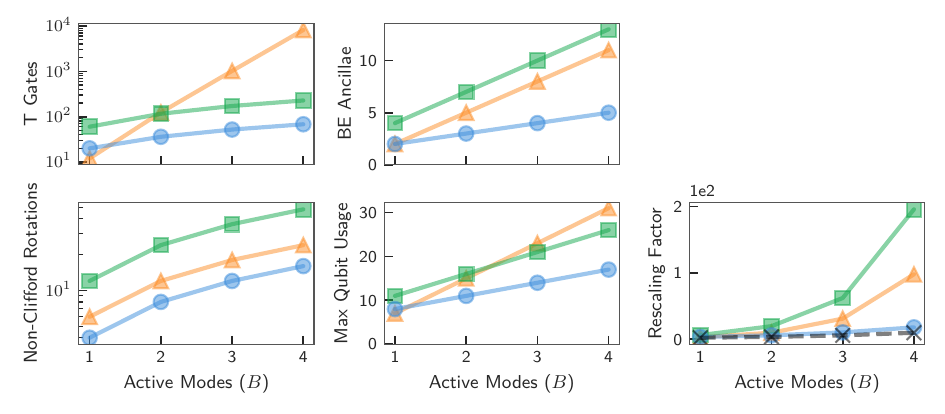}
    \caption{
        \textbf{Space-time Cost to Block-Encode $O = a_0 a_1 \hdots a_{B-1} + h.c.$}
        The number of $T$ gates (upper-left), number of non-Clifford rotations (lower-left), block-encoding ancillae (upper-middle), maximum number of qubits used (lower-middle), and rescaling factor (lower-right) are shown as a function of the number of active modes ($B$).
        The bosonic cutoff is fixed to $\Omega = 3$.
        Results for Pauli Expansion are shown as the orange triangles, results for Piecewise Pauli are shown as the green squares, and results for LOBE are shown as the blue circles.
        The L2 norm of the matrix representing the operator is shown as the dashed black crosses.
    }
    \label{fig:bosonic-hc-comparison}
\end{figure*}

The third class of operators we consider are given as a linear combination of a product of bosonic annihilation operators acting on different modes plus its Hermitian conjugate: $(a_0 a_1 \hdots a_{B-1} + h.c.)$.
The LOBE construction is described in subsection \ref{subsec:bosonic-lc}.
The space-time costs when $\Omega = 3$ are shown in Figure \ref{fig:bosonic-hc-comparison}.

The overall time-complexity ($T$ gates and non-Clifford rotations) scales linearly with $B$ for the LOBE and Piecewise Pauli methods, while it scales exponentially for Pauli Expansion.
For the space complexity (block-encoding ancillae and maximum qubit usage), all methods scale linearly with $B$, yet the LOBE constructions have both the smallest prefactor and values.
Finally, LOBE results in the lowest rescaling factors of all operators.

\begin{figure*}[t]
    \includegraphics[width = 16cm]{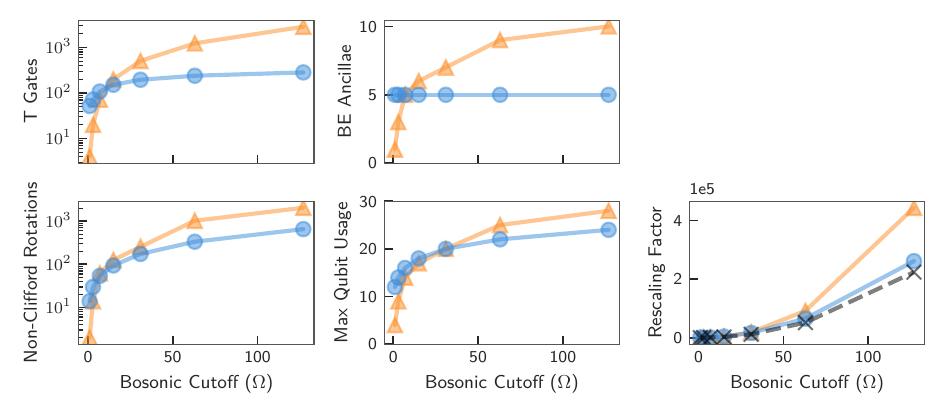}
    \caption{
        \textbf{Quartic Harmonic Oscillator}
        The number of $T$ gates (upper-left), number of non-Clifford rotations (lower-left), block-encoding ancillae (upper-middle), maximum number of qubits used (lower-middle), and rescaling factor (lower-right) are shown as a function of the bosonic occupation cutoff ($\Omega$).
        The parameter $g$ is set to $1$.
        Results for the Pauli Expansion method are shown as the orange triangles and results for LOBE are shown as the blue circles.
        The L2 norm of the Hamiltonian is shown as the dashed black crosses.
    }
    \label{fig:qosc}
\end{figure*}

In most cases we consider here, LOBE leads to the lowest space-time costs.
In certain cases - such as when the number of active modes ($B$) or the bosonic occupation cutoff ($\Omega$) are small - the Pauli Expansion method can lead to the lowest space-time costs.
For all components, Piecewise Pauli has similar asymptotic scalings compared to LOBE, but result in larger numerical costs.
For this reason, the Piecewise Pauli construction will be omitted when benchmarking full systems in the following subsections.

%% file: text/results/qosc.tex
\subsection{Quartic Harmonic Oscillator}
\label{sec:qosc_results}

The quartic harmonic oscillator \cite{bender1969, girgus2024,wojcik2012} is an extension of the standard harmonic oscillator.
This model is of particular interest as a preliminary test of renormalization of Hamiltonians via Gaussian elimination.
In a second-quantized, dimensionless form, the Hamiltonian can be written as:
\begin{equation}
    \label{eq:qosc}
    H = a^\dagger a + g\left(a + a^\dagger \right)^4
\end{equation}
where there is only one bosonic mode.

After expanding the product and normal ordering all terms, this Hamiltonian can be written as a linear combination of $9$ terms consisting of three pairs of operators plus their Hermitian conjugates, two operators that are their own Hermitian conjugates, and a constant offset:
\begin{equation}
    \begin{split}
        H = &(12g + 1) a^\dagger a + 6g a^{\dagger^2} a^2 + 6g \left(a^{\dagger^2} + a^2 \right) \\
        + &4g \left(a^{\dagger^3} a + a^\dagger a^3 \right) + g \left(a^{\dagger^4} + a^4 \right) + 3
    \end{split}
\end{equation}

The space-time costs as a function of the bosonic occupation cutoff ($\Omega$) are shown for LOBE and Pauli Expansion in Figure \ref{fig:qosc}.
For small values of $\Omega$, the Pauli Expansion block-encodings result in lower space-time costs, however, the LOBE constructions have favorable scaling and required fewer resources when $\Omega$ is large.
LOBE requires fewer non-Clifford rotations when $\Omega \geq 7$, fewer block-encoding ancillae when $\Omega > 7$, fewer $T$ gates when $\Omega \geq 15$, and a lower maximum number of qubits and rescaling factor when  $\Omega \geq 31$.

%% file: text/results/static_yukawa.tex
\subsection{Static Yukawa}
\label{sec:static_yukawa}

\begin{figure*}[t]
    \includegraphics[width = 16cm]{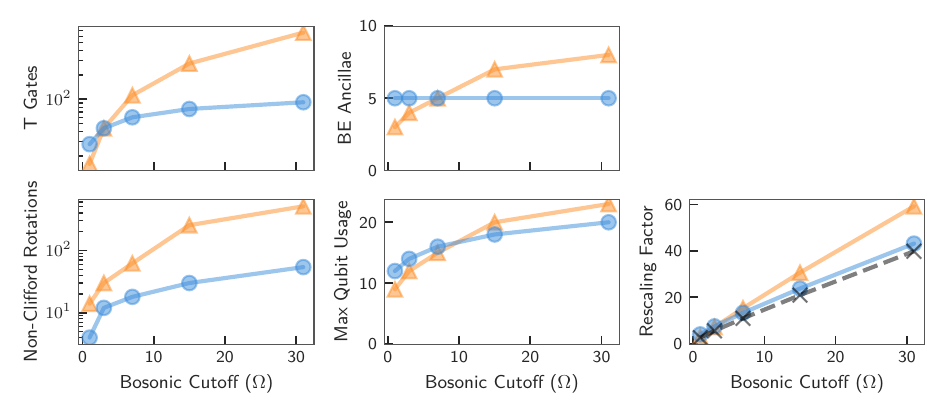}
    \caption{
        \textbf{Static Massive Yukawa}
        The number of $T$ gates (upper-left), number of non-Clifford rotations (lower-left), block-encoding ancillae (upper-middle), maximum number of qubits used (lower-middle), and rescaling factor (lower-right) are shown as a function of the bosonic occupation cutoff ($\Omega$).
        The parameters $C_f$, $C_b$, and $g$ are set to $1$.
        Results for the Pauli Expansion method are shown as the orange triangles and results for LOBE are shown as the blue circles.
        The L2 norm of the Hamiltonian is shown as the dashed black crosses.
    }
    \label{fig:static_yukawa}
\end{figure*}

The next model we consider is a non-relativistic approximation to the Yukawa model called the static Yukawa model \cite{glazek2021}.
This model is taken as the limit of the massive Yukawa model of infinitely heavy fermions, resulting in bosons at rest relative to the fermions that emit or absorb them.
The Renormalization Group Procedure for Effective Particles (RGPEP) \cite{glazek2012} is exactly solvable when applied to the static Yukawa model, making this model an interesting system to consider.

The second-quantized Hamiltonian for the static Yukawa model is given by:
\begin{equation}
    \label{eq:static-yukawa}
    H = C_f b^\dagger b + C_b a^\dagger a + g b^\dagger b \left( a + a^\dagger \right)
\end{equation}
where $C_f$ and $C_b$ are constants derived from the masses of the free fermion and free boson, respectively, and $g$ represents the strength of the fermion-boson interaction.
There is one fermionic mode and one bosonic mode, so the indices on each mode are omitted. 

We compare the space-time costs for block-encoding this Hamiltonian as a function of the bosonic occupation cutoff ($\Omega$).
These results are shown in Figure \ref{fig:static_yukawa}.
Similar to the quartic harmonic oscillator, the Pauli Expansion method requires fewer resources for some metrics when the bosonic cutoff is low.
As $\Omega$ increases, LOBE becomes more favorable.

LOBE requires fewer non-Clifford rotations at all values of $\Omega$, a lower rescaling factor when $\Omega \geq 3$, fewer $T$ gates when $\Omega > 3$, fewer block-encoding ancillae when $\Omega > 7$, and a lower maximum number of qubits when $\Omega \geq 15$.

%% file: text/results/phi4.tex
\subsection{$\phi^4$ Theory}
\label{sec:phi4_results}

$\phi^4$ theory is one of the simplest interacting field theories. 
This theory leads to quartic interactions between scalar particles.
In lightfront coordinates, the $\phi^4$ Hamiltonian can be written as:
\begin{align}
    H = \sum_i &c_i a_i^\dagger a_i + \sum_{ijkl}c_{ijkl} \left(a_i^\dagger a_j^\dagger a_k^\dagger a_l + h.c. \right) + \nonumber\\
    &\sum_{ijkl}\tilde c_{ijkl}a_i^\dagger a_j^\dagger a_k a_l
\end{align}
where the values of the coefficients can be determined analytically \cite{vary2022}.

\begin{figure*}
    \includegraphics[width = 16cm]{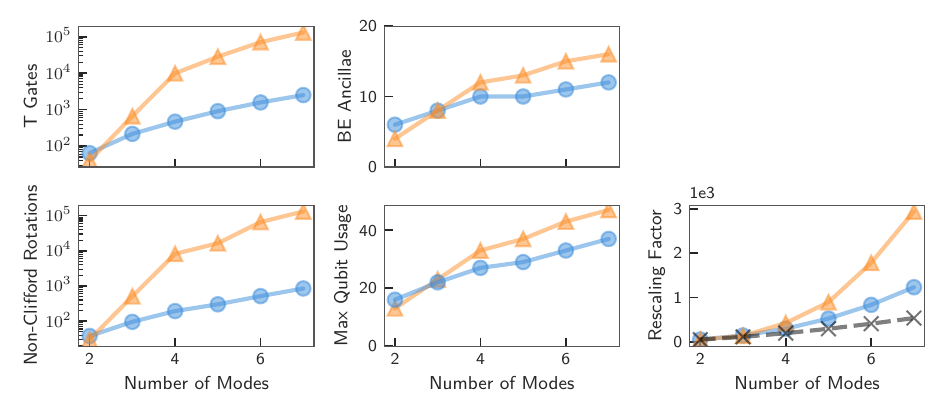}
    \caption{
        \textbf{$\phi^4$ Theory}
        The number of $T$ gates (upper-left), number of non-Clifford rotations (lower-left), block-encoding ancillae (upper-middle), maximum number of qubits used (lower-middle), and rescaling factor (lower-right) are shown as a function of the number of momentum modes.
        The bosonic cutoff is fixed to $\Omega = 3$ and the parameters $g$ and $m_b$ are set to $1$.
        Results for the Pauli Expansion method are shown as the orange triangles and results for LOBE are shown as the blue circles.
        The L2 norm of the Hamiltonian is shown as the dashed black crosses.
    }
    \label{fig:phi4}
\end{figure*}

Unlike the quartic harmonic oscillator and the static Yukawa model, $\phi^4$ theory on the light front is defined based on a discretization of momentum modes.
The results in subsections \ref{sec:components}, \ref{sec:qosc_results}, and \ref{sec:static_yukawa} suggest that LOBE is more efficient when $\Omega$ is large.
Here we analyze the space-time cost as a function of the total number of momentum modes.

In Figure \ref{fig:phi4}, we show the space-time costs for LOBE and Pauli Expansion as a function of the total number of modes.
We fix $\Omega = 3$, since this is the smallest non-trivial bosonic occupation cutoff.
Pauli Expansion is better suited for lower cutoffs, and we expect LOBE to be more efficient when $\Omega$ is large.

When the number of modes is small (low resolution), LOBE and Pauli Expansion require similar space-time costs.
For more modes (higher resolutions), LOBE requires signficantly fewer $T$ gates and non-Clifford rotations.
At high resolution, LOBE also requires fewer block-encoding ancillae and uses fewer total qubits, and has smaller rescaling factors.
Notably, when $7$ momentum modes are used, LOBE uses approximately two orders of magnitude fewer $T$ gates and non-Clifford rotations than Pauli Expansion.

%% file: text/results/full_yukawa.tex
\subsection{Yukawa Theory}
\label{sec:yukawa_results}

Finally, we compute the space-time cost to block-encode the massive Yukawa model.
This theory includes interactions between fermionic, antifermionic, and bosonic modes, and can be used as a model of the strong nuclear force between hadrons.  

The Hamiltonian in second quantization on the light front is given by:
\begin{align}
    \begin{split}
        H = &\sum_i c_i b_i^\dagger b_i + \sum_i \bar c_i d_i^\dagger d_i + \sum_i \tilde c_i a_i^\dagger a_i + \\
        &\sum_{ijk}c_{ijk}\left(b_i^\dagger b_j a_k^\dagger + h.c. \right) + \\
        &\sum_{ijk}\bar c_{ijk}\left(d_i^\dagger d_j a_k^\dagger + h.c. \right) + \\
        &\sum_{ijk}\tilde c_{ijk}\left(b_i^\dagger d_j^\dagger a_k + h.c. \right) + \\
        &\sum_{ijkl}c_{ijkl}b_i^\dagger b_j a_k^\dagger a_l + \\
        &\sum_{ijkl}\bar c_{ijkl}d_i^\dagger d_j a_k^\dagger a_l + \\
        &\sum_{ijkl}\tilde c_{ijkl}\left(b_i^\dagger d_j^\dagger a_k a_l + h.c. \right)
    \end{split}
\end{align}
where the values of the coefficients can be classically computed \cite{pauli_and_brodsky}.

\begin{figure*}
    \includegraphics[width = 16cm]{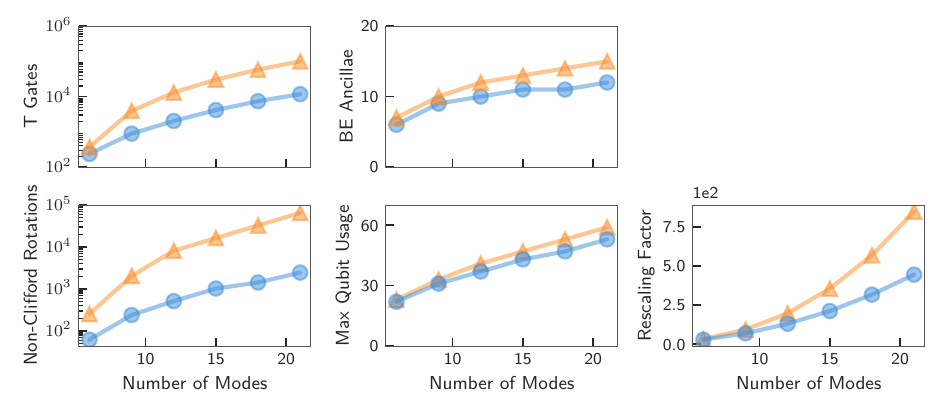}
    \caption{
        \textbf{Massive Yukawa}
        The number of $T$ gates (upper-left), number of non-Clifford rotations (lower-left), block-encoding ancillae (upper-middle), maximum number of qubits used (lower-middle), and rescaling factor (lower-right) are shown as a function of the number of momentum modes.
        The bosonic cutoff is fixed to $\Omega = 3$ and the parameters $m_f$, $m_b$, and $g$ are set to $1$.
        Results for the Pauli Expansion method are shown as the orange triangles and results for LOBE are shown as the blue circles.
    }
    \label{fig:full-yukawa}
\end{figure*}

In Figure \ref{fig:full-yukawa}, the space-time costs are shown as a function of the number of momentum modes.
The bosonic cutoff is fixed to the smallest non-trivial value ($\Omega = 3$), and we note that LOBE will be more efficient for larger cutoffs.
Even for $\Omega = 3$, LOBE results in fewer required quantum resources for all metrics.
Notably, the number of $T$ gates and number of non-Clifford rotations required for LOBE is smaller than those required by the Pauli Expansion by a constant factor.
These relative improvements will increase for larger cutoffs.

%% file: text/conclusions.tex
\section{Conclusions}
\label{sec:conclusions}

In this work, we detail a block-encoding framework that we refer to as LOBE (Ladder Operator Block-Encoding).
In this framework, quantum circuits to block-encode operators are constructed directly from the ladder operator representation.
We give explicit circuit compilations for operators written as linear combinations of products of ladder operators acting on both fermionic and bosonic modes and detail how these constructions can be generalized.
We provide an implementation of LOBE via an open-source library, allowing users to build block-encoding circuits following the LOBE framework \cite{lobe}.

LOBE explicitly avoids expanding ladder operators in the Pauli basis, which is often done in other block-encoding frameworks.
Block-encoding circuits are designed to update the system directly based on the action of the ladder operators, and rotate any unwanted effects outside of the encoded subspace.
This results in more efficient block-encoding circuits compared to techniques that require these operator transformations.

We compare LOBE to frameworks that require expressing the operators in the Pauli basis.
We provide analytical and numerical space-time costs for the relevant quantum resources required by LOBE.
Our numerical results show that LOBE often produces block-encodings that require significantly fewer $T$ gates, non-Clifford rotations, block-encoding ancillae and total number of qubits, and have smaller rescaling factors.
In addition, LOBE has favorable scaling with respect to the maximum occupation of bosonic modes, the number of momentum modes, and the locality of the operator.
In certain cases, such as when block-encoding a product of fermionic operators, LOBE results in exponentially fewer non-Clifford operations with respect to the locality.

Expanding operators in the Pauli basis can lead to operator cancellations that are not accounted for in LOBE.
This can make methods that expand operators in the Pauli basis more favorable, and it may be useful to incorporate these operator cancellations into the LOBE framework.
A more thorough comparison between these frameworks for operators with this form, such as those arising in quantum chemistry, is needed, and we leave this for future work.

The framework presented in this work allows for the construction of quantum circuits that generate block-encodings of operators written in terms of fermionic, antifermionic, and bosonic ladder operators.
Many operators arising in quantum mechanics, including quantum chemistry and quantum field theories, can be expressed efficiently in the basis of ladder operators.
This makes LOBE particularly relevant for constructing efficient quantum algorithms to simulate quantum systems.
By constructing more efficient circuits to block-encode these operators, this work opens another avenue to reduce resource estimates for quantum simulation algorithms.

%% file: text/acknowledgements.tex
\section{Acknowledgements}

We thank William Kirby, Michael Kreshchuk, Mason Rhodes, James Vary, Pieter Maris, Chao Yang and Weijie Du for productive discussions.
William A. Simon is supported by the Department of Defense (DoD) through the National Defense Science \& Engineering Graduate (NDSEG) Fellowship Program.
Carter Gustin was supported by the {\bf EXCL}usives via {\bf A}rtificial {\bf I}ntelligence and {\bf M}achine Learning (EXCLAIM) collaboration, DOE grant DE-SC0024644.
Kamil Serafin, Peter Love and  Gary Goldstein were supported by US DOE Grant DE-SC0023707 under the Office of Nuclear Physics Quantum Horizons program for the ``{\bf Nu}clei and {\bf Ha}drons with {\bf Q}uantum computers (NuHaQ)" project.
Alexis Ralli was supported by the STAQ project under award NSF-PHY-232580.
This material is based upon work supported by the U.S. Department of Energy (DOE), Office of Science, National Quantum Information Science Research Centers, Quantum Systems Accelerator.

%% file: text/appendices/glossary.tex
\section{Glossary}
\label{sec:glossary}

\begin{itemize}
    \item \textit{active mode}: A fermionic or bosonic mode upon which a ladder operator is being non-trivially applied. 
    \item \textit{block-encoding ancillae}: A register of qubits that give additional degrees of freedom to produce a block-encoding for the operator in a larger Hilbert space.
    \item \textit{encoded subspace}: The chosen subspace of the block-encoding ancillae that denotes the subspace in which the non-unitary operator is encoded. Typically, this is the subspace where all block-encoding ancillae are in the $\ket{0}$ state.
    \item \textit{clean ancillae}: A register of qubits that are promised to begin in the $\ket{0}$ and are returned to the $\ket{0}$ state at the end of a particular operation. 
    \item \textit{all-zero state}: A state of a register where all qubits are in the $\ket{0}$ state.
    \item $L$: The number of terms in the Hamiltonian.
    \item $\alpha_l$: The coefficient of the $l^\text{th}$ term in a linear combination. Assumed to be real and positive unless otherwise stated.
    \item $O_l$: The $l^\text{th}$ operator in a linear combination.
    \item $\Omega$: The occupation cutoff for the bosonic modes. $\Omega$ gives the maximum number of bosons that can be present in a single mode. 
    \item $I$: The number of fermionic or bosonic modes. The subscripts $b$ and $a$ will be used to denote the number of fermionic and bosonic modes respectively.
    \item $b_i$: Fermionic annihilation (creation - $b_i^\dagger$) operator acting on mode $i$.
    \item $d_i$: Antifermionic annihilation (creation - $d_i^\dagger$) operator acting on mode $i$.
    \item $a_i$: Bosonic annihilation (creation - $a_i^\dagger$) operator acting on mode $i$.
    \item $n_i$: The number of fermions occupying the $i^{th}$ mode.
    \item $\omega_{i}$: The number of bosons occupying the $i^{th}$ bosonic mode.
    \item $N_A$: The dimension of a matrix $A$.
    \item $U_A$: A unitary matrix providing a block-encoding of the matrix $A$.
    \item $\beta_\psi$: The amplitude of a state that is outside of the encoded subspace after a block-encoding unitary is applied to $\ket{\psi}$.
    \item $\lambda$: The rescaling factor imposed on the operator for a given block-encoding. This is also referred to as the subnormalization factor.
    \item $s$: The sparsity of a matrix.
    \item $\mathcal{S}$: An integer power-of-two for the sparsity of a matrix, obtained by treated zero-valued entries as nonzero. $\mathcal{S} = 2^{\lceil \log_2 s \rceil}$
    \item $W$: The number of qubits storing the occupation state of bosonic modes. $W = \lceil \log_2(\Omega+1) \rceil$. 
    \item $w$: The index of the $w^\text{th}$ least-significant bit in a binary encoding.
    \item $B$: The number of active modes within an operator.
    \item $C$: The number of active modes within an operator \textit{excluding} modes upon which a number operator is being applied.
    \item $S_{l, i}$: The exponent of bosonic annihilation operators acting on the $i^{th}$ bosonic mode within the $l^{th}$ term.
    \item $R_{l, i}$: The exponent of bosonic creation operators acting on the $i^{th}$ bosonic mode within the $l^{th}$ term.
    \item $P$: The sum of the exponents of bosonic ladder operators acting on all modes within a single term.
\end{itemize}

%% file: text/appendices/usp.tex
\section{Uniform State Preparation}
\label{sec:usp}

In the interests of keeping this paper self-contained, we present an algorithm for the preparation of the uniform superposition of $n$ logical basis states $\ket{l}$ for $0\leq l\leq n-1$. That is, we obtain a quantum circuit for the operator $S_n$:
\begin{equation}
S_n\ket{0}^{\otimes b} =\frac{1}{\sqrt{n}}\sum_{l=0}^{n-1} \ket{l}. 
\end{equation}
If $n=2^b$ it is well known that this can be accomplished by the application of $b$ Hadamard gates~\cite{nielsen2001quantum}:
\begin{equation}\label{had}
\frac{1}{\sqrt{2^b}}\sum_{l=0}^{2^b-1}\ket{l} = \left(H\ket{0}\right)^{\otimes b}.
\end{equation}
The superposition for $n\neq 2^b$ can also be prepared efficiently by the method of~\cite{kitaev1995quantum}. First, we take $b=\lceil \log_2 n\rceil$. Write $n=2^{b-1} +n_1$ and proceed recursively:
\begin{equation}
\frac{1}{\sqrt{n}}\sum_{l=0}^{n-1} \ket{l} = \frac{1}{\sqrt{n}}\sum_{l=0}^{2^{b-1}-1} \ket{l}+\frac{1}{\sqrt{n}}\sum_{l=2^{b-1}}^{n-1} \ket{l}
\end{equation}
the highest of the $b$ bits is always zero in the first sum, and always one in the second sum. Hence we may write:
\begin{equation}
\frac{1}{\sqrt{n}}\sum_{l=0}^{n-1} \ket{l} = \sqrt{\frac{2^{b-1}}{n}}\ket{0}\frac{1}{\sqrt{2^{b-1}}}\sum_{l=0}^{2^{b-1}-1} \ket{l}+\sqrt{\frac{n_1}{n}}\ket{1}\frac{1}{\sqrt{n_1}}\sum_{l=0}^{n_1-1} \ket{l}.
\end{equation}
This gives a recursive procedure for the definition of $S_n$ in terms of $S_{n_1}$:

\begin{equation}
\begin{split}
S_n\ket{0}^{\otimes b} &=\sqrt{\frac{2^{b-1}}{n}}\ket{0} \left(H\ket{0}\right)^{\otimes {b-1}} + \sqrt{\frac{n_1}{n}}\ket{1}S_{n_1}\ket{0}^{\otimes b-1}\\
&= \left[\Lambda_0(H^{\otimes b-1}) +\Lambda_1(S_{n_1})\right]\biggl[ \sqrt{\frac{2^{b-1}}{n}}\ket{0} + \sqrt{\frac{n_1}{n}}\ket{1} \biggr]\ket{0}^{\otimes b-1}\\
\end{split}
\end{equation}

\begin{figure}
\begin{center}
\includegraphics[width=4in]{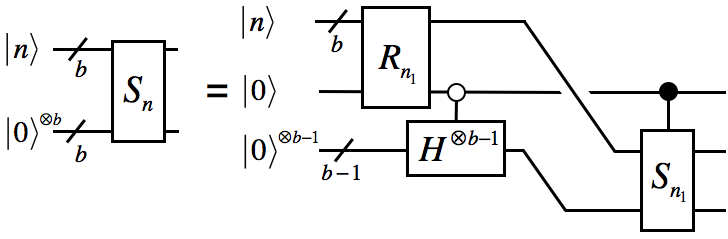}
\caption{\label{S} The quantum circuits showing the recursive definition of the algorithm to prepare a uniform superposition of the first $n$ logical basis states. Here $n_1=n-2^{\lfloor \log_2 n \rfloor}$, $R_{n_1}$ is the rotation defined in equation~\ref{rn} where $n$ is a parameter for the rotation, and $H$ is the Hadamard gate. This recursive procedure is applied until $n_1$ is a power of two.}
\end{center}
\end{figure}
So that we may write:
\begin{equation}
S_n = \biggl[\Lambda_0(H^{\otimes b-1}) +\Lambda_1(S_{n_1})\biggr]\biggl[R_n\otimes 1^{\otimes b-1}\biggr]
\end{equation}
where the single qubit rotation $R_n$ is defined by:
\begin{equation}\label{rn}
R_n = \frac{1}{\sqrt{n}}\begin{pmatrix} \sqrt{2^{b-1}}& \sqrt{n_1}\\ \sqrt{n_1}& \sqrt{2^{b-1}}\end{pmatrix}
\end{equation}

This procedure, which is represented by the circuit in Figure~\ref{S}, is repeated until $n_1$ is a power of two, in which case the method of eq.~\ref{had} is used. The cost of this procedure is maximal when $n=2^a-1$, in which case at each stage of the recursion $n_1$ is also of the form $2^{a'}-1$, and asymptotically the cost scales as $\mathcal{O}(b^2)$. In addition to this, if $n$ is provided in a second register, there will be an additional ancilla cost due to the need to coherently compute the angles of the rotations $R_{n_1}$ so that they can be applied by phase kickback.

%% file: text/appendices/compiling_toffolis.tex
\section{Compiling Toffoli Gates}
\label{sec:elbows}

\begin{figure*}
    \begin{center}
    \begin{quantikz}[column sep=0.25cm, row sep=0.25cm, font=\scriptsize]
        && \ghost{H} & \ctrl{1} &&\\
        && \ghost{H} & \ctrl{1} &&\\
        && \ghost{H} & \ctrl{3} &&\\
        &\ghost{H}\wireoverride{}&\wireoverride{}&\wireoverride{}&\wireoverride{}&\wireoverride{} \\
        &\ghost{H}\wireoverride{}&\wireoverride{}&\wireoverride{}&\wireoverride{}&\wireoverride{} \\
        \lstick{$\psi$}& \qwbundle{} & \ghost{H} & \gate{U} &&
    \end{quantikz} = 
    \begin{quantikz}[column sep=0.25cm, row sep=0.25cm, font=\scriptsize]
        && \ghost{H} & \ctrl{1} && \ctrl{1} &&&\\
        && \ghost{H} & \ctrl{1} && \ctrl{1} &&&\\
        && \ghost{H} & \ctrl{1} && \ctrl{1} &&&\\
        &\ghost{H}\wireoverride{}&\wireoverride{}\lstick{$\ket{0}$}& \targ{} & \ctrl{2} & \targ{} &\rstick{$\ket{0}$}& \wireoverride{}& \wireoverride{}\\
        &\ghost{H}\wireoverride{}&\wireoverride{}&\wireoverride{}&\wireoverride{}&\wireoverride{}&\wireoverride{}&\wireoverride{}& \wireoverride{}\\
        \lstick{$\psi$}& \qwbundle{} & \ghost{H} && \gate{U} &&&&
    \end{quantikz} = 
    \begin{quantikz}[column sep=0.25cm, row sep=0.25cm, font=\scriptsize]
        && \ghost{H} & \ctrl{1} &&&& \ctrl{1} &&&\\
        && \ghost{H} & \ctrl{2} &&&& \ctrl{2} &&&\\
        && \ghost{H} &          & \ctrl{1} &  & \ctrl{1} & &&&\\
        &\wireoverride{}&\wireoverride{}\lstick{$\ket{0}$} & \targ{}  & \ctrl{1} & \ctrl{2} & \ctrl{1} & \targ{} & \rstick{$\ket{0}$}&\wireoverride{}&\wireoverride{}\\
        &\wireoverride{}& \wireoverride{} & \wireoverride{}\lstick{$\ket{0}$}  & \targ{} & \ctrl{1} & \targ{} & \rstick{$\ket{0}$}&\wireoverride{}&\wireoverride{}&\wireoverride{}\\
        \lstick{$\psi$}& \qwbundle{} & \ghost{H} &&  & \gate{U} &&&&&
    \end{quantikz}
    \end{center}
    \caption{
        \textbf{Multi-Controlled Operations}
        Operations with multiple controls are referred to as \textit{multi-controlled operations}.
        An $N$-controlled operation can be decomposed into $2(N-1)$ Toffoli gates using $N-1$ clean ancillae.
    }
    \label{fig:multi-controlled-op}
\end{figure*}
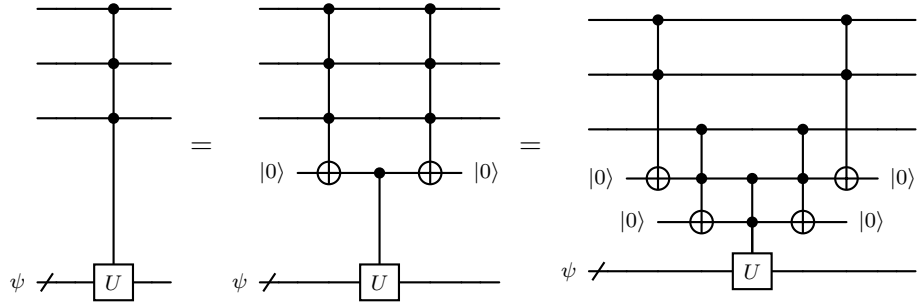

The Toffoli gate is a ubiquitous operation in quantum algorithms and is often a significant contributor to the number of non-Clifford resources.
In many algorithms - including those described in this work - multi-controlled Toffoli gates are used to control an operation on the logical-AND of several variables.
Even small reductions in the cost of compiling a multi-controlled Toffoli gate can have a significant impact on the overall space-time cost of an algorithm.

In an effort to keep this work self-contained, we review a specific compilation technique that can reduce the number of $T$ gates required to synthesize a pair of $N$-controlled Toffoli gates.
Mulit-controlled Toffoli gates can be decomposed as a series of Toffoli gates (Figure \ref{fig:multi-controlled-op}).
More efficient schemes have been recently proposed \cite{gosset2025}, which specifically apply to multi-controlled Toffoli gates, though we do not review these techniques here. 

In Nielsen and Chuang \cite{nielsen2001quantum}, a method for synthesizing a single Toffoli gate from $7$ $T$ gates is shown, which is likely derived from the methods proposed by Barenco et al. \cite{barenco1995elementary}.
Selinger \cite{selinger2013quantum} and Jones \cite{jones2013low} reduce this cost for a single Toffoli gate from $7$ $T$ gates to $4$ using one clean ancilla.
To the best of our knowledge, this is the most T-gate efficient compilation for a single Toffoli gate.

\begin{figure}
    \begin{center}
        \includegraphics[width=8cm]{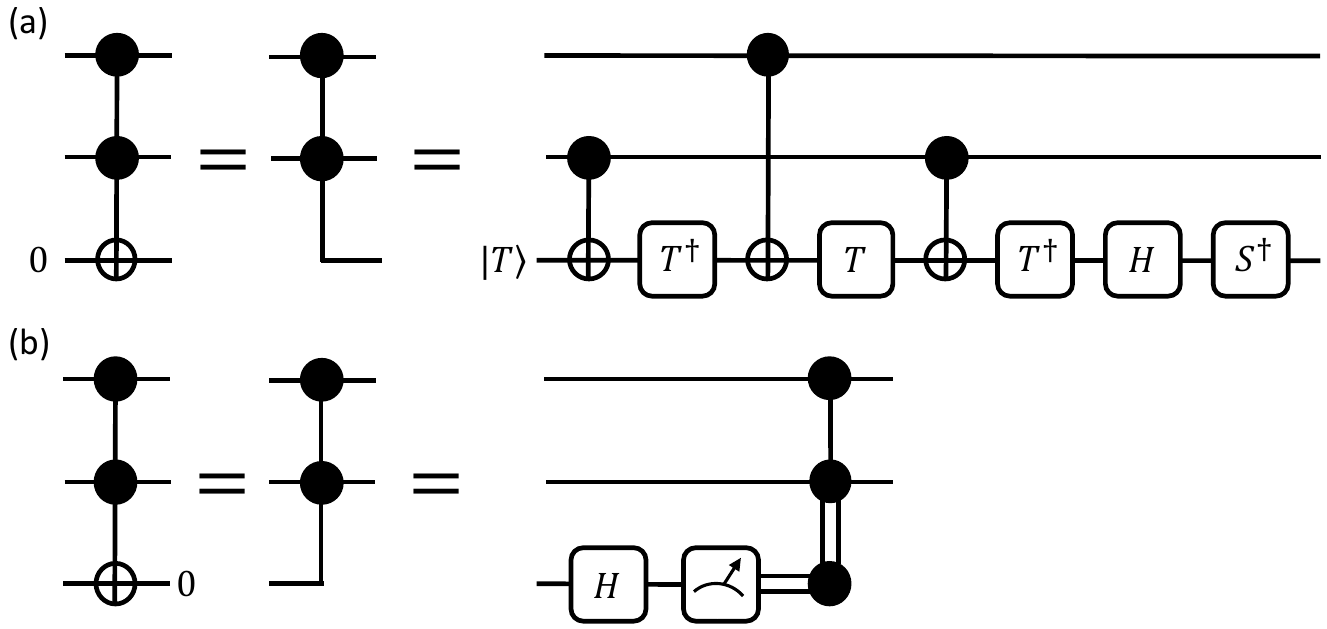}
    \end{center}
    \caption{
        \textbf{Elbows}
        In (a), a compilation scheme for a Toffoli gate acting on a clean ancillae is shown that uses four $T$ gates.
        This is sometimes referred to as a ``left-elbow''.
        In (b), a compilation scheme for a Toffoli gate that uncomputes a clean ancillae is shown that uses a measurement and classically conditioned operation.
        This is sometimes referred to as a ``right-elbow''.
    }
    \label{fig:elbows}
\end{figure}

Gidney \cite{gidney2018halving} showed that a pair of Toffoli gates which subsequently compute and uncompute the logical-AND of two variables can be synthesized using $4$ $T$ gates, a measurement, and a classically conditioned operation (Figure 3 of \cite{gidney2018halving}).
This compilation scheme is sometimes referred to as ``elbows'' and is depicted in Figure \ref{fig:elbows}.

Following the methods shown by Barenco et al. \cite{barenco1995elementary} and Gidney \cite{gidney2018halving}, a pair of $N$-controlled Toffoli gates that are used to compute and uncompute the logical-AND of the same variables can be composed using $4(N-1)$ $T$ gates and $N-1$ clean ancilla using a series of elbows.
It is important to note that this technique does not always apply to neighboring Toffoli gates, but sufficient conditions are given in Figure 5 of \cite{gidney2018halving}.
The space-time costs for the block-encoding circuits used in this work assume this strategy when applicable.

\begin{figure}
    \begin{center}
    \begin{quantikz}[column sep=0.25cm, row sep=0.4cm, font=\scriptsize]
        \lstick{ctrl}  &             &\ghost{H}& \ctrl{3} &&\\
        \lstick{l$_1$} &             &\ghost{H}& \gate[2]{In_l} &&\\
        \lstick{l$_0$} &             &\ghost{H}&    &&\\
        \lstick{$\psi$}& \qwbundle{} &\ghost{H}& \gate{U_l} &&
    \end{quantikz} = 
    \begin{quantikz}[column sep=0.25cm, row sep=0.4cm, font=\scriptsize]
        &\ghost{H}& \ctrl{1} & \ctrl{1} & \ctrl{1} & \ctrl{1} &&\\
        &\ghost{H}& \ctrl[open]{1} & \ctrl[open]{1} & \ctrl{1} & \ctrl{1} &&\\
        &\ghost{H}& \ctrl[open]{1} & \ctrl{1} & \ctrl[open]{1} & \ctrl{1} &&\\
        &\ghost{H}& \gate{U_0} & \gate{U_1} & \gate{U_2} & \gate{U_3} &&
    \end{quantikz}
    \caption{
        \textbf{Uniformly Controlled Operations}
        A series of operations that are subsequently controlled on the computational basis states of a register are referred to as \textit{uniformly controlled operations}.
    }
    \label{fig:uniformly-controlled-ops}
    \end{center}
\end{figure}
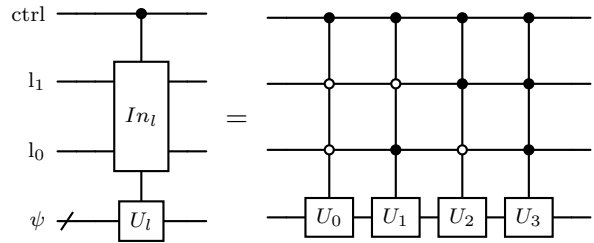

Additionally, there are several optimizations that can reduce the cost of neighboring pairs of elbows which are shown in Figure 6 of Babbush et al. \cite{babbush2018encoding}.
These optimizations are particularly useful when compiling uniformly controlled operations (Figure \ref{fig:uniformly-controlled-ops}).
With these optimizations, the control structure for applying $L$ uniformly controlled operations uses $4(L-1)$ $T$ gates.
The space-time costs quoted throughout this work assume this strategy when applicable, including in cases such as Figures \ref{fig:lc-bosonic} and \ref{fig:be-term-example}.

%% file: text/appendices/addition.tex
\section{Addition by a Known Classical Value}
\label{sec:addition}

Adding a (known) classical integer ($m$) to a quantum register encoded in binary is a required operation throughout this work:
\begin{equation}
    \label{eq:addition-by-classical-value}
    \ket{n} \rightarrow \ket{n + m}
\end{equation}
In an effort to keep this work self-contained and pedagogical, we review some methods for constructing this operation.
Since $m$ is known, the cost of all methods can be determined during compilation and the most favorable method can be chosen.

\begin{figure}
    \centering
    \includegraphics[width=8cm]{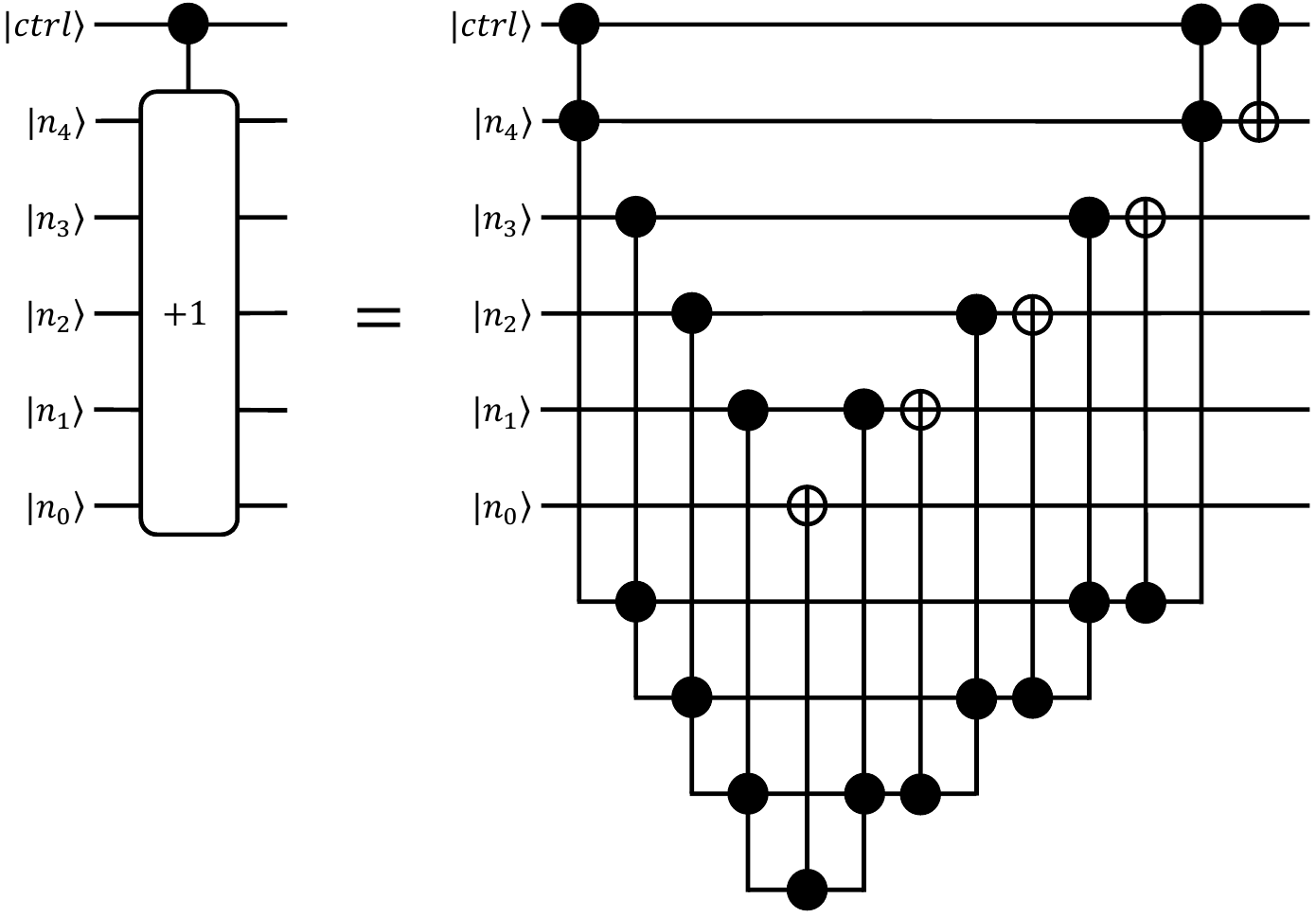}
    \caption{
        \textbf{Controlled Incrementer} 
        A circuit diagram implementing a controlled incrementer (mod $32$) is shown.
        The controlled incrementer performs the operation $\ket{x} \rightarrow \ket{x + 1}$ when the control qubit is in the $\ket{1}$ state.
        Decrementing the register by $1$ can be achieved by applying Pauli-X gates on each qubit before and after the operation.
    }
    \label{fig:incrementer}
\end{figure}

One option is to use a series of controlled incrementer ($+1$) circuits.
An efficient implementation of an incrementer circuit given by Gidney \cite{Gidney_2015} is shown in Figure \ref{fig:incrementer}.
If $N$ is the number of qubits in the register being incremented, this implementation requires $4(N-1)$ $T$ gates and $N-1$ clean ancillae.
Naively, a series of $m$ incrementers will result in increasing the value of the register by $m \mod 2^N$.

\begin{figure}
    \begin{center}
    \begin{quantikz}[column sep=0.25cm, row sep=0.15cm, font=\scriptsize]
        \lstick{ctrl}  && \ghost{H} & \ctrl{1} &&\\
        \lstick{n$_4$} && \ghost{H} & \gate[5]{+11} &&\\
        \lstick{n$_3$} && \ghost{H} &  &&\\
        \lstick{n$_2$} && \ghost{H} &  &&\\
        \lstick{n$_1$} && \ghost{H} &  &&\\
        \lstick{n$_0$} && \ghost{H} &    &&
    \end{quantikz} = 
    \begin{quantikz}[column sep=0.25cm, row sep=0.15cm, font=\scriptsize]
        \lstick{ctrl}  & \ghost{H} & \ctrl{1} & \ctrl{1} & \ctrl{1} &&\\
        \lstick{n$_4$} & \ghost{H} & \gate[2]{+1} & \gate[4]{+1} & \gate[5]{+1} &&\\
        \lstick{n$_3$} & \ghost{H} &&&&&\\
        \lstick{n$_2$} & \ghost{H} &&&&&\\
        \lstick{n$_1$} & \ghost{H} &&&&&\\
        \lstick{n$_0$} & \ghost{H} &&&&&
    \end{quantikz}
    \end{center}
    \caption{
        \textbf{Addition via Incrementers} 
        An implementation of addition by a classical value ($11$) (mod $32$) using a series of incrementers is shown.
        An incrementer applied onto a register excluding the least-significant qubit implements a bit-shifted incrementer.
        This effectively increases the value of the register by $2$.
        Addition by the classical value $11$ can be constructed by bit-shifted incrementers adding the values $+8$, $+2$, and $+1$.
        Subtraction by the same value can be achieved by applying Pauli-X gates on each qubit before and after the operation.
    }
    \label{fig:addition-via-incrementers}
\end{figure}
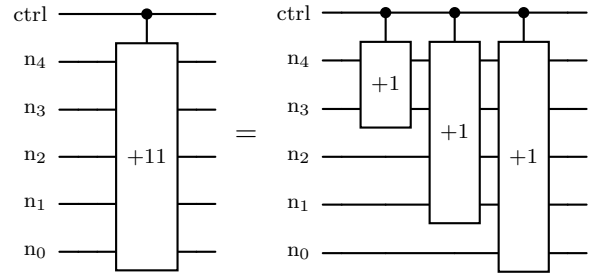

However, an incrementer circuit can also be used to perform addition by a power of $2$ by acting on only the most-significant qubits.
For example, adding the value $8$ can be achieved using an incrementer circuit that disregards the $3$ least-significant qubits.
A circuit adding any classical value can be constructed based on the binary representation of the classical number (Figure \ref{fig:addition-via-incrementers}).
In the worst-case, this construction requires $N$ incrementer circuits, totaling $4 \sum_{i=0}^{N - 2} (N - i - 1)$ $T$ gates and $N - 1$ clean ancillae.

When performing modular addition, the same effect can be achieved by subtracting the value $2^N - m$.
As an example, if the classical value is $31$ and $N = 5$, then this can be accomplished using a single decrementer circuit.
A decrementer can be constructed by conjugating an incrementer with X gates acting on the qubits encoding $\ket{n}$.

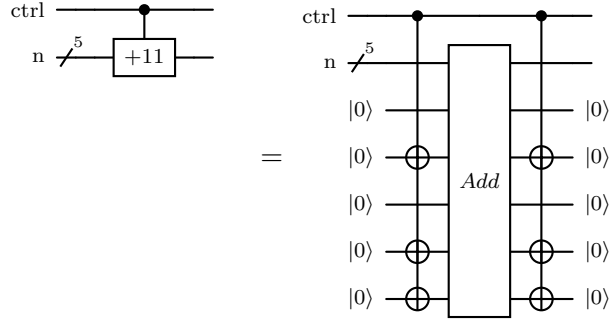
\begin{figure}
    \begin{center}
    \begin{quantikz}[column sep=0.25cm, row sep=0.15cm, font=\scriptsize]
        \lstick{ctrl}  && \ghost{H} & \ctrl{1} &&\\
        \lstick{n} & \qwbundle{5} & \ghost{H} & \gate[1]{+11} &&\\
        & \ghost{H}\wireoverride{}&\wireoverride{} & \wireoverride{}  &\wireoverride{}& \wireoverride{} & \wireoverride{} & \wireoverride{}\\
        & \ghost{H}\wireoverride{}&\wireoverride{} & \wireoverride{}  &\wireoverride{}& \wireoverride{} & \wireoverride{} & \wireoverride{}\\
        & \ghost{H}\wireoverride{}&\wireoverride{} & \wireoverride{}  &\wireoverride{}& \wireoverride{} & \wireoverride{} & \wireoverride{}\\
        & \ghost{H}\wireoverride{}&\wireoverride{} & \wireoverride{}  &\wireoverride{}& \wireoverride{} & \wireoverride{} & \wireoverride{}\\
        & \ghost{H}\wireoverride{}&\wireoverride{} & \wireoverride{}  &\wireoverride{}& \wireoverride{} & \wireoverride{} & \wireoverride{}\\
    \end{quantikz} = 
    \begin{quantikz}[column sep=0.25cm, row sep=0.15cm, font=\scriptsize]
        \lstick{ctrl}  &                  & \ghost{H} & \ctrl{6} && \ctrl{6} &&\\
        \lstick{n}     & \qwbundle{5}     & \ghost{H} &          & \gate[6]{Add} &&&\\
        & \ghost{H}\wireoverride{}&\wireoverride{}\lstick{$\ket{0}$} &          &&& \rstick{$\ket{0}$} & \wireoverride{}\\
        & \ghost{H}\wireoverride{}&\wireoverride{}\lstick{$\ket{0}$} & \targ{}  && \targ{} & \rstick{$\ket{0}$} & \wireoverride{}\\
        & \ghost{H}\wireoverride{}&\wireoverride{}\lstick{$\ket{0}$} &   &&  & \rstick{$\ket{0}$} & \wireoverride{}\\
        & \ghost{H}\wireoverride{}&\wireoverride{}\lstick{$\ket{0}$} &   \targ{}       &&\targ{}& \rstick{$\ket{0}$} & \wireoverride{}\\
        & \ghost{H}\wireoverride{}&\wireoverride{}\lstick{$\ket{0}$} & \targ{}  && \targ{} & \rstick{$\ket{0}$} & \wireoverride{}
    \end{quantikz}
    \end{center}
    \caption{
        \textbf{Time-Efficient Controlled Addition of $11$}
        Increasing the value of a quantum register by a known classical value can be implemented using clean ancillae and an uncontrolled quantum addition circuit.
        The known classical value is loaded into a clean ancilla register using a series of CNO$T$ gates corresponding to the binary representation of the classical value.
        Then an uncontrolled quantum addition circuit is applied to the two registers.
        Finally, the loading of the classical value is uncomputed.
    }
    \label{fig:addition-gate-efficient}
\end{figure}

Another option is to load the classical value ($m$) into an clean ancilla register, controlled on the control qubit, and perform \textit{uncontrolled} addition on the two registers.
After the addition, the classical value can be ``unloaded''.
If the control is off, then the classical value is not loaded and the addition simply adds the value $0$, leaving the register unchanged.
An example diagram for this construction depicting adding the value $m = 11$ to a register with $N = 5$ qubits is shown in Figure \ref{fig:addition-gate-efficient}.

Loading the classical value into an ancilla register with $N$ qubits can be achieved using at most $N$ CNOTs.
If the $p$ least significant bits of $m$ are zero, then only $N - p$ qubits are required and the $p$ least-significant qubits of the register storing $\ket{n}$ can be omitted from the addition.
Uncontrolled addition of two registers can be performed using $4(N-1)$ $T$ gates and $N - 1$ clean ancillae using the construction for addition shown in Figure 1 of \cite{gidney2018halving}.
In total, this compilation requires $4(N - p - 1)$ $T$ gates and $2(N - p) - 1$ clean ancillae.

When $m$ is a power of $2$, then compilation using incrementer circuits uses the same number of $T$ gates, but fewer clean ancillae.
When $m$ is not a power of $2$, then the compilation using uncontrolled quantum addition uses fewer $T$ gates at the expense of more clean ancillae.

\begin{figure*}
    \centering
    \includegraphics[width=12cm]{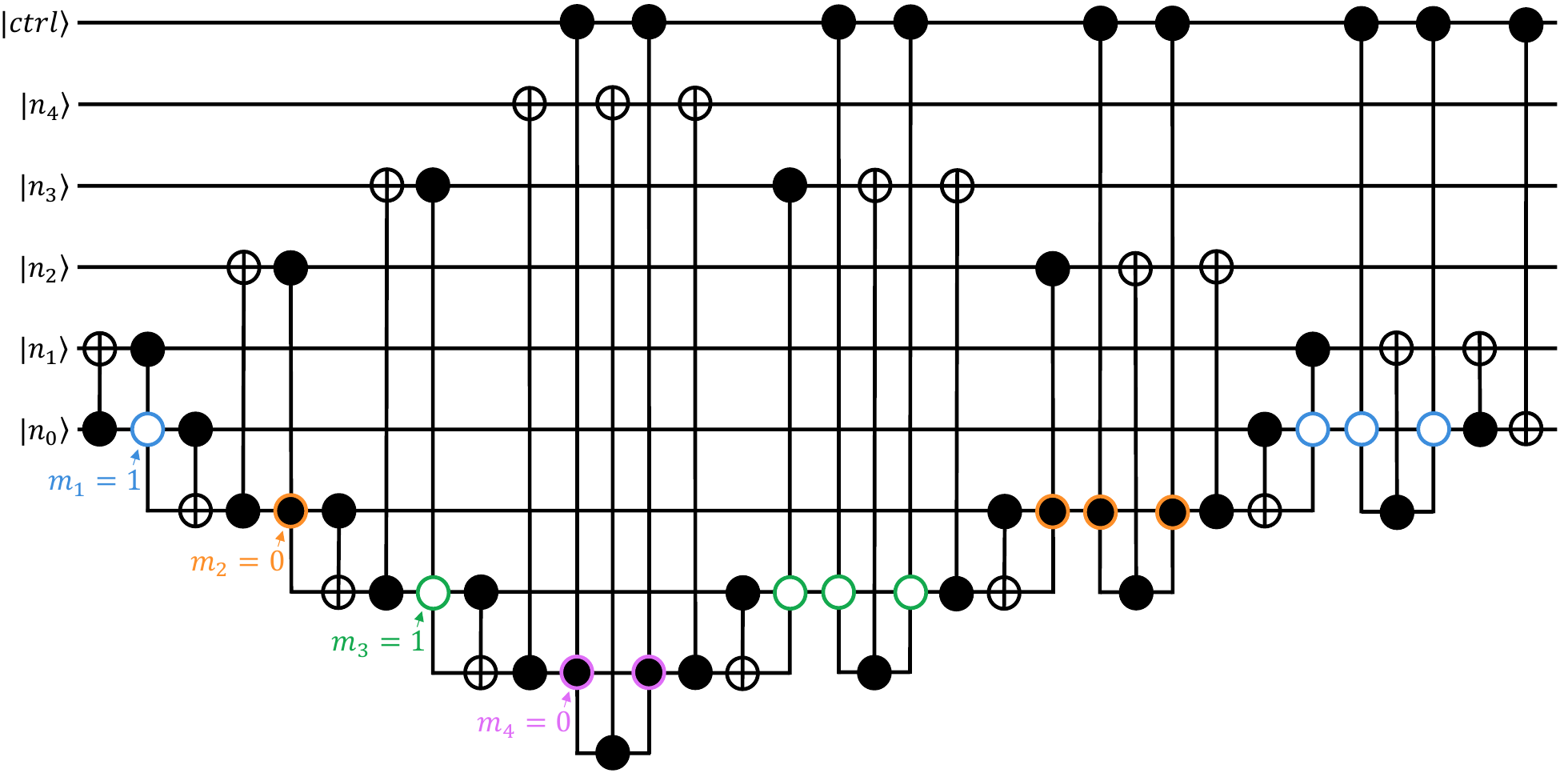}
    \caption{
        \textbf{Space-Efficient Controlled Addition of 11}
        An implementation for increasing the value of a register by a known classical value is shown for the case when the known value is $11$ and the number of qubits in the register is $5$.
        The binary representation of $11$ is $01011$ with the left-most bit being the most-significant.
        The values of these $M$ classical bits can be propagated into the control structure of the controlled quantum addition.
        If the value of the $i^\text{th}$ bit of $M$ is $0$ ($1$), the corresponding control in the circuit is controlled on the $\ket{1}$  ($\ket{0}$) state.
    }
    \label{fig:addition-qubit-efficient-11}
\end{figure*}

\begin{figure}
    \centering
    \includegraphics[width=12cm]{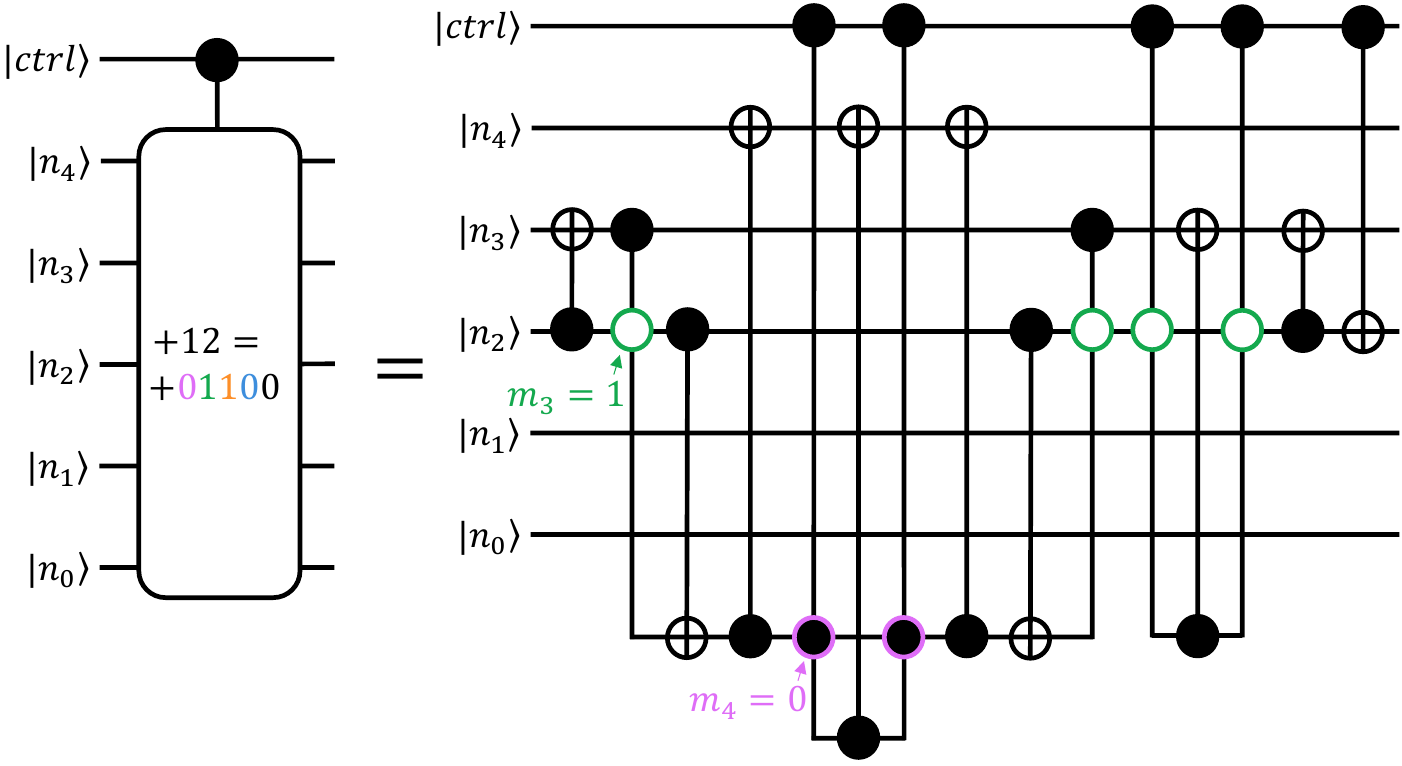}
    \caption{
        \textbf{Space-Efficient Controlled Addition of 12} 
        An implementation for increasing the value of a register by a known classical value is shown for the case when the known value is $12$ ($01100$ in binary) and the number of qubits in the register is $5$.
        When the least-significant bits of $M$ are $0$, the circuit can be bit-shifted, resulting in a lower cost implementation.
        In this case, the two least-significant bits are $0$, so the circuit can be bit-shifted twice.
    }
    \label{fig:addition-qubit-efficient-12}
\end{figure}

Another implementation, which uses the classical information about $m$ to modify the circuit for \textit{controlled} quantum addition, can be used to reduce the number of clean ancillae.
Controlled addition of two registers can be performed using $4(2N - 3)$ $T$ gates and $2N - 1$ clean ancillae using the construction for addition shown in Figure 4 of \cite{gidney2018halving}.
This circuit can be modified by propagating the classical information about the binary encoding of $m$ into the control structure of the adder circuit, reducing the number of clean ancillae by $N$.
An example diagram showing this propagation when $m = 11$ and $N = 5$ is shown in Figure \ref{fig:addition-qubit-efficient-11}.

If the $p$ least-significant bits of $m$ are zero, the addition can be performed beginning with the first non-zero bit of $m$.
If the $p$ least significant bits of $m$ are zero, then this circuit uses $4(2(N - p) - 3)$ $T$ gates and $N - p - 1$ clean ancillae.
An example circuit diagram for the case where $m=12$ ($01100$ in binary) and $N = 5$ is shown in Figure \ref{fig:addition-qubit-efficient-12}.

%% file: text/appendices/pauli_transform.tex
\section{Pauli Expansion of Fermionic and Bosonic Ladder Operators}
\label{sec:SB}

In this section, we review the operator transformations which express fermionic and bosonic ladder operators in the Pauli basis.

\subsection{Jordan-Wigner Transformation}

The fermionic ladder operators are mapped under the Jordan-Wigner transform \cite{jordan-wigner}.
The corresponding qubit operators that accomplish the same action as a ladder operator on a given mode are given by $\frac12 \left(X \mp iY \right)_i$ for creation ($-$) and annihilation ($+$) operators on mode $i$. 
This is not the entire transformation; however, because it does not pick up the property parity exchange rules for fermions.
In order to pick up the potential sign flip ($p(n)$), a string of Pauli $Z$ operators are included, which act on qubits indexed before the mode that the ladder operator acts on.

This transformation is given by:
\begin{align}
    b_i &= \frac12 \left(X_i + iY_i\right) \otimes Z_{i - 1} \otimes \dots \otimes Z_0 \\ \nonumber
    b_i^\dagger &= \frac12 \left(X_i - iY_i\right) \otimes Z_{i - 1} \otimes \dots \otimes Z_0\\
\end{align}

\subsection{Standard Binary}

In this work, the occupation of a bosonic mode is stored in binary.
The Standard Binary encoding \cite{standard-binary} provides a method to expand bosonic ladder operators in the Pauli basis under this encoding.

The expansion in the Pauli basis begins by noting the following definitions of the bosonic creation and annihilation operators:
\begin{align}
    \label{eq:sb-expansion}
    a^\dagger &= \sum_{s = 0}^{\Omega - 1}\sqrt{s + 1}|s + 1\rangle \langle s| = \ket{1}\bra{0} + \sqrt{2}\ket{2}\bra{1} + \dots\\
    a &= \sum_{s = 0}^{\Omega - 1}\sqrt{s + 1}|s\rangle \langle s + 1| = \ket{0}\bra{1} + \sqrt{2}\ket{1}\bra{2} + \dots
\end{align}

Each state in the expansion of $a$ and $a^\dagger$ is converted into binary, e.g. $|3\rangle = |0\hdots011\rangle$, where the preceeding (most-significant) qubits are zeros and the total length of the register is given by $W \equiv \lceil \log_2 (\Omega+1) \rceil$.

Each outer product in Eq. \eqref{eq:sb-expansion} can be expanded by computing the outer product of the individual qubit values tensored together.
For example, with $\Omega = 3$ the outer product of $\ket{2}$ and $\ket{3}$ is given by:
\begin{equation}
    \ket{2}\bra{3} = \ket{10}\bra{11} = \ket{1}\bra{1} \otimes \ket{0}\bra{1}.
\end{equation}

Each single-qubit outer product can be mapped onto the Pauli basis following:
\begin{equation}
    \ket{0}\bra{0} = \frac12 \left(I + Z \right)
\end{equation}
\begin{equation}
    \ket{0}\bra{1} = \frac12 \left(X + iY \right)
\end{equation}
\begin{equation}
    \ket{1}\bra{0} = \frac12 \left(X - iY \right)
\end{equation}
\begin{equation}
    \ket{1}\bra{1} = \frac12 \left(I - Z \right) 
\end{equation}

Finally, these single-qubit outer products can be tensored together to give the expansion of the multi-qubit outer products.
As an example, when $\Omega = 3$ the bosonic creation operator is given by: 
\begin{equation}
\begin{split}
    a_0^\dagger = &0.683IX -0.183 ZX - 0.683\mathrm{i} IY \\
                + &0.183\mathrm{i} ZY +0.354 XX - 0.354\mathrm{i} YX \\
                + &0.354\mathrm{i} XY + 0.354 YY
\end{split}
\end{equation}

%% file: text/appendices/uniformly_controlled_rotations.tex
\section{Uniformly Controlled Rotations}
\label{sec:multiplexed-rotations}

Implementing a series of uniformly controlled rotations is a common subroutine used in this work.
In this section, we discuss the cost and explicit circuit compilation for a series of uniformly controlled rotations about the same axis, but with different angles:
\begin{equation}
    \sum_{l=0}^{L - 1} \ket{l} \ket{\phi} \rightarrow \sum_{l=0}^{L - 1} \ket{l} R_a (\alpha_l) \ket{\phi}
\end{equation}

Möttönen et. al \cite{mottonen2004transformation}, provide a construction for \textit{uncontrolled} uniformly controlled rotations.
This construction is defined when the number of rotations ($L$) is a power of $2$, but this can be achieved for any integer $L$ by padding with zero-angle rotations.
The rotation angles are classically preprocessed based on the Gray code:
\begin{equation}
    \begin{bmatrix}
        \theta_{0} \\
        \theta_{1} \\
        \vdots \\
        \theta_{L - 1}
    \end{bmatrix} = M \begin{bmatrix}
        \alpha_{0} \\
        \alpha_{1} \\
        \vdots \\
        \alpha_{L - 1}
    \end{bmatrix}
\end{equation}
where $M$ is a matrix transformation defined by:
\begin{equation}
    M_{i, j} = L^{-1} (-1)^{b_{j} . g_{i}}
\end{equation}
where $b_j$ is the binary representation of the integer $j$, $g_i$ is the Gray code representation of the integer $i$, and $b_{j} . g_{i}$ is the bitwise inner product of $b_{j}$ and $g_{i}$.

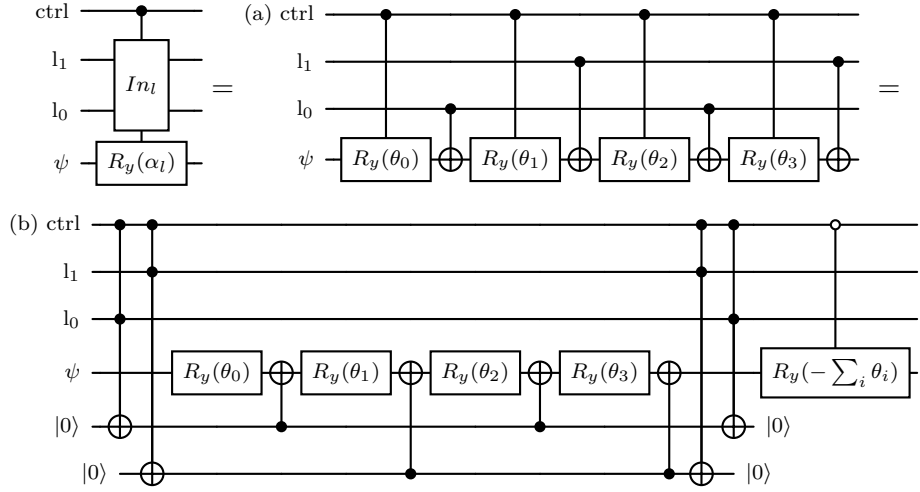
\begin{figure*}
    \begin{center}
    \begin{quantikz}[column sep=0.1cm, row sep=0.15cm, font=\scriptsize]
        \lstick{ctrl}  & \ghost{H}& \ctrl{3} &&\\
        \lstick{l$_1$} & \ghost{H}& \gate[2]{In_l} &&\\
        \lstick{l$_0$} & \ghost{H}&    &&\\
        \lstick{$\psi$}& \ghost{H}& \gate{R_y(\alpha_l)} &&
    \end{quantikz}=\begin{quantikz}[column sep=0.1cm, row sep=0.15cm, font=\scriptsize]
        \lstick{(a) ctrl}  & \ghost{H}& \ctrl{3} & & \ctrl{3} && \ctrl{3} && \ctrl{3} &&\\
        \lstick{l$_1$} & \ghost{H}& & & & \ctrl{2} &&&& \ctrl{2} &\\
        \lstick{l$_0$} & \ghost{H}& & \ctrl{1} & & && \ctrl{1} &&&\\
        \lstick{$\psi$}& \ghost{H}& \gate{R_y(\theta_0)} & \targ{} & \gate{R_y(\theta_1)} & \targ{} & \gate{R_y(\theta_2)} & \targ{} & \gate{R_y(\theta_3)} & \targ{} &
    \end{quantikz} =
    
    \begin{quantikz}[column sep=0.1cm, row sep=0.15cm, font=\scriptsize]
        \lstick{(b) ctrl}  & \ghost{H}& \ctrl{4} & \ctrl{5} &&&&&&&&&\ctrl{5} & \ctrl{4} &&\ctrl[open]{3} &\\
        \lstick{l$_1$} & \ghost{H}&& \ctrl{4} &&&&&&&&&\ctrl{4}&&&&\\
        \lstick{l$_0$} & \ghost{H}&\ctrl{2}&&&&&&&&&&&\ctrl{2}&&&\\
        \lstick{$\psi$}& \ghost{H}&&& \gate{R_y(\theta_0)} & \targ{} & \gate{R_y(\theta_1)} & \targ{} & \gate{R_y(\theta_2)} &\targ{}&\gate{R_y(\theta_3)} &\targ{}&&&&\gate{R_y(-\sum_{i}\theta_i)} &\\
        \lstick{$\ket{0}$} & \ghost{H} & \targ{} &&& \ctrl{-1}&&&&\ctrl{-1}&&&&\targ{}&\rstick{$\ket{0}$}&\wireoverride{}&\wireoverride{}\\
        & \ghost{H}\wireoverride{} &\wireoverride{} \lstick{$\ket{0}$} & \targ{} &&&& \ctrl{-2} &&&& \ctrl{-2}&\targ{}&\rstick{$\ket{0}$}&\wireoverride{}&\wireoverride{}&\wireoverride{}
    \end{quantikz}
    \end{center}
    \caption{
        \textbf{Controlled Uniformly Controlled Rotations}
        Two implementations for controlling a series of uniformly controlled rotations are shown.
        In (a), a naive implementation is shown, which doubles the number of arbitrary rotations.
        The implementation shown in (b) uses only one additional controlled rotation (two uncontrolled rotations) and $\log_2 L$ Toffoli gates, but requires $\log_2 L$ clean ancillae.
    }
    \label{fig:controlled-multiplexed-rotations}
\end{figure*}

In this work, we require the use of a \textit{controlled} series of uniformly controlled rotations.
Naively, this can be implemented by controlling each arbitrary rotation (subfigure \ref{fig:controlled-multiplexed-rotations}a).
Each controlled rotation can be implemented by two uncontrolled rotations, meaning this compilation strategy uses $2L$ uncontrolled arbitrary rotations.

An alternative approach that uses $4 \log_2 L$ $T$ gates, $L + 2$ arbitrary rotations, and $\log_2 L$ clean ancillae is shown in subfigure \ref{fig:controlled-multiplexed-rotations}b.
In this construction, the temporary logical-AND of each qubit in the index register and the control qubit is computed using $\log_2 L$ Toffoli gates.
CNOTs from these clean ancillae are interspersed between each of the arbitrary rotations, which are left uncontrolled.
When the control is on, this recovers the construction given by Möttönen et. al.
When the control is off, the uncontrolled arbitrary rotations are applied, resulting an undesired rotation of angle $\sum_{i} (\theta_i)$.
This undesired rotation can then be undone using one $0$-controlled rotation of angle $- \sum_{i} (\theta_i)$.

%% file: text/appendices/grover_rudolph.tex
\section{Grover-Rudolph State Preparation}
\label{sec:grover-rudolph}

In this section, we describe the Grover-Rudolph state preparation routine \cite{grover2002creating} in the context that it is used in this work.
The Grover-Rudolph state preparation algorithm constructs quantum circuits that prepare states of the form given by:
\begin{equation}
    \ket{0^{\otimes \lceil \log_2{L} \rceil}} \rightarrow_{\textit{Grover-Rudolph}} \sum_{l=0}^L \sqrt{p(l)} \ket{l}
\end{equation}
where $p(l)$ is a probability distribution with the constraint that $\sum_l p(l) = 1$.

Preparing probability distributions can be used to construct the $Prepare$ oracle (Eq. \eqref{eq:prep-state}).
The probability distribution is defined by the normalized magnitudes of the coefficients of the terms in the linear combination: $p(l) = |\alpha_l| / \lambda$.

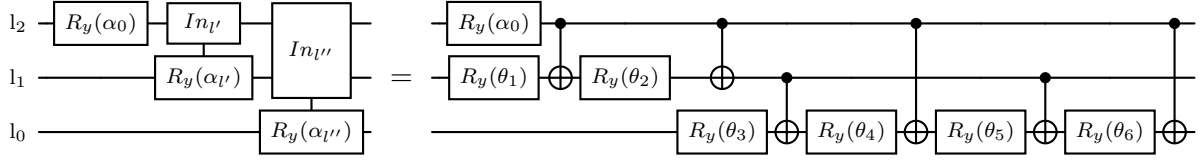
\begin{figure*}
    \begin{center}
    \begin{quantikz}[column sep=0.1cm, row sep=0.15cm, font=\scriptsize]
        \lstick{l$_2$} & \ghost{H}& \gate[1]{R_y(\alpha_0)} &  \gate[1]{In_{l^\prime}}{1} & \gate[2]{In_{l^{\prime \prime}}}{1} &\\
        \lstick{l$_1$} & \ghost{H}&& \gate[1]{R_y(\alpha_{l^\prime})} \wire[u]{q} &&\\
        \lstick{l$_0$} & \ghost{H}&    & & \gate[1]{R_y(\alpha_{l^{\prime \prime}})} \wire[u]{q} &
    \end{quantikz} = 
    \begin{quantikz}[column sep=0.1cm, row sep=0.15cm, font=\scriptsize]
        &\ghost{H}& \gate[1]{R_y(\alpha_0)} & \ctrl{1} &&\ctrl{1} &&& \ctrl{2} &&&& \ctrl{2} &\\
        &\ghost{H}& \gate[1]{R_y(\theta_1)} & \targ{} & \gate[1]{R_y(\theta_2)} & \targ{} & \ctrl{1} &&&& \ctrl{1} &&&\\
        &\ghost{H}&&&& \gate[1]{R_y(\theta_3)} & \targ{} & \gate[1]{R_y(\theta_4)} & \targ{} & \gate[1]{R_y(\theta_5)} & \targ{} & \gate[1]{R_y(\theta_6)} & \targ{} &
    \end{quantikz}
    \end{center}
    \caption{
        \textbf{Grover-Rudolph Circuit Compilation.} 
        An implementation of the Grover-Rudolph algorithm using several series of uniformly controlled rotations is shown when $L = 8$.
        This circuit requires $L - 1$ rotations when $L$ is a power of $2$ and the angles of the rotations are changed ($\theta_i \rightarrow \alpha_i$) based on classical preprocessing.
    }
    \label{fig:grover-rudolph}
\end{figure*}

The Grover-Rudolph algorithm sequentially sums the probability distribution to the left and right of a given index and then performs a rotation controlled on the current index.
This can be thought of as performing several series of uniformly controlled rotations (Figure \ref{fig:grover-rudolph}), using $L-1$ rotations.
For example, given the (noramlized) probabilities $\alpha_0$, $\alpha_1$, $\alpha_2$, and $\alpha_3$, the Grover-Rudolph algorithm proceeds as follows:
\begin{enumerate}
    \item Perform a Pauli-Y rotation on the top (left-most) qubit in the register by the angle: $\theta = 2 \cos^{-1}\big( \sqrt{\alpha_0 + \alpha_1} \big)$.
    \item Perform a Pauli-Y rotation on the second qubit in the register, controlled on the first qubit being in the state $\ket{0}$ by the angle: $\theta = 2 \cos^{-1}\big( \sqrt{\frac{\alpha_0}{\alpha_0 + \alpha_1}} \big)$
    \item Perform a Pauli-Y rotation on the second qubit in the register, controlled on the first qubit being in the state $\ket{1}$ by the angle: $\theta = 2 \cos^{-1}\big( \sqrt{\frac{\alpha_2}{\alpha_2 + \alpha_3}} \big)$
\end{enumerate}

The evolution of the quantum state is given by:
\begin{equation}
    \begin{split}
        \ket{00} &\rightarrow_{\textit{(i)}} \sqrt{\alpha_0 + \alpha_1} \ket{00} + \sqrt{\alpha_2 + \alpha_3} \ket{10} \\
        &\rightarrow_{\textit{(ii)}} \sqrt{\alpha_0} \ket{00} + \sqrt{\alpha_1} \ket{01} + \sqrt{\alpha_2 + \alpha_3} \ket{10} \\
        &\rightarrow_{\textit{(iii)}} \sqrt{\alpha_0} \ket{00} + \sqrt{\alpha_1} \ket{01} + \sqrt{\alpha_2} \ket{10} + \sqrt{\alpha_3} \ket{11}
    \end{split}
\end{equation}

%% file: text/appendices/qft.tex
\section{Quantum Field Theory Hamiltonians}
\label{subsec:qft-hamiltonians}

The two quantum field theory models studied in section \ref{sec:results} are the $\phi^4$ and Yukawa model.
In performing a Legendre transformation from the Lagrangian to the Hamiltonian, $\mathcal{L} \rightarrow H$, one must choose an explicit coordinate system.
Any field theory calculation done in a Hamiltonian approach benefits from using \textit{front form} (light front) coordinates \cite{Dirac1949}.
Lightfront coordinates are defined as $x^\mu = \left(x^+, x^-, x^1, x^2 \right)$, where $x^+ \equiv x^0 + x^3$ acts as the space coordinate and $x^- \equiv x^0 - x^3$ acts as the time coordinate.
In each model studied above, the transverse spacial coordinates $x^1, x^2$ are discarded, and the only independent coordinates are $x^+$ and $x^-$.

One main benefit of using this set of coordinates is that the Hamiltonian eigenvalue equation, whose eigenvectors are bound states of the theory, is simpler than in instant (equal time) coordinates.
This is because the Hamiltonian operator in instant form coordinates is $H = \sqrt{\vec{P}^2 - m^2}$, where the presence of the square root leads to an ambiguity.
In front form coordinates, the Hamiltonian operator is given via the quantized lightfront `energy' $$\hat P^- = \frac{\left(\vec{P}^\perp\right)^2 + m^2}{P^+},$$ where the square root is absent.

In one light front space and time dimension, the Hamiltonian eigenvalue equation is given by $\hat P^-|\psi\rangle = \frac{m^2}{P^+}|\psi\rangle$.
Any state can be expanded as $$\ket{\psi} = \sum_n \int d[\mu_n] \ket{\mu_n}\braket{\mu_n}{\psi},$$ where $\{\ket{\mu_n}\}$ is a set of Fock states with $n$ particles, i.e. $\ket{f}, \ket{ff}, \ket{b}, \ket{fb}, \dots$.

The light front `momentum' $P^+$ is a conserved quantity which every constituent's longitudinal momentum in a given Fock state $\in \{\ket{\mu_n}\}$ must sum to $P^+$.
The `modes' of the ladder operators correspond to discrete longitudinal momentum quantum numbers $k$, where the discretized approximation to a continuous constituent momentum $p^+ = \frac{\pi k}{L}$, where $L$ truncates $x^-$ to a finite region.
Thus, the conservation of $P^+$ is analogous to $\sum k = K$, labeled the \textit{resolution} \cite{pauli_and_brodsky}.
Increasing the resolution leads to a larger Hilbert space of states, with a more refined set of momentum values these states can describe, and thus the number of terms in the Hamiltonian that act non-trivially on this set of states also expands.
This is why increasing $K$ is a metric for `number of terms in the Hamiltonian'.

%% file: main.bbl
\begin{thebibliography}{92}
\providecommand{\natexlab}[1]{#1}
\providecommand{\url}[1]{\texttt{#1}}
\expandafter\ifx\csname urlstyle\endcsname\relax
  \providecommand{\doi}[1]{doi: #1}\else
  \providecommand{\doi}{doi: \begingroup \urlstyle{rm}\Url}\fi

\bibitem[Feynman(1982)]{feynman2018simulating}
Richard~P. Feynman.
\newblock Simulating physics with computers.
\newblock \emph{International Journal of Theoretical Physics}, 21\penalty0 (6):\penalty0 467--488, 1982.
\newblock \doi{10.1007/BF02650179}.
\newblock URL \url{https://doi.org/10.1007/BF02650179}.

\bibitem[Lloyd(1996)]{lloyd1996universal}
Seth Lloyd.
\newblock Universal quantum simulators.
\newblock \emph{Science}, 273\penalty0 (5278):\penalty0 1073--1078, 1996.
\newblock \doi{10.1126/science.273.5278.1073}.
\newblock URL \url{https://www.science.org/doi/abs/10.1126/science.273.5278.1073}.

\bibitem[Meyer(1996)]{meyer1996quantum}
David~A. Meyer.
\newblock From quantum cellular automata to quantum lattice gases.
\newblock \emph{Journal of Statistical Physics}, 85\penalty0 (5):\penalty0 551--574, 1996.
\newblock \doi{10.1007/BF02199356}.
\newblock URL \url{https://doi.org/10.1007/BF02199356}.

\bibitem[Boghosian and Taylor(1997)]{boghosian1997quantum}
Bruce~M. Boghosian and Washington Taylor.
\newblock Quantum lattice-gas models for the many-body schrödinger equation.
\newblock \emph{International Journal of Modern Physics C}, 08\penalty0 (04):\penalty0 705--716, 1997.
\newblock \doi{10.1142/S0129183197000606}.
\newblock URL \url{https://doi.org/10.1142/S0129183197000606}.

\bibitem[Abrams and Lloyd(1999)]{abrams1999quantum}
Daniel~S. Abrams and Seth Lloyd.
\newblock Quantum algorithm providing exponential speed increase for finding eigenvalues and eigenvectors.
\newblock \emph{Phys. Rev. Lett.}, 83:\penalty0 5162--5165, Dec 1999.
\newblock \doi{10.1103/PhysRevLett.83.5162}.
\newblock URL \url{https://link.aps.org/doi/10.1103/PhysRevLett.83.5162}.

\bibitem[Lidar and Wang(1999)]{lidar1999calculating}
Daniel~A. Lidar and Haobin Wang.
\newblock Calculating the thermal rate constant with exponential speedup on a quantum computer.
\newblock \emph{Phys. Rev. E}, 59:\penalty0 2429--2438, Feb 1999.
\newblock \doi{10.1103/PhysRevE.59.2429}.
\newblock URL \url{https://link.aps.org/doi/10.1103/PhysRevE.59.2429}.

\bibitem[Terhal and DiVincenzo(2000)]{terhal2000problem}
Barbara~M. Terhal and David~P. DiVincenzo.
\newblock Problem of equilibration and the computation of correlation functions on a quantum computer.
\newblock \emph{Phys. Rev. A}, 61:\penalty0 022301, Jan 2000.
\newblock \doi{10.1103/PhysRevA.61.022301}.
\newblock URL \url{https://link.aps.org/doi/10.1103/PhysRevA.61.022301}.

\bibitem[Wu et~al.(2002)Wu, Byrd, and Lidar]{wu2002polynomial}
L.-A. Wu, M.~S. Byrd, and D.~A. Lidar.
\newblock Polynomial-time simulation of pairing models on a quantum computer.
\newblock \emph{Phys. Rev. Lett.}, 89:\penalty0 057904, Jul 2002.
\newblock \doi{10.1103/PhysRevLett.89.057904}.
\newblock URL \url{https://link.aps.org/doi/10.1103/PhysRevLett.89.057904}.

\bibitem[Aspuru-Guzik et~al.(2005)Aspuru-Guzik, Dutoi, Love, and Head-Gordon]{aspuru2005simulated}
Alán Aspuru-Guzik, Anthony~D. Dutoi, Peter~J. Love, and Martin Head-Gordon.
\newblock Simulated quantum computation of molecular energies.
\newblock \emph{Science}, 309\penalty0 (5741):\penalty0 1704--1707, 2005.
\newblock \doi{10.1126/science.1113479}.
\newblock URL \url{https://www.science.org/doi/abs/10.1126/science.1113479}.

\bibitem[Kassal et~al.(2008)Kassal, Jordan, Love, Mohseni, and Aspuru-Guzik]{kassal2008polynomial}
Ivan Kassal, Stephen~P. Jordan, Peter~J. Love, Masoud Mohseni, and Alán Aspuru-Guzik.
\newblock Polynomial-time quantum algorithm for the simulation of chemical dynamics.
\newblock \emph{Proceedings of the National Academy of Sciences}, 105\penalty0 (48):\penalty0 18681--18686, 2008.
\newblock \doi{10.1073/pnas.0808245105}.
\newblock URL \url{https://www.pnas.org/doi/abs/10.1073/pnas.0808245105}.

\bibitem[Suzuki(1976{\natexlab{a}})]{wiese2014towards}
Masuo Suzuki.
\newblock Generalized trotter's formula and systematic approximants of exponential operators and inner derivations with applications to many-body problems.
\newblock \emph{Communications in Mathematical Physics}, 51\penalty0 (2):\penalty0 183--190, 1976{\natexlab{a}}.
\newblock \doi{10.1007/BF01609348}.
\newblock URL \url{https://doi.org/10.1007/BF01609348}.

\bibitem[Jordan et~al.(2012)Jordan, Lee, and Preskill]{jordan2012quantum}
Stephen~P. Jordan, Keith S.~M. Lee, and John Preskill.
\newblock Quantum algorithms for quantum field theories.
\newblock \emph{Science}, 336\penalty0 (6085):\penalty0 1130--1133, 2012.
\newblock \doi{10.1126/science.1217069}.
\newblock URL \url{https://www.science.org/doi/abs/10.1126/science.1217069}.

\bibitem[Cao et~al.(2019)Cao, Romero, Olson, Degroote, Johnson, Kieferov{\'a}, Kivlichan, Menke, Peropadre, Sawaya, Sim, Veis, and Aspuru-Guzik]{cao2019quantum}
Yudong Cao, Jonathan Romero, Jonathan~P. Olson, Matthias Degroote, Peter~D. Johnson, M{\'a}ria Kieferov{\'a}, Ian~D. Kivlichan, Tim Menke, Borja Peropadre, Nicolas P.~D. Sawaya, Sukin Sim, Libor Veis, and Al{\'a}n Aspuru-Guzik.
\newblock Quantum chemistry in the age of quantum computing.
\newblock \emph{Chemical Reviews}, 119\penalty0 (19):\penalty0 10856--10915, 10 2019.
\newblock \doi{10.1021/acs.chemrev.8b00803}.
\newblock URL \url{https://doi.org/10.1021/acs.chemrev.8b00803}.

\bibitem[McArdle et~al.(2020)McArdle, Endo, Aspuru-Guzik, Benjamin, and Yuan]{mcardle2020quantum}
Sam McArdle, Suguru Endo, Al\'an Aspuru-Guzik, Simon~C. Benjamin, and Xiao Yuan.
\newblock Quantum computational chemistry.
\newblock \emph{Rev. Mod. Phys.}, 92:\penalty0 015003, Mar 2020.
\newblock \doi{10.1103/RevModPhys.92.015003}.
\newblock URL \url{https://link.aps.org/doi/10.1103/RevModPhys.92.015003}.

\bibitem[Bauer et~al.(2023)Bauer, Davoudi, Balantekin, Bhattacharya, Carena, de~Jong, Draper, El-Khadra, Gemelke, Hanada, Kharzeev, Lamm, Li, Liu, Lukin, Meurice, Monroe, Nachman, Pagano, Preskill, Rinaldi, Roggero, Santiago, Savage, Siddiqi, Siopsis, Van~Zanten, Wiebe, Yamauchi, Yeter-Aydeniz, and Zorzetti]{bauer2023quantum}
Christian~W. Bauer, Zohreh Davoudi, A.~Baha Balantekin, Tanmoy Bhattacharya, Marcela Carena, Wibe~A. de~Jong, Patrick Draper, Aida El-Khadra, Nate Gemelke, Masanori Hanada, Dmitri Kharzeev, Henry Lamm, Ying-Ying Li, Junyu Liu, Mikhail Lukin, Yannick Meurice, Christopher Monroe, Benjamin Nachman, Guido Pagano, John Preskill, Enrico Rinaldi, Alessandro Roggero, David~I. Santiago, Martin~J. Savage, Irfan Siddiqi, George Siopsis, David Van~Zanten, Nathan Wiebe, Yukari Yamauchi, K\"ubra Yeter-Aydeniz, and Silvia Zorzetti.
\newblock Quantum simulation for high-energy physics.
\newblock \emph{PRX Quantum}, 4:\penalty0 027001, May 2023.
\newblock \doi{10.1103/PRXQuantum.4.027001}.
\newblock URL \url{https://link.aps.org/doi/10.1103/PRXQuantum.4.027001}.

\bibitem[Suzuki(1976{\natexlab{b}})]{suzuki1976generalized}
Masuo Suzuki.
\newblock Generalized trotter's formula and systematic approximants of exponential operators and inner derivations with applications to many-body problems.
\newblock \emph{Communications in Mathematical Physics}, 51\penalty0 (2):\penalty0 183--190, 1976{\natexlab{b}}.
\newblock \doi{10.1007/BF01609348}.
\newblock URL \url{https://doi.org/10.1007/BF01609348}.

\bibitem[Hatano and Suzuki(2005)]{hatano2005finding}
Naomichi Hatano and Masuo Suzuki.
\newblock \emph{Finding Exponential Product Formulas of Higher Orders}, pages 37--68.
\newblock Springer Berlin Heidelberg, Berlin, Heidelberg, 2005.
\newblock ISBN 978-3-540-31515-5.
\newblock \doi{10.1007/11526216_2}.
\newblock URL \url{https://doi.org/10.1007/11526216_2}.

\bibitem[Lie(1893)]{lie1893theorie}
Sophus Lie.
\newblock \emph{Theorie der Transformationsgruppen Abschn. 3}.
\newblock Teubner, 1893.
\newblock URL \url{http://eudml.org/doc/202686}.

\bibitem[Trotter(1959)]{trotter1959product}
Hale~F Trotter.
\newblock On the product of semi-groups of operators.
\newblock \emph{Proceedings of the American Mathematical Society}, 10\penalty0 (4):\penalty0 545--551, 1959.
\newblock \doi{https://doi.org/10.1090/S0002-9939-1959-0108732-6}.

\bibitem[Childs et~al.(2021)Childs, Su, Tran, Wiebe, and Zhu]{childs2021theory}
Andrew~M. Childs, Yuan Su, Minh~C. Tran, Nathan Wiebe, and Shuchen Zhu.
\newblock Theory of trotter error with commutator scaling.
\newblock \emph{Phys. Rev. X}, 11:\penalty0 011020, Feb 2021.
\newblock \doi{10.1103/PhysRevX.11.011020}.
\newblock URL \url{https://link.aps.org/doi/10.1103/PhysRevX.11.011020}.

\bibitem[Lin(2022)]{lin2022lecture}
Lin Lin.
\newblock Lecture notes on quantum algorithms for scientific computation, 2022.
\newblock URL \url{https://arxiv.org/abs/2201.08309}.
\newblock DOI: \url{https://doi.org/10.48550/arXiv.2201.08309}.

\bibitem[Poulin et~al.(2018)Poulin, Kitaev, Steiger, Hastings, and Troyer]{poulin2018quantum}
David Poulin, Alexei Kitaev, Damian~S. Steiger, Matthew~B. Hastings, and Matthias Troyer.
\newblock Quantum algorithm for spectral measurement with a lower gate count.
\newblock \emph{Phys. Rev. Lett.}, 121:\penalty0 010501, Jul 2018.
\newblock \doi{10.1103/PhysRevLett.121.010501}.
\newblock URL \url{https://link.aps.org/doi/10.1103/PhysRevLett.121.010501}.

\bibitem[Low and Chuang(2019)]{low2019hamiltonian}
Guang~Hao Low and Isaac~L. Chuang.
\newblock Hamiltonian {S}imulation by {Q}ubitization.
\newblock \emph{{Quantum}}, 3:\penalty0 163, July 2019.
\newblock ISSN 2521-327X.
\newblock \doi{10.22331/q-2019-07-12-163}.
\newblock URL \url{https://doi.org/10.22331/q-2019-07-12-163}.

\bibitem[Berry and Childs(2012)]{berry2009black}
Dominic~W. Berry and Andrew~M. Childs.
\newblock Black-box hamiltonian simulation and unitary implementation.
\newblock \emph{Quantum Info. Comput.}, 12\penalty0 (1–2):\penalty0 29–62, January 2012.
\newblock ISSN 1533-7146.
\newblock \doi{https://doi.org/10.26421/QIC12.1-2}.

\bibitem[Childs(2009)]{childs2009universal}
Andrew~M. Childs.
\newblock Universal computation by quantum walk.
\newblock \emph{Phys. Rev. Lett.}, 102:\penalty0 180501, May 2009.
\newblock \doi{10.1103/PhysRevLett.102.180501}.
\newblock URL \url{https://link.aps.org/doi/10.1103/PhysRevLett.102.180501}.

\bibitem[Childs and Wiebe(2012)]{childs2012hamiltonian}
Andrew~M. Childs and Nathan Wiebe.
\newblock Hamiltonian simulation using linear combinations of unitary operations.
\newblock \emph{Quantum Info. Comput.}, 12\penalty0 (11–12):\penalty0 901–924, November 2012.
\newblock ISSN 1533-7146.
\newblock \doi{https://doi.org/10.26421/QIC12.11-12}.

\bibitem[Bluvstein et~al.(2024)Bluvstein, Evered, Geim, Li, Zhou, Manovitz, Ebadi, Cain, Kalinowski, Hangleiter, Bonilla~Ataides, Maskara, Cong, Gao, Sales~Rodriguez, Karolyshyn, Semeghini, Gullans, Greiner, Vuleti{\'c}, and Lukin]{bluvstein2024logical}
Dolev Bluvstein, Simon~J. Evered, Alexandra~A. Geim, Sophie~H. Li, Hengyun Zhou, Tom Manovitz, Sepehr Ebadi, Madelyn Cain, Marcin Kalinowski, Dominik Hangleiter, J.~Pablo Bonilla~Ataides, Nishad Maskara, Iris Cong, Xun Gao, Pedro Sales~Rodriguez, Thomas Karolyshyn, Giulia Semeghini, Michael~J. Gullans, Markus Greiner, Vladan Vuleti{\'c}, and Mikhail~D. Lukin.
\newblock Logical quantum processor based on reconfigurable atom arrays.
\newblock \emph{Nature}, 626\penalty0 (7997):\penalty0 58--65, 2024.
\newblock \doi{10.1038/s41586-023-06927-3}.
\newblock URL \url{https://doi.org/10.1038/s41586-023-06927-3}.

\bibitem[{Google Quantum AI and Collaborators}(2025)]{acharya2024quantum}
{Google Quantum AI and Collaborators}.
\newblock Quantum error correction below the surface code threshold.
\newblock \emph{Nature}, 638\penalty0 (8052):\penalty0 920--926, Feb 2025.
\newblock ISSN 1476-4687.
\newblock \doi{10.1038/s41586-024-08449-y}.
\newblock URL \url{https://doi.org/10.1038/s41586-024-08449-y}.

\bibitem[Peruzzo et~al.(2014)Peruzzo, McClean, Shadbolt, Yung, Zhou, Love, Aspuru-Guzik, and O'Brien]{peruzzo2014variational}
Alberto Peruzzo, Jarrod McClean, Peter Shadbolt, Man-Hong Yung, Xiao-Qi Zhou, Peter~J. Love, Al{\'a}n Aspuru-Guzik, and Jeremy~L. O'Brien.
\newblock A variational eigenvalue solver on a photonic quantum processor.
\newblock \emph{Nature Communications}, 5\penalty0 (1):\penalty0 4213, 2014.
\newblock \doi{10.1038/ncomms5213}.
\newblock URL \url{https://doi.org/10.1038/ncomms5213}.

\bibitem[Babbush et~al.(2014)Babbush, Love, and Aspuru-Guzik]{babbush2014adiabatic}
Ryan Babbush, Peter~J. Love, and Al{\'a}n Aspuru-Guzik.
\newblock Adiabatic quantum simulation of quantum chemistry.
\newblock \emph{Scientific Reports}, 4\penalty0 (1):\penalty0 6603, 2014.
\newblock \doi{10.1038/srep06603}.
\newblock URL \url{https://doi.org/10.1038/srep06603}.

\bibitem[O'Malley et~al.(2016)O'Malley, Babbush, Kivlichan, Romero, McClean, Barends, Kelly, Roushan, Tranter, Ding, Campbell, Chen, Chen, Chiaro, Dunsworth, Fowler, Jeffrey, Lucero, Megrant, Mutus, Neeley, Neill, Quintana, Sank, Vainsencher, Wenner, White, Coveney, Love, Neven, Aspuru-Guzik, and Martinis]{o2016scalable}
P.~J.~J. O'Malley, R.~Babbush, I.~D. Kivlichan, J.~Romero, J.~R. McClean, R.~Barends, J.~Kelly, P.~Roushan, A.~Tranter, N.~Ding, B.~Campbell, Y.~Chen, Z.~Chen, B.~Chiaro, A.~Dunsworth, A.~G. Fowler, E.~Jeffrey, E.~Lucero, A.~Megrant, J.~Y. Mutus, M.~Neeley, C.~Neill, C.~Quintana, D.~Sank, A.~Vainsencher, J.~Wenner, T.~C. White, P.~V. Coveney, P.~J. Love, H.~Neven, A.~Aspuru-Guzik, and J.~M. Martinis.
\newblock Scalable quantum simulation of molecular energies.
\newblock \emph{Phys. Rev. X}, 6:\penalty0 031007, Jul 2016.
\newblock \doi{10.1103/PhysRevX.6.031007}.
\newblock URL \url{https://link.aps.org/doi/10.1103/PhysRevX.6.031007}.

\bibitem[Babbush et~al.(2018)Babbush, Gidney, Berry, Wiebe, McClean, Paler, Fowler, and Neven]{babbush2018encoding}
Ryan Babbush, Craig Gidney, Dominic~W. Berry, Nathan Wiebe, Jarrod McClean, Alexandru Paler, Austin Fowler, and Hartmut Neven.
\newblock Encoding electronic spectra in quantum circuits with linear t complexity.
\newblock \emph{Phys. Rev. X}, 8:\penalty0 041015, Oct 2018.
\newblock \doi{10.1103/PhysRevX.8.041015}.
\newblock URL \url{https://link.aps.org/doi/10.1103/PhysRevX.8.041015}.

\bibitem[Quantum et~al.(2020)Quantum, Collaborators*†, Arute, Arya, Babbush, Bacon, Bardin, Barends, Boixo, Broughton, Buckley, Buell, Burkett, Bushnell, Chen, Chen, Chiaro, Collins, Courtney, Demura, Dunsworth, Farhi, Fowler, Foxen, Gidney, Giustina, Graff, Habegger, Harrigan, Ho, Hong, Huang, Huggins, Ioffe, Isakov, Jeffrey, Jiang, Jones, Kafri, Kechedzhi, Kelly, Kim, Klimov, Korotkov, Kostritsa, Landhuis, Laptev, Lindmark, Lucero, Martin, Martinis, McClean, McEwen, Megrant, Mi, Mohseni, Mruczkiewicz, Mutus, Naaman, Neeley, Neill, Neven, Niu, O’Brien, Ostby, Petukhov, Putterman, Quintana, Roushan, Rubin, Sank, Satzinger, Smelyanskiy, Strain, Sung, Szalay, Takeshita, Vainsencher, White, Wiebe, Yao, Yeh, and Zalcman]{google2020hartree}
Google~AI Quantum, Collaborators*†, Frank Arute, Kunal Arya, Ryan Babbush, Dave Bacon, Joseph~C. Bardin, Rami Barends, Sergio Boixo, Michael Broughton, Bob~B. Buckley, David~A. Buell, Brian Burkett, Nicholas Bushnell, Yu~Chen, Zijun Chen, Benjamin Chiaro, Roberto Collins, William Courtney, Sean Demura, Andrew Dunsworth, Edward Farhi, Austin Fowler, Brooks Foxen, Craig Gidney, Marissa Giustina, Rob Graff, Steve Habegger, Matthew~P. Harrigan, Alan Ho, Sabrina Hong, Trent Huang, William~J. Huggins, Lev Ioffe, Sergei~V. Isakov, Evan Jeffrey, Zhang Jiang, Cody Jones, Dvir Kafri, Kostyantyn Kechedzhi, Julian Kelly, Seon Kim, Paul~V. Klimov, Alexander Korotkov, Fedor Kostritsa, David Landhuis, Pavel Laptev, Mike Lindmark, Erik Lucero, Orion Martin, John~M. Martinis, Jarrod~R. McClean, Matt McEwen, Anthony Megrant, Xiao Mi, Masoud Mohseni, Wojciech Mruczkiewicz, Josh Mutus, Ofer Naaman, Matthew Neeley, Charles Neill, Hartmut Neven, Murphy~Yuezhen Niu, Thomas~E. O’Brien, Eric Ostby, Andre Petukhov, Harald
  Putterman, Chris Quintana, Pedram Roushan, Nicholas~C. Rubin, Daniel Sank, Kevin~J. Satzinger, Vadim Smelyanskiy, Doug Strain, Kevin~J. Sung, Marco Szalay, Tyler~Y. Takeshita, Amit Vainsencher, Theodore White, Nathan Wiebe, Z.~Jamie Yao, Ping Yeh, and Adam Zalcman.
\newblock Hartree-fock on a superconducting qubit quantum computer.
\newblock \emph{Science}, 369\penalty0 (6507):\penalty0 1084--1089, 2020.
\newblock \doi{10.1126/science.abb9811}.
\newblock URL \url{https://www.science.org/doi/abs/10.1126/science.abb9811}.

\bibitem[Lee et~al.(2021)Lee, Berry, Gidney, Huggins, McClean, Wiebe, and Babbush]{lee2021even}
Joonho Lee, Dominic~W. Berry, Craig Gidney, William~J. Huggins, Jarrod~R. McClean, Nathan Wiebe, and Ryan Babbush.
\newblock Even more efficient quantum computations of chemistry through tensor hypercontraction.
\newblock \emph{PRX Quantum}, 2:\penalty0 030305, Jul 2021.
\newblock \doi{10.1103/PRXQuantum.2.030305}.
\newblock URL \url{https://link.aps.org/doi/10.1103/PRXQuantum.2.030305}.

\bibitem[Kivlichan et~al.(2020)Kivlichan, Gidney, Berry, Wiebe, McClean, Sun, Jiang, Rubin, Fowler, Aspuru-Guzik, Neven, and Babbush]{kivlichan2020improved}
Ian~D. Kivlichan, Craig Gidney, Dominic~W. Berry, Nathan Wiebe, Jarrod McClean, Wei Sun, Zhang Jiang, Nicholas Rubin, Austin Fowler, Al{\'{a}}n Aspuru-Guzik, Hartmut Neven, and Ryan Babbush.
\newblock Improved {F}ault-{T}olerant {Q}uantum {S}imulation of {C}ondensed-{P}hase {C}orrelated {E}lectrons via {T}rotterization.
\newblock \emph{{Quantum}}, 4:\penalty0 296, July 2020.
\newblock ISSN 2521-327X.
\newblock \doi{10.22331/q-2020-07-16-296}.
\newblock URL \url{https://doi.org/10.22331/q-2020-07-16-296}.

\bibitem[Campbell(2021)]{campbell2021early}
Earl~T Campbell.
\newblock Early fault-tolerant simulations of the hubbard model.
\newblock \emph{Quantum Science and Technology}, 7\penalty0 (1):\penalty0 015007, nov 2021.
\newblock \doi{10.1088/2058-9565/ac3110}.
\newblock URL \url{https://doi.org/10.1088/2058-9565/ac3110}.

\bibitem[Peskin and Schroeder(1995)]{Peskin:1995ev}
Michael~E. Peskin and Daniel~V. Schroeder.
\newblock \emph{{An Introduction to quantum field theory}}.
\newblock Addison-Wesley, Reading, USA, 1995.
\newblock ISBN 978-0-201-50397-5, 978-0-429-50355-9, 978-0-429-49417-8.
\newblock \doi{10.1201/9780429503559}.

\bibitem[Girguś and Głazek(2024)]{girgus2024}
M.~Girguś and S.~D. Głazek.
\newblock Spiral flow of quantum quartic oscillator with energy cutoff, 2024.
\newblock URL \url{https://arxiv.org/abs/2404.17446}.
\newblock DOI: \url{10.48550/arXiv.2404.17446}.

\bibitem[Bender and Wu(1969)]{bender1969}
Carl~M. Bender and Tai~Tsun Wu.
\newblock Anharmonic oscillator.
\newblock \emph{Phys. Rev.}, 184:\penalty0 1231--1260, Aug 1969.
\newblock \doi{10.1103/PhysRev.184.1231}.
\newblock URL \url{https://link.aps.org/doi/10.1103/PhysRev.184.1231}.

\bibitem[Kirby et~al.(2021{\natexlab{a}})Kirby, Hadi, Kreshchuk, and Love]{Kirby_2021}
William~M. Kirby, Sultana Hadi, Michael Kreshchuk, and Peter~J. Love.
\newblock Quantum simulation of second-quantized hamiltonians in compact encoding.
\newblock \emph{Physical Review A}, 104\penalty0 (4), October 2021{\natexlab{a}}.
\newblock ISSN 2469-9934.
\newblock \doi{10.1103/physreva.104.042607}.
\newblock URL \url{http://dx.doi.org/10.1103/PhysRevA.104.042607}.

\bibitem[Camps et~al.(2024)Camps, Lin, Van~Beeumen, and Yang]{camps2024explicit}
Daan Camps, Lin Lin, Roel Van~Beeumen, and Chao Yang.
\newblock Explicit quantum circuits for block encodings of certain sparse matrices.
\newblock \emph{SIAM Journal on Matrix Analysis and Applications}, 45\penalty0 (1):\penalty0 801--827, 2024.
\newblock \doi{10.1137/22M1484298}.
\newblock URL \url{https://doi.org/10.1137/22M1484298}.

\bibitem[Liu et~al.(2025)Liu, Du, Lin, Vary, and Yang]{liu2024efficient}
Diyi Liu, Weijie Du, Lin Lin, James~P. Vary, and Chao Yang.
\newblock An efficient quantum circuit for block encoding a pairing hamiltonian.
\newblock \emph{Journal of Computational Science}, 85:\penalty0 102480, 2025.
\newblock ISSN 1877-7503.
\newblock \doi{https://doi.org/10.1016/j.jocs.2024.102480}.
\newblock URL \url{https://www.sciencedirect.com/science/article/pii/S1877750324002734}.

\bibitem[Rhodes et~al.(2024)Rhodes, Kreshchuk, and Pathak]{rhodes2024exponential}
Mason~L. Rhodes, Michael Kreshchuk, and Shivesh Pathak.
\newblock Exponential improvements in the simulation of lattice gauge theories using near-optimal techniques.
\newblock \emph{PRX Quantum}, 5:\penalty0 040347, Dec 2024.
\newblock \doi{10.1103/PRXQuantum.5.040347}.
\newblock URL \url{https://link.aps.org/doi/10.1103/PRXQuantum.5.040347}.

\bibitem[Hariprakash et~al.(2025)Hariprakash, Modi, Kreshchuk, Kane, and Bauer]{hariprakash2025strategies}
Siddharth Hariprakash, Neel~S. Modi, Michael Kreshchuk, Christopher~F. Kane, and Christian~W. Bauer.
\newblock Strategies for simulating the time evolution of hamiltonian lattice field theories.
\newblock \emph{Phys. Rev. A}, 111:\penalty0 022419, Feb 2025.
\newblock \doi{10.1103/PhysRevA.111.022419}.
\newblock URL \url{https://link.aps.org/doi/10.1103/PhysRevA.111.022419}.

\bibitem[Du and Vary(2025)]{du2024systematicinputschememanyboson}
Weijie Du and James~P. Vary.
\newblock Systematic input scheme for many-boson hamiltonians with applications to the two-dimensional ${\ensuremath{\phi}}^{4}$ theory.
\newblock \emph{Phys. Rev. D}, 111:\penalty0 016013, Jan 2025.
\newblock \doi{10.1103/PhysRevD.111.016013}.
\newblock URL \url{https://link.aps.org/doi/10.1103/PhysRevD.111.016013}.

\bibitem[Halimeh et~al.(2024)Halimeh, Hanada, Matsuura, Nori, Rinaldi, and Schäfer]{halimeh2024universal}
Jad~C. Halimeh, Masanori Hanada, Shunji Matsuura, Franco Nori, Enrico Rinaldi, and Andreas Schäfer.
\newblock A universal framework for the quantum simulation of yang-mills theory, 2024.
\newblock URL \url{https://arxiv.org/abs/2411.13161}.
\newblock DOI: \url{https://doi.org/10.48550/arXiv.2411.13161}.

\bibitem[Peng et~al.(2025)Peng, Su, Claudino, Kowalski, Hao~Low, and Roetteler]{Peng_2025}
Bo~Peng, Yuan Su, Daniel Claudino, Karol Kowalski, Guang Hao~Low, and Martin Roetteler.
\newblock Quantum simulation of boson-related hamiltonians: techniques, effective hamiltonian construction, and error analysis.
\newblock \emph{Quantum Science and Technology}, 10\penalty0 (2):\penalty0 023002, mar 2025.
\newblock \doi{10.1088/2058-9565/adbf42}.
\newblock URL \url{https://doi.org/10.1088/2058-9565/adbf42}.

\bibitem[Jordan and Wigner(1928)]{jordan-wigner}
P.~Jordan and E.~Wigner.
\newblock Über das paulische Äquivalenzverbot.
\newblock \emph{Zeitschrift für Physik}, 47\penalty0 (9), 1928.

\bibitem[Bravyi and Kitaev(2002)]{bravyi2002fermionic}
Sergey~B. Bravyi and Alexei~Yu. Kitaev.
\newblock Fermionic quantum computation.
\newblock \emph{Annals of Physics}, 298\penalty0 (1):\penalty0 210--226, 2002.
\newblock ISSN 0003-4916.
\newblock \doi{https://doi.org/10.1006/aphy.2002.6254}.
\newblock URL \url{https://www.sciencedirect.com/science/article/pii/S0003491602962548}.

\bibitem[Seeley et~al.(2012)Seeley, Richard, and Love]{seeley2012bravyi}
Jacob~T. Seeley, Martin~J. Richard, and Peter~J. Love.
\newblock The bravyi-kitaev transformation for quantum computation of electronic structure.
\newblock \emph{The Journal of Chemical Physics}, 137\penalty0 (22):\penalty0 224109, 12 2012.
\newblock ISSN 0021-9606.
\newblock \doi{10.1063/1.4768229}.
\newblock URL \url{https://doi.org/10.1063/1.4768229}.

\bibitem[Somma(2005)]{somma2005quantum}
Rolando~D. Somma.
\newblock Quantum computation, complexity, and many-body physics, 2005.
\newblock URL \url{https://arxiv.org/abs/quant-ph/0512209}.
\newblock DOI: \url{https://doi.org/10.48550/arXiv.quant-ph/0512209}.

\bibitem[Sawaya et~al.(2020)Sawaya, Menke, Kyaw, Johri, Aspuru-Guzik, and Guerreschi]{standard-binary}
Nicolas P.~D. Sawaya, Tim Menke, Thi~Ha Kyaw, Sonika Johri, Al{\'a}n Aspuru-Guzik, and Gian~Giacomo Guerreschi.
\newblock Resource-efficient digital quantum simulation of d-level systems for photonic, vibrational, and spin-s hamiltonians.
\newblock \emph{npj Quantum Information}, 6\penalty0 (1):\penalty0 49, 2020.
\newblock \doi{10.1038/s41534-020-0278-0}.
\newblock URL \url{https://doi.org/10.1038/s41534-020-0278-0}.

\bibitem[Berezin(1966)]{berezin1966method}
F.A. Berezin.
\newblock Introduction.
\newblock In \emph{The Method of Second Quantization}, volume~24 of \emph{Pure and Applied Physics}, pages 1--8. Elsevier, 1966.
\newblock \doi{https://doi.org/10.1016/B978-0-12-089450-5.50006-7}.
\newblock URL \url{https://www.sciencedirect.com/science/article/pii/B9780120894505500067}.

\bibitem[Schwartz(2013)]{Schwartz_2013}
Matthew~D. Schwartz.
\newblock \emph{Quantum Field Theory and the Standard Model}.
\newblock Cambridge University Press, 2013.
\newblock \doi{https://doi.org/10.1017/9781139540940}.

\bibitem[Pauli(1925)]{pauli1925zusammenhang}
W.~Pauli.
\newblock {\"U}ber den zusammenhang des abschlusses der elektronengruppen im atom mit der komplexstruktur der spektren.
\newblock \emph{Zeitschrift f{\"u}r Physik}, 31\penalty0 (1):\penalty0 765--783, 1925.
\newblock \doi{10.1007/BF02980631}.
\newblock URL \url{https://doi.org/10.1007/BF02980631}.

\bibitem[Kreshchuk et~al.(2022)Kreshchuk, Kirby, Goldstein, Beauchemin, and Love]{kreshchuk2022quantum}
Michael Kreshchuk, William~M. Kirby, Gary Goldstein, Hugo Beauchemin, and Peter~J. Love.
\newblock Quantum simulation of quantum field theory in the light-front formulation.
\newblock \emph{Phys. Rev. A}, 105:\penalty0 032418, Mar 2022.
\newblock \doi{10.1103/PhysRevA.105.032418}.
\newblock URL \url{https://link.aps.org/doi/10.1103/PhysRevA.105.032418}.

\bibitem[Chakraborty et~al.(2019)Chakraborty, Gily\'{e}n, and Jeffery]{chakraborty2019power}
Shantanav Chakraborty, Andr\'{a}s Gily\'{e}n, and Stacey Jeffery.
\newblock {The Power of Block-Encoded Matrix Powers: Improved Regression Techniques via Faster Hamiltonian Simulation}.
\newblock In Christel Baier, Ioannis Chatzigiannakis, Paola Flocchini, and Stefano Leonardi, editors, \emph{46th International Colloquium on Automata, Languages, and Programming (ICALP 2019)}, volume 132 of \emph{Leibniz International Proceedings in Informatics (LIPIcs)}, pages 33:1--33:14, Dagstuhl, Germany, 2019. Schloss Dagstuhl -- Leibniz-Zentrum f{\"u}r Informatik.
\newblock ISBN 978-3-95977-109-2.
\newblock \doi{10.4230/LIPIcs.ICALP.2019.33}.
\newblock URL \url{https://drops.dagstuhl.de/entities/document/10.4230/LIPIcs.ICALP.2019.33}.

\bibitem[Gily\'{e}n et~al.(2019)Gily\'{e}n, Su, Low, and Wiebe]{gilyen2019quantum}
Andr\'{a}s Gily\'{e}n, Yuan Su, Guang~Hao Low, and Nathan Wiebe.
\newblock Quantum singular value transformation and beyond: exponential improvements for quantum matrix arithmetics.
\newblock In \emph{Proceedings of the 51st Annual ACM SIGACT Symposium on Theory of Computing}, STOC 2019, page 193–204, New York, NY, USA, 2019. Association for Computing Machinery.
\newblock ISBN 9781450367059.
\newblock \doi{10.1145/3313276.3316366}.
\newblock URL \url{https://doi.org/10.1145/3313276.3316366}.

\bibitem[O'Brien et~al.(2022)O'Brien, Streif, Rubin, Santagati, Su, Huggins, Goings, Moll, Kyoseva, Degroote, Tautermann, Lee, Berry, Wiebe, and Babbush]{obrien2022}
Thomas~E. O'Brien, Michael Streif, Nicholas~C. Rubin, Raffaele Santagati, Yuan Su, William~J. Huggins, Joshua~J. Goings, Nikolaj Moll, Elica Kyoseva, Matthias Degroote, Christofer~S. Tautermann, Joonho Lee, Dominic~W. Berry, Nathan Wiebe, and Ryan Babbush.
\newblock Efficient quantum computation of molecular forces and other energy gradients.
\newblock \emph{Phys. Rev. Res.}, 4:\penalty0 043210, Dec 2022.
\newblock \doi{10.1103/PhysRevResearch.4.043210}.
\newblock URL \url{https://link.aps.org/doi/10.1103/PhysRevResearch.4.043210}.

\bibitem[Berry et~al.(2015{\natexlab{a}})Berry, Childs, and Kothari]{berry2015hamiltonian}
Dominic~W. Berry, Andrew~M. Childs, and Robin Kothari.
\newblock Hamiltonian simulation with nearly optimal dependence on all parameters.
\newblock In \emph{Proceedings of the 2015 IEEE 56th Annual Symposium on Foundations of Computer Science (FOCS)}, FOCS '15, page 792–809, USA, 2015{\natexlab{a}}. IEEE Computer Society.
\newblock ISBN 9781467381918.
\newblock \doi{10.1109/FOCS.2015.54}.
\newblock URL \url{https://doi.org/10.1109/FOCS.2015.54}.

\bibitem[Berry et~al.(2015{\natexlab{b}})Berry, Childs, Cleve, Kothari, and Somma]{berry2015simulating}
Dominic~W. Berry, Andrew~M. Childs, Richard Cleve, Robin Kothari, and Rolando~D. Somma.
\newblock Simulating hamiltonian dynamics with a truncated taylor series.
\newblock \emph{Phys. Rev. Lett.}, 114:\penalty0 090502, Mar 2015{\natexlab{b}}.
\newblock \doi{10.1103/PhysRevLett.114.090502}.
\newblock URL \url{https://link.aps.org/doi/10.1103/PhysRevLett.114.090502}.

\bibitem[Low and Chuang(2017)]{low2017optimal}
Guang~Hao Low and Isaac~L. Chuang.
\newblock Optimal hamiltonian simulation by quantum signal processing.
\newblock \emph{Phys. Rev. Lett.}, 118:\penalty0 010501, Jan 2017.
\newblock \doi{10.1103/PhysRevLett.118.010501}.
\newblock URL \url{https://link.aps.org/doi/10.1103/PhysRevLett.118.010501}.

\bibitem[Childs et~al.(2017)Childs, Kothari, and Somma]{childs2017quantum}
Andrew~M. Childs, Robin Kothari, and Rolando~D. Somma.
\newblock Quantum algorithm for systems of linear equations with exponentially improved dependence on precision.
\newblock \emph{SIAM Journal on Computing}, 46\penalty0 (6):\penalty0 1920--1950, 2017.
\newblock \doi{10.1137/16M1087072}.
\newblock URL \url{https://doi.org/10.1137/16M1087072}.

\bibitem[Camps and Van~Beeumen(2022)]{camps2022fable}
Daan Camps and Roel Van~Beeumen.
\newblock Fable: Fast approximate quantum circuits for block-encodings.
\newblock In \emph{2022 IEEE International Conference on Quantum Computing and Engineering (QCE)}, pages 104--113, 2022.
\newblock \doi{10.1109/QCE53715.2022.00029}.

\bibitem[Sanavio et~al.(2025)Sanavio, Mauri, and Succi]{sanavio2024explicit}
Claudio Sanavio, Enea Mauri, and Sauro Succi.
\newblock Explicit quantum circuit for simulating the advection–diffusion–reaction dynamics.
\newblock \emph{IEEE Transactions on Quantum Engineering}, 6:\penalty0 1--12, 2025.
\newblock \doi{10.1109/TQE.2025.3544839}.

\bibitem[Grover and Rudolph(2002)]{grover2002creating}
Lov Grover and Terry Rudolph.
\newblock Creating superpositions that correspond to efficiently integrable probability distributions, 2002.
\newblock URL \url{https://arxiv.org/abs/quant-ph/0208112}.
\newblock DOI: \url{https://doi.org/10.48550/arXiv.quant-ph/0208112}.

\bibitem[Jennings et~al.(2025)Jennings, Lostaglio, Pallister, Sornborger, and Subaşı]{jennings2023efficient}
David Jennings, Matteo Lostaglio, Sam Pallister, Andrew~T Sornborger, and Yiğit Subaşı.
\newblock Randomized adiabatic quantum linear solver algorithm with optimal complexity scaling and detailed running costs, 2025.
\newblock URL \url{https://arxiv.org/abs/2305.11352}.
\newblock DOI: \url{https://doi.org/10.48550/arXiv.2305.11352}.

\bibitem[Selinger(2013)]{selinger2013quantum}
Peter Selinger.
\newblock Quantum circuits of $t$-depth one.
\newblock \emph{Phys. Rev. A}, 87:\penalty0 042302, Apr 2013.
\newblock \doi{10.1103/PhysRevA.87.042302}.
\newblock URL \url{https://link.aps.org/doi/10.1103/PhysRevA.87.042302}.

\bibitem[Jones(2013)]{jones2013low}
Cody Jones.
\newblock Low-overhead constructions for the fault-tolerant toffoli gate.
\newblock \emph{Phys. Rev. A}, 87:\penalty0 022328, Feb 2013.
\newblock \doi{10.1103/PhysRevA.87.022328}.
\newblock URL \url{https://link.aps.org/doi/10.1103/PhysRevA.87.022328}.

\bibitem[Barenco et~al.(1995)Barenco, Bennett, Cleve, DiVincenzo, Margolus, Shor, Sleator, Smolin, and Weinfurter]{barenco1995elementary}
Adriano Barenco, Charles~H. Bennett, Richard Cleve, David~P. DiVincenzo, Norman Margolus, Peter Shor, Tycho Sleator, John~A. Smolin, and Harald Weinfurter.
\newblock Elementary gates for quantum computation.
\newblock \emph{Phys. Rev. A}, 52:\penalty0 3457--3467, Nov 1995.
\newblock \doi{10.1103/PhysRevA.52.3457}.
\newblock URL \url{https://link.aps.org/doi/10.1103/PhysRevA.52.3457}.

\bibitem[Gosset et~al.(2025)Gosset, Kothari, and Zhang]{gosset2025}
David Gosset, Robin Kothari, and Chenyi Zhang.
\newblock Multi-qubit toffoli with exponentially fewer t gates, 2025.
\newblock URL \url{https://arxiv.org/abs/2510.07223}.
\newblock DOI: \url{https://doi.org/10.48550/arXiv.2510.07223}.

\bibitem[Gidney(2015)]{Gidney_2015}
Craig Gidney.
\newblock Constructing large increment gates, Jun 2015.
\newblock URL \url{https://algassert.com/circuits/2015/06/12/Constructing-Large-Increment-Gates.html}.

\bibitem[Gidney(2018)]{gidney2018halving}
Craig Gidney.
\newblock Halving the cost of quantum addition.
\newblock \emph{{Quantum}}, 2:\penalty0 74, June 2018.
\newblock ISSN 2521-327X.
\newblock \doi{10.22331/q-2018-06-18-74}.
\newblock URL \url{https://doi.org/10.22331/q-2018-06-18-74}.

\bibitem[Kitaev(1997)]{Kitaev_1997}
A~Yu Kitaev.
\newblock Quantum computations: algorithms and error correction.
\newblock \emph{Russian Mathematical Surveys}, 52\penalty0 (6):\penalty0 1191, dec 1997.
\newblock \doi{10.1070/RM1997v052n06ABEH002155}.
\newblock URL \url{https://doi.org/10.1070/RM1997v052n06ABEH002155}.

\bibitem[Dawson and Nielsen(2006)]{dawson_2006}
Christopher~M. Dawson and Michael~A. Nielsen.
\newblock The solovay-kitaev algorithm.
\newblock \emph{Quantum Info. Comput.}, 6\penalty0 (1):\penalty0 81–95, January 2006.
\newblock ISSN 1533-7146.
\newblock DOI: \url{https://doi.org/10.48550/arXiv.quant-ph/0505030}.

\bibitem[Wójcik(2012)]{wojcik2012}
Krzysztof~Piotr Wójcik.
\newblock Application of a numerical renormalization group procedure to an elementary anharmonic oscillator, 2012.
\newblock URL \url{https://arxiv.org/abs/1210.1703}.
\newblock DOI: \url{https://doi.org/10.48550/arXiv.1210.1703}.

\bibitem[G\l{}azek(2021)]{glazek2021}
Stanis\l{}aw~D. G\l{}azek.
\newblock Elementary example of exact effective-hamiltonian computation.
\newblock \emph{Phys. Rev. D}, 103:\penalty0 014021, Jan 2021.
\newblock \doi{10.1103/PhysRevD.103.014021}.
\newblock URL \url{https://link.aps.org/doi/10.1103/PhysRevD.103.014021}.

\bibitem[Vary et~al.(2022)Vary, Huang, Jawadekar, Sharaf, Harindranath, and Chakrabarti]{vary2022}
James~P. Vary, Mengyao Huang, Shreeram Jawadekar, Mamoon Sharaf, Avaroth Harindranath, and Dipankar Chakrabarti.
\newblock Critical coupling for two-dimensional ${\ensuremath{\phi}}^{4}$ theory in discretized light-cone quantization.
\newblock \emph{Phys. Rev. D}, 105:\penalty0 016020, Jan 2022.
\newblock \doi{10.1103/PhysRevD.105.016020}.
\newblock URL \url{https://link.aps.org/doi/10.1103/PhysRevD.105.016020}.

\bibitem[Pauli and Brodsky(1985)]{pauli_and_brodsky}
Hans-Christian Pauli and Stanley~J. Brodsky.
\newblock Solving field theory in one space and one time dimension.
\newblock \emph{Phys. Rev. D}, 32:\penalty0 1993--2000, Oct 1985.
\newblock \doi{10.1103/PhysRevD.32.1993}.
\newblock URL \url{https://link.aps.org/doi/10.1103/PhysRevD.32.1993}.

\bibitem[Simon et~al.(2024)Simon, Gustin, , Serafin, and Ralli]{lobe}
William~A. Simon, Carter Gustin, , Kamil Serafin, and Alexis Ralli.
\newblock {LOBE}.
\newblock \url{https://github.com/Tufts-QIQC/LOBE}, 2024.

\bibitem[Developers(2022)]{cirq}
Cirq Developers.
\newblock Cirq, December 2022.
\newblock URL \url{https://doi.org/10.5281/zenodo.7465577}.
\newblock DOI: \url{https://doi.org/10.5281/zenodo.7465577}.

\bibitem[Gustin et~al.(2024)Gustin, Simon, Guo, Serafin, and Ralli]{openparticle}
Carter Gustin, William~A. Simon, Ella Guo, Kamil Serafin, and Alexis Ralli.
\newblock {OpenParticle}.
\newblock \url{https://github.com/cgustin99/OpenParticle}, 2024.

\bibitem[Weaving et~al.(2025)Weaving, Ralli, Love, Succi, and Coveney]{Weaving_2025}
Tim Weaving, Alexis Ralli, Peter~J. Love, Sauro Succi, and Peter~V. Coveney.
\newblock Contextual subspace variational quantum eigensolver calculation of the dissociation curve of molecular nitrogen on a superconducting quantum computer.
\newblock \emph{npj Quantum Information}, 11\penalty0 (1), February 2025.
\newblock ISSN 2056-6387.
\newblock \doi{10.1038/s41534-024-00952-4}.
\newblock URL \url{http://dx.doi.org/10.1038/s41534-024-00952-4}.

\bibitem[Ralli et~al.(2023)Ralli, Weaving, Tranter, Kirby, Love, and Coveney]{PhysRevResearch.5.013095}
Alexis Ralli, Tim Weaving, Andrew Tranter, William~M. Kirby, Peter~J. Love, and Peter~V. Coveney.
\newblock Unitary partitioning and the contextual subspace variational quantum eigensolver.
\newblock \emph{Phys. Rev. Res.}, 5:\penalty0 013095, Feb 2023.
\newblock \doi{10.1103/PhysRevResearch.5.013095}.
\newblock URL \url{https://link.aps.org/doi/10.1103/PhysRevResearch.5.013095}.

\bibitem[Weaving et~al.(2023)Weaving, Ralli, Kirby, Tranter, Love, and Coveney]{doi:10.1021/acs.jctc.2c00910}
Tim Weaving, Alexis Ralli, William~M. Kirby, Andrew Tranter, Peter~J. Love, and Peter~V. Coveney.
\newblock A stabilizer framework for the contextual subspace variational quantum eigensolver and the noncontextual projection ansatz.
\newblock \emph{Journal of Chemical Theory and Computation}, 19\penalty0 (3):\penalty0 808--821, 2023.
\newblock \doi{10.1021/acs.jctc.2c00910}.
\newblock URL \url{https://doi.org/10.1021/acs.jctc.2c00910}.
\newblock PMID: 36689668.

\bibitem[Kirby et~al.(2021{\natexlab{b}})Kirby, Tranter, and Love]{Kirby2021contextualsubspace}
William~M. Kirby, Andrew Tranter, and Peter~J. Love.
\newblock Contextual {S}ubspace {V}ariational {Q}uantum {E}igensolver.
\newblock \emph{{Quantum}}, 5:\penalty0 456, May 2021{\natexlab{b}}.
\newblock ISSN 2521-327X.
\newblock \doi{10.22331/q-2021-05-14-456}.
\newblock URL \url{https://doi.org/10.22331/q-2021-05-14-456}.

\bibitem[Peduri and Ramacciotti(2024)]{grover-rudolph-github}
Anurudh Peduri and Debora Ramacciotti.
\newblock grover-rudolph, 2024.
\newblock URL \url{https://github.com/qubrabench/grover-rudolph}.

\bibitem[Glazek(2012)]{glazek2012}
Stanislaw~D. Glazek.
\newblock {Perturbative formulae for relativistic interactions of effective particles}.
\newblock \emph{Acta Phys. Polon. B}, 43:\penalty0 1843--1862, 2012.
\newblock \doi{10.5506/APhysPolB.43.1843}.

\bibitem[Nielsen and Chuang(2010)]{nielsen2001quantum}
Michael~A. Nielsen and Isaac~L. Chuang.
\newblock \emph{Quantum Computation and Quantum Information: 10th Anniversary Edition}.
\newblock Cambridge University Press, 2010.
\newblock DOI: \url{https://doi.org/10.1017/CBO9780511976667}.

\bibitem[Kitaev(1995)]{kitaev1995quantum}
A.~Yu. Kitaev.
\newblock Quantum measurements and the abelian stabilizer problem, 1995.
\newblock URL \url{https://arxiv.org/abs/quant-ph/9511026}.
\newblock DOI: \url{https://doi.org/10.48550/arXiv.quant-ph/9511026}.

\bibitem[M\"{o}tt\"{o}nen et~al.(2005)M\"{o}tt\"{o}nen, Vartiainen, Bergholm, and Salomaa]{mottonen2004transformation}
Mikko M\"{o}tt\"{o}nen, Juha~J. Vartiainen, Ville Bergholm, and Martti~M. Salomaa.
\newblock Transformation of quantum states using uniformly controlled rotations.
\newblock \emph{Quantum Info. Comput.}, 5\penalty0 (6):\penalty0 467–473, September 2005.
\newblock ISSN 1533-7146.
\newblock DOI: \url{https://doi.org/10.48550/arXiv.quant-ph/0407010}.

\bibitem[Dirac(1949)]{Dirac1949}
P.~A.~M. Dirac.
\newblock Forms of relativistic dynamics.
\newblock \emph{Rev. Mod. Phys.}, 21:\penalty0 392--399, Jul 1949.
\newblock \doi{10.1103/RevModPhys.21.392}.
\newblock URL \url{https://link.aps.org/doi/10.1103/RevModPhys.21.392}.

\end{thebibliography}
